\documentclass[letterpaper,titlepage,11pt]{article}

\usepackage{amssymb,amsmath,amsfonts}
\usepackage{epsfig}
\usepackage{ccaption}
\usepackage{graphicx}

\usepackage[
      dvipdfm,
      colorlinks=true,
      linkcolor=blue,
      urlcolor=blue,
      filecolor=black,
      citecolor=red,
      pdfstartview=FitV,
      pdftitle={},
        pdfauthor={Ryotaku Suzuki, Kentaro Tanabe},
        pdfsubject={},
        pdfkeywords={},
        pdfpagemode=None,
        bookmarksopen=true
      ]{hyperref}

\def\clock{{\count0=\time
           \divide\count0 60
           \ifnum\count0<10 0\fi\the\count0
           \multiply\count0 -60 \advance\count0 \time
           :\ifnum\count0<10 0\fi \the\count0
         }}
\newcommand{\timestamp}{{\small\vbox{\hbox{\tt\jobname.tex}
\hbox{\the\day/\the\month/\the\year, \clock}}}}

\newcommand{\beq}{\begin{equation}}
\newcommand{\eeq}{\end{equation}}
\newcommand{\bea}{\begin{eqnarray}}
\newcommand{\eea}{\end{eqnarray}}
\newcommand{\beqa}{\begin{eqnarray}}
\newcommand{\eeqa}{\end{eqnarray}}

\newcommand{\sR}{\mathsf{R}}

\newcommand{\sV}{\mathsf{V}}
\newcommand{\sD}{\mathsf{D}}
\newcommand{\cR}{\mathcal{R}}

\newcommand{\tH}{\text{H}}

\numberwithin{equation}{section}
\begin{document}

\begin{titlepage}
%\timestamp
\rightline{KEK-TH-1818, OCU-PHYS-423, AP-GR-123} 
%\leftline{}{\timestamp}
\vskip 1.cm
\centerline{\LARGE \bf Stationary black holes: Large $D$ analysis} 
%\medskip
\vskip 1.cm
\centerline{\bf Ryotaku Suzuki$^{a}$, Kentaro Tanabe$^{b}$}
\vskip 0.5cm
\centerline{\sl $^{a}$Department of Physics, Osaka City University, Osaka 558-8585, Japan}
\smallskip
\centerline{\sl $^{b}$Theory Center, Institute of Particles and Nuclear Studies, KEK,}
\centerline{\sl  Tsukuba, Ibaraki, 305-0801, Japan}
\vskip 0.5cm
\centerline{\small\tt ryotaku@sci.osaka-cu.ac.jp,\, ktanabe@post.kek.jp}

\vskip 1.cm
\centerline{\bf Abstract} \vskip 0.2cm 
\noindent 
We consider the effective theory of large $D$ stationary black holes. By solving the Einstein equations with a cosmological 
constant using the $1/D$ expansion in near zone of the black hole we obtain the effective equation for the stationary black hole. 
The effective equation describes the Myers-Perry black hole, bumpy black holes and, possibly, the black ring solution as its solutions. 
In this effective theory the black hole is represented as an embedded membrane in the background, e.g., Minkowski
or Anti-de Sitter spacetime and its mean curvature is given by the surface gravity redshifted by the background gravitational field 
and the local Lorentz boost. The local Lorentz boost property of the effective equation is observed also in the metric itself. In fact 
we show that the leading order metric of the Einstein equation in the $1/D$ expansion is generically regarded as a Lorentz boosted
Schwarzschild black hole. We apply this Lorentz boost property of the stationary black hole solution to solve perturbation equations. 
As a result we obtain an analytic formula for quasinormal modes of the singly rotating Myers-Perry black hole in the $1/D$ expansion.

\end{titlepage}
\pagestyle{empty}
\small
\tableofcontents
\normalsize
\newpage
\pagestyle{plain}
\setcounter{page}{1}

\section{Introduction}

The infinite dimensional limit of General Relativity gives not only various interesting pictures of black hole physics, 
but also new useful analytic method to solve gravitational problem \cite{Asnin:2007rw,Emparan:2013moa,Emparan:2013xia,
Emparan:2013oza,Emparan:2014cia,Emparan:2014jca,Emparan:2014aba,Emparan:2015rva,Emparan:2015hwa,Bhattacharyya:2015dva}. 
The perturbation problem, such as quasinormal modes (QNMs), has been solved for general static black holes 
\cite{Emparan:2014aba,Emparan:2015rva} and the Myers-Perry black hole with equal spin \cite{Emparan:2014jca}. 
As for non-linear problem the effective theory for the large $D$ black hole has been considered recently in 
\cite{Emparan:2015hwa,Bhattacharyya:2015dva} and some non-trivial static solutions were constructed.
In this paper we study the effective theory of the large $D$ stationary black hole in asymptotically 
flat or Anti-de Sitter (AdS) background.   

The reason for the simplification of the gravity at large $D$ is simple \cite{Emparan:2013moa}. The radial gradient 
of the gravitational field around black holes becomes very large with $O(D/r_{0})$ at large $D$. $r_{0}$ is the black hole size. 
So the  tangential dynamics along the horizon becomes sub-dominant, and the system reduces to the ordinarily differential 
equation system with respect to the radial direction. In addition the appearance of the hierarchical sale, $r_{0}/D$ and $r_{0}$, 
gives two separated excitations around the black hole, that is, the non-decoupled mode with the frequency $\omega r_{0}=O(D)$ and 
decoupled mode $\omega r_{0}=O(D^{0})$. While the non-decoupled mode is universal for black holes \cite{Emparan:2014cia}, 
the decoupled mode has a particular and interesting feature for each black hole. For instance the instability of the 
ultraspinning back holes and black branes are belonging to this decoupled sector \cite{Emparan:2013moa,Emparan:2014jca}. 
These two facts, the dominance of the radial dynamics and appearance of two hierarchical scales, make General Relativity simple and 
analytically solvable, but still non-trivial at large $D$. 

Achievements in QNMs by large $D$ method motivate us to construct the effective theory for general black hole solutions in the $1/D$
expansion as a next step. The gravitational field of the black hole is localized in the near region of the horizon due to its large 
gradient at $D\rightarrow\infty$. Then the black hole can be regarded as a membrane $\Sigma_{B}$ in the background spacetime like the 
membrane paradigm \cite{membrane,Price:1986yy,Emparan:2009at} at this limit. The physical feature of $\Sigma_{B}$ is determined by 
solving Einstein equations in near zone of the black hole. If we consider the decoupled mode dynamics of the black hole, the Einstein 
equations can be solved consistently only in near zone. The obtained solution gives the boundary condition for $\Sigma_{B}$ 
as physical properties. Then we can construct the effective theory of the large $D$ black hole as the membrane $\Sigma_{B}$ in the 
background spacetime. This construction of the effective theory has been considered in \cite{Emparan:2015hwa} for static black holes 
and for more general solutions with time dependences in \cite{Bhattacharyya:2015dva}. 

Here we study the effective theory for general large $D$ stationary black holes in the Minkowski or AdS background such as
the (AdS)Myers-Perry black hole, bumpy black holes and black ring. These solutions have been known as exact solutions 
\cite{Myers:1986un,Hawking:1998kw,Gibbons:2004uw} or numerical solutions \cite{Dias:2014cia,Emparan:2014pra,Kleihaus:2012xh}. 
We obtain the general analytic metric for the stationary black holes in the $1/D$ expansion, which describes these solutions in the 
unified manner. The solution gives the equation for the embedding of $\Sigma_{B}$ in the background. This embedding determines the 
horizon topology and its geometric shape. Then the effective equation for the embedding of $\Sigma_{B}$ is given by the equation 
for the mean curvature $K$ of $\Sigma_{B}$ as
%============<Equation>=============%
%
\begin{eqnarray}
\gamma^{-1}K\Big|_{\Sigma_{B}}=2\kappa, \label{EFTp}
\end{eqnarray}
%
%==================================%
where $\gamma$ is a redshift factor containing both the gravitational redshift on the membrane from the background geometry 
and the Lorentz redshift from the local rotation along the membrane, {\it i.e.},
%============<Equation>=============%
%
\begin{eqnarray}
\gamma^{-1}=\sqrt{-g_{tt}(1-\Omega_H^2\cR_\phi^2)}.
\label{gammadef}
\end{eqnarray}
%
%==================================%
$\kappa$ and $\Omega_{H}$ are the surface gravity and horizon angular velocity of the black hole respectively. 
$\cR_{\phi}$ is the rotation radius of $\Sigma_{B}$ defined by
%============<Equation>=============%
%
\begin{eqnarray}
\cR_{\phi}=\sqrt{-\frac{g_{\phi\phi}}{g_{tt}}}\Bigg|_{\Sigma_{B}}.
\end{eqnarray}
%
%==================================%. 
A solution of this effective equation gives the black hole metric which solves the leading order Einstein equation in the $1/D$ expansion. 
This effective equation is one of main results of this paper. In the static case this effective equation reduces to one of
static black holes \cite{Emparan:2015hwa}.  

The effective equation implies that the mean curvature of $\Sigma_{B}$ is regarded as the surface gravity redshifted 
by the local Lorentz boost. This local Lorentz boost property of the effective equation turns out to be also the property of the leading order
metric itself. Actually we find that the leading order metric in near zone is obtained as a Lorentz boosted Schwarzschild black hole. 
This property of the leading order solution would not hold if we go to the higher order structure in the $1/D$ expansion. However 
it can still play a crucial role in solving the Einstein equations. As one interesting example, we show that perturbation equations 
on the obtained leading order general stationary black hole can be explicitly solved by using the solution of one of the Schwarzschild 
black hole. This perturbation solution, of course, contains the solution of perturbation equations of the Myers-Perry black hole. 
Going to the higher order of perturbation equations in the $1/D$ expansion we obtain an analytic formula for QNMs frequency of the 
singly rotating Myers-Perry black hole both for the axisymmetric and non-axisymmetric perturbations. As a result the threshold angular 
momentum of the ultraspinning instability of the singly rotating Myers-Perry black hole is found to be 
%============<Equation>=============%
%
\begin{eqnarray}
\frac{a^{2}}{r^{2}_{0}}=2k_{S}-1,~~k_{S}=1,2,3,...
\end{eqnarray}
%
%==================================%$
where $r_{0}$ is the horizon radius and $k_{S}$ is a positive integer. This result can be obtained in a much easier and simpler manner 
by examining the effective equation (\ref{EFTp}) for the embedding of the bumpy black hole. Note that there is no dynamical instability 
at the threshold angular momentum of the case $k_{S}=1$. The threshold for $k_{S}=1$ does not imply any existence of a bifurcation zero-mode. 
There is only a trivial perturbation along the Myers-Perry solution branch. 

The remaining of this paper is organized as follows. In section \ref{2} we consider the effective theory for stationary black holes 
at large $D$. We give the leading order solution and its boost representation. The effective equation for the large $D$ stationary
black holes is derived from this leading order metric. We study a simple solution and non-trivial solutions of the effective equation there. 
In section \ref{3} we apply the boost property of the leading order solution to solve perturbation equations on the general stationary 
solutions. The section \ref{4} pursues the consideration for the perturbation in the previous section in more detail. We concentrate 
on the perturbation of the Myers-Perry black hole and obtain the QNMs frequency analytically. We close this paper by discussion 
and outlook in section \ref{5}. The appendices contain some technical details of calculations for the main part.

\section{Large $D$ stationary black holes}\label{2}

We study the $D$ dimensional stationary black hole solution and its large $D$ effective theory. 
The following analysis is performed in the similar way with static case \cite{Emparan:2015hwa}
\footnote{
But the index notation is changed from \cite{Emparan:2015hwa}
}. We use the small expansion parameter $1/n$ where
%============<Equation>=============%
%
\begin{eqnarray}
n=D-3,
\end{eqnarray}
%
%==================================%
instead of $1/D$.

\subsection{Set up}

The large $D$ black hole has a very large gradient along the radial direction, $\rho$, with $O(n)$. Then our metric ansatz
for the $D$ dimensional stationary black hole is given by
%============<Equation>=============%
%
\begin{eqnarray}
ds^{2} = \frac{N(\rho,z)^{2}}{n^{2}}d\rho^{2} +g_{ab}dx^{a}dx^{b}.
\end{eqnarray}
%
%==================================%
$z$ is one of $D-1$ dimensional coordinate $x^{a}$. In this paper we consider the metric which has only one inhomogeneous 
coordinate, $z$, other than $\rho$ for simplicity. The generalization to several inhomogeneous coordinate case would be 
straightforward as done in \cite{Emparan:2015hwa}. Then the Einstein equation with a cosmological constant
%============<Equation>=============%
%
\begin{eqnarray}
\Lambda=-\frac{(D-1)(D-2)}{L^{2}}
\end{eqnarray}
%
%==================================%
is decomposed on $\rho=\text{constant}$ surface to 
%============<Equation>=============%
%
\begin{eqnarray}
&&-R+K^2-K^a_{~b} K^b_{~a}-\frac{(D-1)(D-2)}{L^{2}}=0,\label{Hconst} \\
&&\nabla_{a}K^{a}_{~b}-\nabla_{b}K=0, \label{Mconst} \\
&&\frac{n}{N}\partial_\rho K^a_{~b}=-K K^a_{~b} + R^a_{~b}+\frac{D-1}{L^{2}}\delta^{a}_{~b} 
 -\frac{1}{N}\nabla^a\nabla_{b} N, \label{evolEq} 
\end{eqnarray}
%
%==================================%
and
%============<Equation>=============%
%
\begin{eqnarray}
K^a_{~b} =\frac{n}{2N(\rho,z)}g^{ac}\partial_\rho g_{cb}.
\label{Kdef}
\end{eqnarray}
%
%==================================%
$R_{ab}$ and $\nabla_{a}$ is the Ricci curvature and covariant derivative of $D-1$ dimensional metric $g_{ab}$ respectively. 
$K_{ab}$ is the extrinsic curvature of $D-1$ dimensional $\rho=\text{constant}$ surface. 

We consider stationary solutions with a rotation. Then the $D-1$ dimensional metric ansatz for $g_{ab}$ is 
%============<Equation>=============%
%
\begin{eqnarray}
ds^{2}_{D-1} &=& g_{ab}dx^{a}dx^{b} \notag \\
&=& -A(\rho,z)^{2} dt^{2} +F(\rho,z)^{2}(d\phi-W(\rho,z)dt)^{2} \notag \\
&&~~~~~~~~~~~~~~~~~~~~~~~~~~~
+G(\rho,z)^{2}dz^{2} +H(\rho,z)^{2}q_{AB}dx^{A}dx^{B},
\end{eqnarray}
%
%==================================%
where $x^{A}$ and $q_{AB}$ are a coordinate and the standard metric on $S^{D-4}$ respectively. 
We solve the Einstein equations for $K_{ab}$ and metric functions $A(\rho,z)$, $F(\rho,z)$, $W(\rho,z)$, $G(\rho,z)$ and
$H(\rho,z)$ by using the $1/D$ expansion. In the following we specify large $D$ behaviors of these functions and 
boundary conditions as a set up. 

\paragraph{Large $D$ behavior}

At first each metric functions should be $O(n^{0})$. In particular
%============<Equation>=============%
%
\begin{eqnarray}
g_{zz}=O(n^{0})
\end{eqnarray}
%
%==================================% $g_{zz}=O(D^{0})$ 
implies that the derivative with respect to $z$ is $O(n^{0})$. The hierarchical scaling between $\rho-$ and $z-$ dependence 
makes the leading order equation at $D\rightarrow\infty$ be ordinary differential equations with respect to $\rho$. 
Next we assume that the leading order of $G(\rho,z)$ and $H(\rho,z)$ has only $z$-dependence and no $\rho$-dependence. 
This assumption is motivated by the exact solution of the Myers-Perry black hole (see Appendix \ref{A}) and the fact 
that the large dimensionality of the sphere gives another $D$ factor in $\rho$-dependent terms through the trace on $S^{D-4}$. 
Thus we assume
%============<Equation>=============%
%
\begin{eqnarray}
K^{z}_{~z}, K^{A}_{~B}=O(n^{0}),
\end{eqnarray}
%
%==================================%
while the leading order of $A(\rho,z)$, $F(\rho,z)$ and $W(\rho,z)$ has $\rho$-dependences as
%============<Equation>=============%
%
\begin{eqnarray}
K^{t}_{~t},K^{t}_{~\phi},K^{\phi}_{~\phi},K=O(n).
\end{eqnarray}
%
%==================================%
From these assumptions we can derive the scaling law of various geometric quantities at large $D$. For example, 
the Ricci scalar $R$ of $g_{ab}$ is $O(n^{2})$. 

\paragraph{Boundary condition}
The boundary conditions are imposed on the horizon and at the asymptotic infinity of near zone defined by $n\gg \rho\gg 1$
\footnote{
The effective theory of the large $D$ black holes is described as the membrane physics seen from the far zone \cite{Emparan:2015hwa}. 
In the overlap zone $n\gg \rho\gg 1$ we can obtain physical properties of the membrane by the near zone solution as the boundary condition.
}. At the horizon the extrinsic curvature has a singular behavior essentially only in one component. This condition guarantees the 
regularity of following quantities 
%============<Equation>=============%
%
\begin{eqnarray}
K-K^{t}_{~t}-K^{\phi}_{~\phi},~
K^{z}_{~z},~
K^{A}_{~B}, \label{BCHo}
\end{eqnarray}
%
%==================================% 
on the horizon. For the static case $K^{\phi}_{~\phi}$ should be regular on the horizon. In this paper we consider the asymptotically 
flat or AdS black hole. Then the boundary condition at asymptotic infinity of near zone is
%============<Equation>=============%
%
\begin{eqnarray}
A(\rho,z)=\sV_{0}(z)+O(1/n,e^{-2\rho}),~~W(\rho,z)=O(e^{-2\rho}). \label{BCIn}
\end{eqnarray}
%
%==================================%
Note that the Newton potential of the black hole is $O(e^{-2\rho})$ in our gauge. So this asymptotic boundary condition means 
the leading order metric becomes the flat or AdS metric other than the potential term at $\rho \gg 1$.  
While $g_{t\phi}$ should be damping exponentially in $\rho$ at all order of $1/D$, $g_{tt}$ has some terms which does not decay 
at the asymptotic infinity in the higher order solution in the $1/D$ expansion when we consider the asymptotically AdS case 
(See Appendix \ref{A}). Since we solve the leading order equations in this section, we do not explore this structure in detail. 
The boundary condition (\ref{BCIn}) is enough for the leading order solutions.

\subsection{Solving the leading order equation} 

We solve the leading order equation of the Einstein equation in the $1/D$ expansion by performing $\rho$ integrations. 
This integration gives free functions of $z$ as integration functions. They consist of the effective theory of large $D$ stationary 
black holes.

Our large $D$ scaling assumptions give following large $D$ behaviors
%============<Equation>=============%
%
\begin{eqnarray}
&N(\rho,z)=N_{0}(z)+O(1/n), \\
&G(\rho,z)=1+O(1/n), \\
&H(\rho,z)=\cR_{\text{H}}(z)+O(1/n),
\end{eqnarray}
%
%==================================% 
where $\rho$-dependence of the leading order of $N(\rho,z)$ and $z$-dependence of the leading order of 
$G(\rho,z)$ have been absorbed into the definition of $\rho$ and $z$ coordinate respectively. 
The leading order of the Ricci scalar of $g_{ab}$ can be calculated from this ansatz as
%============<Equation>=============%
%
\begin{eqnarray}
R=n^{2}\left(\frac{1}{L^{2}}+\frac{1}{\cR_{\tH}(z)^{2}}-\frac{\cR'_{\tH}(z)^{2}}{\cR_{\tH}(z)^{2}}\right),
\end{eqnarray}
%
%==================================%
where here and in the following we omit $O(1/n)$ symbols showing the existence of subleading corrections in $1/D$ for the simplicity 
in the representation. Then the trace part of eq. (\ref{evolEq}) is integrated to
%============<Equation>=============%
%
\begin{eqnarray}
K=\frac{n}{r_{0}(z)}\coth{\left(\frac{N_{0}(z)}{r_{0}(z)}(\rho-\rho_{0}(z)) \right)},
\end{eqnarray}
%
%==================================%
where we defined $r_{0}(z)$ by
%============<Equation>=============%
%
\begin{eqnarray}
\frac{1}{r_{0}(z)^{2}}=\frac{1}{L^{2}}+\frac{1}{\cR_{\tH}(z)^{2}}-\frac{\cR'_{\tH}(z)^{2}}{\cR_{\tH}(z)^{2}}.
\label{r0def}
\end{eqnarray}
%
%==================================%
$\rho_{0}(z)$ is the horizon position $\rho=\rho_{0}(z)$, so we can set to $\rho_{0}(z)=0$ by a gauge choice. The residual gauge 
in $\rho$ coordinate can be used to have
%============<Equation>=============%
%
\begin{eqnarray}
N_{0}(z)=2r_{0}(z).
\end{eqnarray}
%
%==================================%
Note that this choice for $N_{0}(z)$ is different from \cite{Emparan:2015hwa} by factor two. This is just a gauge choice and it
does not affect physical properties. The equations for $K^{t}_{~t}$, $K^{t}_{~\phi}$ and $K^{\phi}_{~\phi}$ in eq. (\ref{evolEq}) 
are solved by
%============<Equation>=============%
%
\begin{eqnarray}
K^{t}_{~t}=n\frac{C_{tt}(z)}{r_{0}(z)\sinh{2\rho}},~
K^{t}_{~\phi}=n\frac{C_{t\phi}(z)}{r_{0}(z)\sinh{2\rho}},~
K^{\phi}_{~\phi}=n\frac{C_{\phi\phi}(z)}{r_{0}(z)\sinh{2\rho}}.
\end{eqnarray}
%
%==================================%
$C_{tt}(z)$, $C_{t\phi}(z)$ and $C_{\phi\phi}(z)$ are integration functions with respect to the $\rho$ integration. The boundary 
condition on the horizon (\ref{BCHo}) gives 
%============<Equation>=============%
%
\begin{eqnarray}
1-C_{tt}(z)-C_{\phi\phi}(z)=0.
\end{eqnarray}
%
%==================================%
In the following we eliminate $C_{\phi\phi}(z)$ by this condition. Then we can integrate other components of eqs. (\ref{evolEq})
and (\ref{Kdef}), and obtain the following leading order solutions
%============<Equation>=============%
%
\begin{eqnarray}
&&
A(\rho,z)^{2} = \frac{A_{0}(z)^{2}F_{0}(z)^{2}\tanh^{2}{\rho}}{F_{0}(z)^{2}-A_{0}(z)^{2}C_{t\phi}(z)^{2}\tanh^{2}{\rho}},\\
&&
F(\rho,z)^{2} =F_{0}(z)^{2}-A_{0}(z)C_{t\phi}(z)^{2}\tanh^{2}{\rho},\\
&&
W(\rho,z) = \frac{F_{0}(z)^{2}(1-C_{tt}(z))-A_{0}(z)^{2}C_{t\phi}(z)^{2}C_{tt}(z)\tanh^{2}{\rho}}
{C_{t\phi}(z)F_{0}(z)^{2}-A_{0}(z)^{2}C_{t\phi}(z)^{3}\tanh^{2}{\rho}},
\end{eqnarray}
%
%==================================% 
and
%============<Equation>=============%
%
\begin{eqnarray}
&&
G(\rho,z) = 1 - \frac{2r_{0}(z)^{2}}{n}\left(\frac{\cR_{\tH}''(z)}{\cR_{\tH}(z)}-\frac{1}{L^{2}}\right)\log{(\cosh{\rho})}, \label{Gsol}  \\
&&
H(\rho,z) = \cR_{\tH}(z)\left( 1 +\frac{2}{n}\log{(\cosh{\rho})} \right). \label{Hsol}
\end{eqnarray}
%
%==================================%
$A_{0}(z)$ and $F_{0}(z)$ are integration functions. The integration functions of $G(\rho,z)$ and $H(\rho,z)$ are absorbed into 
the $O(1/n)$ redefinition of $z$ coordinate and $\cR_{\tH}(z)$. The scalar constraint equation (\ref{Hconst}) could be satisfied by 
choosing the integration function in $W(\rho,z)$ properly. We can see that these solutions reduce to the static solution 
\cite{Emparan:2015hwa} if we set to 
%============<Equation>=============%
%
\begin{eqnarray}
C_{tt}(z)=1,~~
C_{t\phi}=0.
\end{eqnarray}
%
%==================================%$K_{tt}(z)=-1$ and $K_{t\phi}(z)=0$. 
We could solve the scalar constraint equation (\ref{Hconst}) and all evolution equations of eqs. (\ref{evolEq}) and (\ref{Kdef}). 
The remaining equation is the vector constraint equation (\ref{Mconst}). The vector constraint equation gives an additional condition 
between integration functions as
%============<Equation>=============%
%
\begin{eqnarray}
\frac{d}{dz}\log{\left( \frac{A_{0}(z)C_{tt}(z)}{r_{0}(z)} \right)}-(1-C_{tt}(z))\frac{d}{dz}\log{\left(\frac{C_{tt}(z)}{C_{t\phi}(z)}\right)}
=0. \label{vconstLO}
\end{eqnarray}
%
%==================================%
This equation, at first glance, seems to be a non-trivial condition. However we can make this equation trivial in the following sense.
To realize this we observe the surface gravity $\kappa$ and angular velocity $\Omega_{H}$ of the horizon of the leading order solution. 
These quantities are read as
%============<Equation>=============%
%
\begin{eqnarray}
\kappa = \frac{n}{2}\frac{A_{0}(z)}{r_{0}(z)},~~
\Omega_{H} = \frac{1-C_{tt}(z)}{C_{t\phi}(z)}.
\label{kOccond}
\end{eqnarray}
%
%==================================% 
These quantities should be constant on the horizon \cite{Bardeen:1973gs}. Then, using the condition that $\kappa$ and $\Omega_{H}$ 
are constant, we can see that the constraint equation (\ref{vconstLO}) is satisfied automatically. Thus the vector constraint equation 
becomes trivial under the constancy condition of the surface gravity and angular velocity of the horizon. This is the same situation 
with the static case \cite{Emparan:2015hwa}. In the static case the vector constraint is equivalent to the constancy condition of the 
surface gravity. In stationary case the vector constraint can be satisfied if we use the constancy condition of not only the surface gravity 
but also the horizon angular velocity. One may feel that this statement is strange since originally the constancy of the surface 
gravity and angular velocity of the horizon was shown by using the Einstein equations \cite{Bardeen:1973gs}. It means that the constancy
condition of the surface gravity and horizon angular velocity should be derived conditions, not additionally imposed one to satisfy the 
Einstein equation. In fact the constancy conditions (\ref{kOccond}) can be derived if we use eq. (\ref{vconstLO}) and another equation 
which is obtained at the next-to-leading order of $1/D$ expansions. 
The regularity condition of the next-to-leading order correction of $K^{z}{}_{z}$ on the horizon requires one additional condition 
between $A_{0}(z)$ and $F_{0}(z)$ (see eq. (\ref{Feqapp})). Then these two equations give the constancy 
condition of the surface gravity and horizon angular velocity as eq. (\ref{kOccond}). The detail of this argument will be given
in the appendix \ref{Am}. 
So, here, we assume the constancy conditions (\ref{kOccond}) on the leading order solution in advance. If constancy conditions 
(\ref{kOccond}) are not satisfied, the next-to-leading order solutions becomes singular at the horizon although the leading order 
solutions still regular. So we can replace $A_{0}(z)$ and $C_{t\phi}(z)$ by $C_{tt}(z)$, $\kappa$ and $\Omega_{\tH}$ 
using eq. (\ref{kOccond}). 

Finally we impose the boundary condition at the asymptotic infinity (\ref{BCIn}) on the leading order solutions. Then we find
%============<Equation>=============%
%
\begin{eqnarray}
C_{tt}(z)=\frac{\sV_{0}(z)^{2}}{4\hat{\kappa}^{2}r_{0}(z)^{2}},~~
F_{0}(z)^{2}=\frac{\sV_{0}(z)^{2}-4\hat{\kappa}^{2}r_{0}(z)^{2}}{4\hat{\kappa}^{2}\Omega_{H}^{2}r_{0}(z)^{2}},
\end{eqnarray}
%
%==================================%
where we defined the reduced surface gravity $\hat{\kappa}$ by 
%============<Equation>=============%
%
\begin{eqnarray}
\kappa=n\hat{\kappa}.
\end{eqnarray}
%
%==================================%$\kappa=D\hat{\kappa}$. 
As a result we obtain regular leading order solutions as
%============<Equation>=============%
%
\begin{eqnarray}
&&
A(\rho,z)^{2} = \frac{4\hat{\kappa}^{2}r_{0}(z)^{2}\sV_{0}(z)^{2}\sinh^{2}{\rho}}{\sV_{0}(z)^{2}+4\hat{\kappa}^{2}r_{0}(z)^{2}\sinh^{2}{\rho}},\\
&&
F(\rho,z)^{2} = (\sV_{0}(z)^{2}-4\hat{\kappa}^{2}r_{0}(z)^{2})~
\frac{\sV_{0}(z)^{2}+4\hat{\kappa}^{2}r_{0}(z)^{2}\sinh^{2}{\rho}}{4\Omega_{H}^{2}\hat{\kappa}^{2}r_{0}(z)^{2}\cosh^{2}{\rho}},\\
&&
W(\rho,z) = \frac{\Omega_{H}}{\sV_{0}(z)^{2}+4\hat{\kappa}^{2}r_{0}(z)^{2}\sinh^{2}{\rho}},
\end{eqnarray}
%
%==================================%  
and eqs. (\ref{Gsol}) and (\ref{Hsol}) for $G(\rho,z)$ and $H(\rho,z)$.
One can check that the (AdS)Myers-Perry black hole is described by this leading order solution as seen in Appendix \ref{A}.

\subsection{Effective equation}

In the same spirit with \cite{Emparan:2015hwa} we can obtain the effective equation for the stationary black hole from the 
leading order solution of near zone obtained above. The near zone metric is matched with the far zone metric at overlap zone
$n\gg \rho \gg 1$ under the boundary condition (\ref{BCIn}). The near zone metric on $\rho=\text{constant}$ surface $\Sigma_{B}$ 
at overlap zone is  
%============<Equation>=============%
%
\begin{eqnarray}
&&
ds^{2}|_{\Sigma_{B}}=-\sV_{0}(z)^{2}dt^{2} +\frac{\sV_{0}(z)^{2}-4\hat{\kappa}^{2}r_{0}(z)^{2}}{\Omega^{2}_{H}}d\phi^{2}
\notag \\
&&~~~~~~~~~~~~~~~~~~~~~~~~~~~~~~~~~~~~~
+dz^{2}
+\cR_{\tH}(z)^{2}q_{AB}dx^{A}dx^{B},
\label{Match}
\end{eqnarray}
%
%==================================%
where we neglect the potential terms $O(e^{-2\rho})$. Notice that the trace of the extrinsic curvature, $K$, is read as
%============<Equation>=============%
%
\begin{eqnarray}
K\Big|_{\Sigma_{B}}=\frac{n}{r_{0}(z)},
\label{Kval}
\end{eqnarray}
%
%==================================%
where $r_{0}(z)$ is given in eq. (\ref{r0def}). The gravitational redshift factor of $\Sigma_{B}$ by the background spacetime, 
$g_{tt}$, is 
%============<Equation>=============%
%
\begin{eqnarray}
\sqrt{-g_{tt}}|_{\Sigma_{B}}=\sV_{0}(z).
\label{Krel}
\end{eqnarray}
%
%==================================%
Using eqs. (\ref{Kval}) and (\ref{Krel}), we can see that the leading order solution in near zone gives the following embedding equation for 
$\Sigma_{B}$ at overlap zone as
%============<Equation>=============%
%
\begin{eqnarray}
\sqrt{-g_{tt}(1-\Omega_{H}^{2}\cR_{\phi}^{2})}~K\Big|_{\Sigma_{B}}=2\kappa, \label{EFT}
\end{eqnarray}
%
%==================================%
where we defined the rotational radius $\cR_{\phi}$ by
%============<Equation>=============%
%
\begin{eqnarray}
\cR_{\phi}=\sqrt{-\frac{g_{\phi\phi}}{g_{tt}}}~\Bigg|_{\Sigma_{B}}.
\end{eqnarray}
%
%==================================%
The effective equation (\ref{EFT}) gives the mean curvature of $\Sigma_{B}$ as the surface gravity redshifted
by the gravitational effect and the local Lorentz boost effect with the boost velocity $v_{\phi}$ defined by
%============<Equation>=============%
%
\begin{eqnarray}
v_{\phi}=\Omega_{H}\cR_{\phi}.
\end{eqnarray}
%
%==================================%
The regular leading order solution contains arbitrary functions, $\sV_{0}(z)$ and $\cR_{\text{H}}(z)$. These functions
can be determined by specifying the embedding into the background. If one consider the embedding into the Minkowski background, 
we have $\sV_{0}(z)=1$ and there is only one remaining function $\cR_{\text{H}}(z)$ which should satisfy eq. (\ref{EFT}). 
On the other hand, if one considers the embedding into AdS background, the embedding would give a non-trivial redshift factor 
$\sV_{0}(z)$ of $\Sigma_{B}$. Thus the embedding into the AdS background should satisfy eqs. (\ref{Krel}) and (\ref{EFT})
for $\sV_{0}(z)$ and $\cR_{\text{H}}(z)$.

The effective equation (\ref{EFT}) for the stationary black hole can be still interpreted by a soap-film equation 
as one for the static case \cite{Emparan:2015hwa}
\footnote{
This interpretation of the effective equation (\ref{EFT}) is due to Roberto Emparan. We thank him for sharing it with us.
}. The membrane (soap-film) between two fluids satisfies the following 
equation for the mean curvature $K$ of the membrane (see eq. (2.15) in \cite{Caldarelli:2008mv})
%============<Equation>=============%
%
\begin{eqnarray}
K= \frac{\gamma T\Delta s-\Delta\rho}{\sigma},
\label{EFTsoap}
\end{eqnarray}
%
%==================================%
where $\sigma$ is the surface tension, $T$ is the temperature, $\Delta s$ and $\Delta\rho$ are the differences between 
the entropy density and the energy density of the fluids at both sides of the membrane. $\gamma$ is the same redshift 
factor as in eq. (\ref{gammadef}): in eq. (2.14) in \cite{Caldarelli:2008mv}, one just has to extract $\xi^2=-g_{tt}$.
If we now consider a case in which $\Delta\rho$ is negligible, then the equation (\ref{EFTsoap}) becomes
%============<Equation>=============%
%
\begin{eqnarray}
\gamma^{-1}K = \frac{ T\Delta s }{\sigma},
\end{eqnarray}
%
%==================================%
so if we identify $2\kappa= T\Delta s/\sigma$, then this is the same as eq. (\ref{EFT}) (of course in this identification 
$T$, $\Delta s$, and $\sigma$ are not independently determined, instead only their combination into $\kappa$ is fixed). 

There are two remarks on the metric (\ref{Match}) at $\Sigma_{B}$. One is that the metric (\ref{Match}) does not need to 
have a component of $g_{t\phi}$. The metric (\ref{Match}) describes the background metric where the solution is embedded, 
and it does not need to be rotating itself although the solution is rotating. Thus the metric (\ref{Match}) does not need 
to have $g_{t\phi}$ component. Another remark is about the static limit. The static limit corresponds to the zero horizon 
angular velocity $\Omega_{H}=0$. Then the metric (\ref{Match}) requires
%============<Equation>=============%
%
\begin{eqnarray}
\sV_{0}(z)^{2}=4\hat{\kappa}^{2}r_{0}(z)^{2}
\label{STcond}
\end{eqnarray}
%
%==================================%  
at the static limit. One may think that the appearance of this additional condition is curious. However one can immediately find
that the condition (\ref{STcond}) is nothing but the effective equation of general static solutions \cite{Emparan:2015hwa}
%============<Equation>=============%
%
\begin{eqnarray}
\sqrt{-g_{tt}}K\Big|_{\Sigma_{B}}=2\kappa,
\end{eqnarray}
%
%==================================%
by using eqs. (\ref{Kval}) and (\ref{Krel}). This embedding equation can be obtained also by the static limit of eq. (\ref{EFT}). 
So there is no appearance of new additional condition at the static limit. 

The embedding in the Minkowski background has a constant redshift factor as
%============<Equation>=============%
%
\begin{eqnarray}
\sV_{0}(z)=1.
\end{eqnarray}
%
%==================================%
For the static embedding in this Minkowski background we have only one unique embedding by a round sphere corresponding 
to the Schwarzschild black hole \cite{Emparan:2015hwa,hsiang}. This fact is consistent with the uniqueness theorem 
of the static solution \cite{Gibbons:2002av}. On the other hand, however, the embedding of the stationary solution allows 
various embeddings as seen below. This is due to the appearance of new degree of freedom in the horizon deformation 
described by $\Omega_{\tH}\cR_{\phi}$ in the effective equation.

\subsubsection{Ellipsoidal embedding}

As one application of the effective equation (\ref{EFT}) we consider an ellipsoidal embedding in the Minkowski background.
We embed the leading order solution by $r=r(\theta)$ in the Minkowski background using the ellipsoidal coordinate. The flat 
metric in the ellipsoidal coordinate is
%============<Equation>=============%
%
\begin{eqnarray}
&&
ds^{2} = -dt^{2}+\frac{r^{2}+a^{2}\cos^{2}{\theta}}{r^{2}+a^{2}}dr^{2}+ 
(r^{2}+a^{2}\cos^{2}{\theta})d\theta^{2}  \notag \\
&&~~~~~~~~~~~~~~~~~~~~~~~~~~~~~
+(r^{2}+a^{2})\sin^{2}{\theta}d\phi^{2} +r^{2}\cos^{2}{\theta}d\Omega^{2}_{D-4}
.\label{ellip}
\end{eqnarray}
%
%==================================%
$a$ is a parameter describing the oblateness of the ellipsoidal. We set
%============<Equation>=============%
%
\begin{eqnarray}
\hat{\kappa} = \frac{1}{2},
\end{eqnarray}
%
%==================================%
by using the normalization of near zone $t$-coordinate for simplicity. The relation $z$ and $\theta$ can be seen from 
eqs. (\ref{Match}) and (\ref{ellip}) with $r=r(\theta)$ as
%============<Equation>=============%
%
\begin{eqnarray}
\frac{dz}{d\theta} = \sqrt{\frac{(\cR_{\tH}(\theta)^{2} +a^{2}\cos^{4}{\theta})(\cR_{\tH}(\theta)^{2}+(\cR'_{\tH}(\theta)+\tan{\theta}\cR_{\tH}(\theta))^{2}+a^{2}\cos^{2}{\theta})}{\cos^{2}{\theta}(\cR_{\tH}(\theta)^{2}+a^{2}\cos^{2}{\theta})}},
\end{eqnarray}
%
%==================================%
where we used the relation 
%============<Equation>=============%
%
\begin{eqnarray}
r(\theta)\cos{\theta}=\cR_{\tH}(\theta)
\end{eqnarray}
%
%==================================%$r(\theta)\cos{\theta}=\cR_{\tH}(\theta)$ 
in the leading order solution.
Then the effective equation (\ref{EFT}) becomes
%============<Equation>=============%
%
\begin{eqnarray} 
&&
\Omega_{H}^{2}\tan^{2}{\theta}(a^{2}\cos^{4}{\theta}+\cR_{\tH}(\theta)^{2}+\cos{\theta}\sin{\theta}\cR_{\tH}(\theta)\cR'_{\tH}(\theta))^{2}
(a^{2}\cos^{2}{\theta}+\cR_{\tH}(\theta)^{2}) \notag \\
&&~~~~~~
+(\cR_{\tH}(\theta)^{2}-1)(a^{2}\cos^{4}{\theta}+\cR_{\tH}(\theta)^{2})^{2} \notag \\
&&~~~~~~
+2\cos{\theta}\sin{\theta}\cR_{\tH}(\theta)\cR'_{\tH}(\theta)(\cR_{\tH}(\theta)^{2}-1)(a^{2}\cos^{4}{\theta}+\cR_{\tH}(\theta)^{2})\notag \\
&&~~~~~~
+\cos^{2}{\theta}\cR_{\tH}(\theta)^{2}\cR'_{\tH}(\theta)^{2}(a^{2}\cos^{4}{\theta}+\cR_{\tH}(\theta)^{2}-\sin^{2}{\theta})=0.
\label{SPEeq}
\end{eqnarray}
%
%==================================%
We solve this equation for $\cR_{\tH}(\theta)$ in some cases below. 

\paragraph{Myers-Perry solution}

The Myers-Perry solution is described by the embedding
%============<Equation>=============%
%
\begin{eqnarray}
\cR_{\tH}(\theta) =\cos{\theta},~~
\Omega_{H}=\frac{a}{1+a^{2}},
\end{eqnarray}
%
%==================================%
in the ellipsoidal embedding for arbitrary $a$.

\paragraph{Bumpy black holes}
The Myers-Perry black hole has marginally stable quasinormal modes $\omega=0$ in the ultraspinning region 
\cite{Emparan:2003sy,Dias:2009iu,Dias:2010maa}. The marginally stable modes suggest the existences of new deformed solution 
branching off the Myers-Perry solution branch. These solutions, so called bumpy black holes, have been constructed numerically 
in \cite{Dias:2014cia,Emparan:2014pra}. Here we consider this solution by using the effective equation (\ref{SPEeq}). 
To study this we perform the perturbative analysis of the effective equation (\ref{SPEeq}) around the Myers-Perry black hole. 
We expand the embedding function $\cR_{\tH}(\theta)$ around the Myers-Perry black hole as
%============<Equation>=============%
%
\begin{eqnarray}
\cR_{\tH}(\theta)=\cos{\theta}\Bigl[1 +\hat{\cR}(\theta)\epsilon +O\left(\epsilon^{2}\right)\Bigr], 
\end{eqnarray}
%
%==================================%
where
%============<Equation>=============%
%
\begin{eqnarray}
\epsilon=\Omega_{H}-\frac{a}{1+a^{2}}.
\end{eqnarray}
%
%==================================%
Then, perturbing eq. (\ref{SPEeq}) with respect to $\epsilon$, we find the perturbative solution
%============<Equation>=============%
%
\begin{eqnarray}
\hat{\cR}(\theta)= \frac{a(1+a^{2})^{2}\sin^{2}{\theta}}{(1-a^{2})(1+a^{2}\cos^{2}{\theta})}
+A\frac{\sin^{1+a^{2}}{\theta}}{1+a^{2}\cos^{2}{\theta}},
\label{Psol}
\end{eqnarray}
%
%==================================%
where $A$ is an integration constant describing the perturbation amplitude. The first part describes just the trivial perturbation adding the angular 
momentum along the Myers-Perry black hole solution branch. The second part is the non-trivial deformation of the solution into the bumpy black hole 
solution branch. For general $a$ and $A\neq 0$, we have a non-analytic behavior in the solution at $\theta=0$. The regularity of the solution at $\theta=0$ 
requires
\footnote{
Here  the regularity means that any derivatives of $\hat{\cR}(\theta)$ should be finite at $\theta=0$ and $\hat{\cR}(\theta)$ be
an even function around $\theta=0$. To satisfy this requirement $a$ should be discretized by $k$ as given in eq. (\ref{acond}). 
As we can see in appendix \ref{ASP} this regularity condition at $\theta=0$ is equivalent to the condition for the spheroidal harmonics 
at large $D$. Furthermore this discretization parameter $k$ corresponds to the angular momentum number $\ell$.    
} 
%============<Equation>=============%
%
\begin{eqnarray}
a_{c}^{2}=2k-1,~~~ k=1,2,3,... \label{acond}
\end{eqnarray}
%
%==================================%
For $k=1$, {\it i.e.}, $a_{c}=1$, the solution is same with the first part of eq. (\ref{Psol}), so it corresponds to the 
Myers-Perry black hole with a different angular momentum. The value $a_{c}$ for $k=1$ is known analytically in all $D$ by the 
thermodynamic argument \cite{Emparan:2003sy} as
%============<Equation>=============%
%
\begin{eqnarray}
\left(\frac{a}{r_{0}}\right)^{D-3}=\frac{D-3}{2(D-4)}\left( \frac{D-3}{D-5} \right)^{(D-5)/2}. 
\label{aexact}
\end{eqnarray}
%
%==================================%
$r_{0}$ is the horizon radius of the Myers-Perry black hole (see Appendix \ref{A}). 
At the large $D$ limit the formula (\ref{aexact}) gives same value with eq. (\ref{acond}) for $k=1$. 
For $k>1$ we have non-trivial bumpy black hole solutions by eq. (\ref{Psol}). Eq. (\ref{acond}) gives the threshold
angular momentum of the ultraspinning instability of the Myers-Perry black hole. In Table \ref{Table1} we show the comparison of eq. (\ref{acond}) 
with numerical results in \cite{Dias:2010maa}. Actually we have good agreements within the expected error $O(1/D)$. This threshold angular momentum will be reproduced 
also by observing the quasinormal mode frequency directly in section \ref{4}. 

In Appendix \ref{A} we generalize the above analysis to the AdS Myers-Perry black hole. Then we obtain the threshold angular momentum for the
ultraspinning instability of the AdS Myers-Perry black hole in the $1/D$ expansion. 

%============<Equation>=============%
%
\begin{table}[t]
\begin{center}
  \begin{tabular}{|c||c|c|c|c|} \hline
     ~ & $a_{c}|_{k=1}$ & $a_{c}|_{k=2}$ & $a_{c}|_{k=3}$ & $a_{c}|_{k=4}$\\ \hline \hline
    eq. (\ref{acond}) & $1$ & $\sqrt{3}=1.732$ & $\sqrt{5}=2.236$ & $\sqrt{7}=2.646$  \\ \hline
    numerical $(D=6)$ & 1.097 & $1.572$ & $1.849$ & $2.036$  \\
    numerical $(D=7)$ & 1.075 & $1.714$ & $2.141$ & $2.487$ \\
    numerical $(D=8)$ & 1.061 & $1.770$ & $2.275$ & $2.725$ \\
    numerical $(D=9)$ & 1.051 & $1.792$ & $2.337$ & $2.807$ \\
    numerical $(D=10)$ & 1.042 & $1.795$ & $2.361$ & $2.855$ \\
    numerical $(D=11)$ & 1.035 & $1.798$ & $2.373$ & $2.879$ \\ \hline
  \end{tabular}
\end{center}
\caption{Comparison of our analytic formula (\ref{acond}) for the threshold of the instability of the Myers-Perry black hole with numerical results in \cite{Dias:2010maa}. 
The analytic formula shows quite good agreements with numerical results within the expected error $O(1/D)$. The value of $a_{c}$ for $k=1$
by \cite{Dias:2010maa} has perfect agreements with eq. (\ref{aexact}).} 
\label{Table1}
\end{table}
%
%==================================%

\subsection{Boost representation}

One important property of the leading order solution in near zone is its local Lorentz boost representation. The effective equation (\ref{EFT}) has the interpretation 
that the embedded surface $\Sigma_{B}$ is the locally Lorentz boosted membrane. This local Lorentz boost property holds also for the metric itself 
in near zone. To see this we rewrite the $(t,\phi)$ part of the obtained leading order metric in near zone as
%============<Equation>=============%
%
\begin{eqnarray}
&&
ds^{2}_{(t,\phi)} =
-\sV_{0}(z)^{2}dt^{2} +\frac{\sV_{0}(z)^{2}-4\hat{\kappa}^{2}r_{0}(z)^{2}}{\Omega_{H}^{2}}d\phi^{2} \notag \\
&&~~~~~~~~~~~~~~~~~~~~~
+\frac{1}{4\hat{\kappa}^{2}r_{0}(z)^{2}\cosh^{2}{\rho}}\left( \sV_{0}(z)^{2}dt-\frac{\sV_{0}(z)^{2}-4\hat{\kappa}^{2}r_{0}^{2}}{\Omega_{H}}d\phi \right)^{2}. 
\label{tphimetric}
\end{eqnarray}
%
%==================================%
By introducing the boost parameter $\sigma(z)$ defined by
%============<Equation>=============%
%
\begin{eqnarray}
\cosh{\sigma(z)}=\frac{\sV_{0}(z)}{2\hat{\kappa}r_{0}(z)}
\end{eqnarray}
%
%==================================%
and new local frame $dx^{p}$ by
%============<Equation>=============%
%
\begin{eqnarray}
d x^{p}(z)=\left( \sV_{0}(z)dt,~\frac{\sqrt{\sV_{0}(z)^{2}-4\hat{\kappa}^{2}r_{0}(z)^{2}}}{\Omega_{H}}d\phi \right),
\end{eqnarray}
%
%==================================% 
the leading order metric is written in a very simple form as
%============<Equation>=============%
%
\begin{eqnarray}
ds^{2}= \frac{4r_{0}(z)^{2}}{n^{2}}d\rho^{2} 
+ \left(\eta_{pq}+\frac{u_{p}u_{q}}{\cosh^{2}{\rho}}\right)dx^{p}d x^{q}
+dz^{2} +\cR_{\tH}(z)^{2}d\Omega^{2}_{D-4},
\label{LOsol}
\end{eqnarray}
%
%==================================%
where $\eta_{pq}$ is the flat metric on two dimensional spacetime. The fluid velocity $u_{p}$ is 
%============<Equation>=============%
%
\begin{eqnarray}
u_{p}dx^{p}= \sV_{0}(z)\cosh{\sigma(z)}dt -\frac{\sqrt{\sV_{0}(z)^{2}-4\hat{\kappa}^{2}r_{0}(z)^{2}}}{\Omega_{H}}\sinh{\sigma(z)}d\phi.
\end{eqnarray}
%
%==================================%
Note that the leading order metric of the Schwarzschild black hole is (see Appendix \ref{A})
%============<Equation>=============%
%
\begin{eqnarray}
ds^{2} = \frac{4}{n^{2}}d\rho^{2}-\tanh^{2}{\rho}~dt^{2}+dz^{2}+\sin^{2}{\theta}d\phi^{2}+\cos^{2}{\theta}d\Omega^{2}_{D-4}.
\end{eqnarray}
%
%==================================% 
Thus the stationary solution can be represented as the locally boosted Schwarzschild black hole
under the following boost transformation
%============<Equation>=============%
%
\begin{eqnarray}
dt \rightarrow \sV_{0}(z)dt\cosh{\sigma(z)}-\frac{\sqrt{\sV_{0}(z)^{2}-4\hat{\kappa}^{2}r_{0}(z)^{2}}}{\Omega_{H}}d\phi\sinh{\sigma(z)}, \\
d\phi \rightarrow \frac{\sqrt{\sV_{0}(z)^{2}-4\hat{\kappa}^{2}r_{0}(z)}}{\Omega_{H}}d\phi\cosh{\sigma(z)} -\sV_{0}(z)dt\sinh{\sigma(z)}.
\end{eqnarray}
%
%==================================%
This boost representation is known for the Myers-Perry black hole \cite{Emparan:2013xia,Emparan:2014jca}. However 
our leading order solution covers much wider class 
of solutions including the asymptotically AdS black holes. This boost representation is the universal feature 
for the large $D$ stationary black holes under our ansatz. 

The boost parameter, $\sigma(z)$, is not constant and the boost transformation is inhomogeneous in $z$-direction. 
This $z$-dependence also appears as the horizon radius inhomogeneity in $g_{\rho\rho}$. Thus the identification
with the Schwarzschild black hole by the boost transformation is valid only locally in $z$-direction. 
However this inhomogeneity along $z$-direction is not so crucial since the dynamics along $z$-direction is sub-dominant compared 
with the radial dynamics. This boost transformation is the property of the radial dynamics, which is the leading order 
dynamics of large $D$ black holes. The local effect by $z$-dependence would be introduced at the sub-leading order in $1/D$ expansion
as corrections.  
Such $z$-dependent dynamics in the higher order structure of the $1/D$ expansion contains the essential effects 
by the rotation and horizon topology.

\section{Perturbation} \label{3}

As one interesting application of the boost property in the leading order solution we consider the perturbation problem. By using the boost 
transformation we can obtain the solution of the perturbation equation of the stationary black hole from one of the Schwarzschild black hole. 
In the following we investigate the decoupled mode perturbation, whose the frequency is $\omega=O(D^{0})$ \cite{Emparan:2014aba}, of the asymptotically 
flat black hole given by
%============<Equation>=============%
%
\begin{eqnarray}
\sV_{0}(z)=1,
\end{eqnarray}
%
%==================================%
in the leading order metric in section \ref{2}. The extension to the asymptotically AdS black hole is straightforward.

\subsection{Schwarzschild black hole}

We study the perturbation on the Schwarzschild black hole. Next we perform the boost transformation on the perturbation solutions
to obtain the perturbation solution of the stationary solution.
There are various ways in the representation of the perturbation around the Schwarzschild black hole \cite{Kodama:2003jz,Gorbonos:2004uc}. 
Here we fix the gauge for the perturbation, $h_{\mu\nu}$, to solve the perturbation equation. Especially we use the transverse traceless 
gauge as
%============<Equation>=============%
%
\begin{eqnarray}
\hat{\nabla}^{\mu}h_{\mu\nu}=0,~~
\hat{g}^{\mu\nu}h_{\mu\nu}=0,
\end{eqnarray}
%
%==================================% 
where $\hat{\nabla}_{\mu}$ is the covariant derivative of the $D$ dimensional background metric $\hat{g}_{\mu\nu}$. Then the perturbation equation becomes
equivalent to the Lichnerowicz equation given by
%============<Equation>=============%
%
\begin{eqnarray}
\hat{\nabla}^{\rho}\hat{\nabla}_{\rho}h_{\mu\nu} +2\hat{R}_{\mu\rho\nu\sigma}h^{\rho\sigma} =0,
\end{eqnarray}
%
%==================================%
where $\hat{R}_{\mu\nu\rho\sigma}$ is a background Riemann tensor. The explicit components of the perturbation equation are complicate so we do not
show them here.
The background spacetime is the $D=n+3$ dimensional Schwarzschild black hole given by%============<Equation>=============%
\begin{eqnarray}
ds^{2}=-f(r)^{-1}dt^{2}+f(r)^{-1}dr^{2}+r^{2}\left( 
d\theta^{2}+\sin^{2}{\theta}d\phi^{2}+\cos^{2}{\theta}d\Omega^{2}_{D-4}
\right),
\end{eqnarray}
%
%==================================%
where
%============<Equation>=============%
%
\begin{eqnarray}
f(r)=1-\left(\frac{r_{0}}{r}\right)^{n}\equiv 1-\frac{1}{\sR}.
\end{eqnarray}
%
%==================================%
Here we introduced the new radial coordinate $\sR=(r/r_{0})^{n}$. This radial coordinate $\sR$ is related with the radial coordinate $\rho$
which we used in the previous section by
%============<Equation>=============%
%
\begin{eqnarray}
\sR=\cosh^{2}{\rho}. 
\label{Rrho}
\end{eqnarray}
%
%==================================% 
The perturbation is decomposed into the scalar, vector and tensor type 
with respect to $S^{D-4}$ in the metric
\footnote{
Note that the usual scalar, vector and tensor type perturbation  of the Schwarzschild black hole is based on the decomposition 
on $S^{D-2}$ \cite{Kodama:2003jz}. Then, for example, our scalar type perturbation on $S^{D-4}$ consists of the scalar, vector and tensor type 
perturbation on $S^{D-2}$. }. In this paper we investigate the scalar and vector type perturbation on $S^{D-4}$. 
Using the vielbein 
%============<Equation>=============%
%
\begin{eqnarray}
e^{(0)}_{\text{Sch}}=\sqrt{f(r)}dt,~e^{(1)}_{\text{Sch}}=\frac{dr}{\sqrt{f(r)}},~
e^{(2)}_{\text{Sch}}=rd\theta,~e^{(3)}_{\text{Sch}}=r\sin{\theta}d\phi,~
e^{(A)}_{\text{Sch}}=r\cos{\theta}\hat{e}^{(A)},
\label{vSch}
\end{eqnarray}
%
%==================================%
where $\hat{e}^{(A)}$ is a vielbein on the unit sphere $S^{D-4}$, the
scalar type perturbation, $h^{(S)}_{\mu\nu}$, on $S^{D-4}$ can be written by
%============<Equation>=============%
%
\begin{eqnarray}
&&
h^{(S)}_{\mu\nu}dx^{\mu}dx^{\nu} = e^{-i\omega t}e^{i m\phi}\Bigl[ f^{(S)\text{Sch}}_{IJ}\mathbb{Y}^{(S)}_{j}e^{(I)}_{\text{Sch}}e^{(J)}_{\text{Sch}} 
+ f^{(S)\text{Sch}}_{I}\sD_{A}\mathbb{Y}^{(S)}_{j}e^{(I)}_{\text{Sch}}e^{(A)}_{\text{Sch}}  \notag \\
&&~~~~~~~~~~~~~~~~~~~~~~~~~~~~~~
+r^{2}\cos^{2}{\theta}(H^{(S)\text{Sch}}_{T}\mathbb{Y}^{(S)}_{j}\gamma_{AB} + H^{(S)\text{Sch}}_{L}\mathbb{Y}^{(S)j}_{AB})
dx^{A}dx^{B} \Bigr],
\end{eqnarray}
%
%==================================%
where $I,J=0,1,2,3$. 
$\mathbb{Y}^{(S)}_{j}$ is the scalar harmonics on $S^{D-4}$ satisfying
%============<Equation>=============%
%
\begin{eqnarray}
(\sD^{2}+\lambda_{S} )\mathbb{Y}^{(S)}_{j}=0,
\end{eqnarray}
%
%==================================%
where the eigenvalue $\lambda_{S}$ is given by
%============<Equation>=============%
%
\begin{eqnarray}
\lambda_{S}=j(j+D-5).
\end{eqnarray}
%
%==================================%$\lambda_{S}=j(j+D-5)$. 
The scalar derived tensor harmonics $\mathbb{Y}^{(S)j}_{AB}$ is defined by
%============<Equation>=============%
%
\begin{eqnarray}
\mathbb{Y}^{(S)j}_{AB} = \sD_{A}\sD_{B}\mathbb{Y}^{(S)}_{j}+\frac{\lambda_{S}}{D-4}\mathbb{Y}^{(S)}_{j}q_{AB}.
\end{eqnarray}
%
%==================================% 
The vector type perturbation, $h^{(V)}_{\mu\nu}$, on $S^{D-4}$ is given by
%============<Equation>=============%
%
\begin{eqnarray}
h^{(V)}_{\mu\nu}dx^{\mu}dx^{\nu}= e^{-i\omega t}e^{i m\phi}\Bigl[
f^{(V)\text{Sch}}_{I}\mathbb{Y}^{(V)j}_{A}e^{(I)}e^{(A)}+H^{(V)}_{L}\mathbb{Y}^{(V)j}_{AB}e^{(A)}e^{(B)}
\Bigr],
\end{eqnarray}
%
%==================================%
where $\mathbb{Y}^{(V)j}_{A}$ and $\mathbb{Y}^{(V)j}_{AB}$ are the vector harmonics on $S^{D-4}$ 
defined by
%============<Equation>=============%
%
\begin{eqnarray}
(\sD^{2}+\lambda_{V} )\mathbb{Y}^{(V)j}_{A}=0,~~
\sD^{A}\mathbb{Y}^{(V)j}_{A}=0,~~
\mathbb{Y}^{(V)j}_{AB}=\sD_{(A}\mathbb{Y}^{(V)j}_{B)},
\end{eqnarray}
%
%==================================%
with the eigenvalue $\lambda_{V}$ given by
%============<Equation>=============%
%
\begin{eqnarray}
\lambda_{V}=j(j+D-5)-1.
\end{eqnarray}
%
%==================================%
The perturbation equation is a coupled PDE system for scalar and vector type perturbations. At large $D$ the situation is
changed and the perturbation equations become decoupled ODE system. We can find decoupled perturbation variables $F^{(S)\text{Sch}}_{IJ}$
for the scalar type perturbation easily as
%============<Equation>=============%
%
\begin{eqnarray}
&&
F^{(S)\text{Sch}}_{00}=f^{(S)\text{Sch}}_{00}+f^{(S)\text{Sch}}_{11},~~
F^{(S)\text{Sch}}_{11}=f^{(S)\text{Sch}}_{00}-f^{(S)\text{Sch}}_{11},~~
F^{(S)\text{Sch}}_{01}=f^{(S)\text{Sch}}_{01},
\notag\\
&&
F^{(S)\text{Sch}}_{02}=f^{(S)\text{Sch}}_{02},~~
F^{(S)\text{Sch}}_{03}=f^{(S)\text{Sch}}_{03},~~ 
F^{(S)\text{Sch}}_{12}=f_{12}^{(S)\text{Sch}},~~
F^{(S)\text{Sch}}_{13}=f_{13}^{(S)\text{Sch}},
\notag \\
&&
F^{(S)\text{Sch}}_{22}=f_{22}^{(S)\text{Sch}},~~
F^{(S)\text{Sch}}_{23}=f_{23}^{(S)\text{Sch}},~~
F^{(S)\text{Sch}}_{33}=f_{33}^{(S)\text{Sch}},
\end{eqnarray}
%
%==================================%
and
%============<Equation>=============%
%
\begin{eqnarray}
F^{(S)\text{Sch}}_{I}=f^{(S)\text{Sch}}_{I},~~F_{T,L}^{(S)\text{Sch}}=H^{(S)\text{Sch}}_{T,L}.
\end{eqnarray}
%
%==================================%
The vector type perturbation has following decoupled perturbation variables at large $D$
%============<Equation>=============%
%
\begin{eqnarray}
F^{(V)\text{Sch}}_{I}=f^{(V)\text{Sch}}_{I},~~
F^{(V)\text{Sch}}_{L}=H^{(V)\text{Sch}}_{L}.
\end{eqnarray}
%
%==================================%
Then each decoupled perturbation variables are expanded at large $D$ as
%============<Equation>=============%
%
\begin{eqnarray}
F^{(S)\text{Sch}}_{ab}=\sum_{k\geq 0}\frac{{}^{(k)}F^{(S)\text{Sch}}_{ab}}{n^{k}},~
F^{(S)\text{Sch}}_{a}=\sum_{k\geq 0}\frac{{}^{(k)}F^{(S)\text{Sch}}_{a}}{n^{k}},~
F_{T,L}^{(S)\text{Sch}}=\sum_{k\geq 0}\frac{{}^{(k)}F_{T,L}^{(S)\text{Sch}}}{n^{k}},
\end{eqnarray}
%
%==================================%
for the scalar type perturbation and
%============<Equation>=============%
%
\begin{eqnarray}
F^{(V)\text{Sch}}_{I}=\sum_{k\geq 0}\frac{{}^{(k)}F^{(V)\text{Sch}}_{I}}{n^{k}},~
F_{L}^{(V)\text{Sch}}=\sum_{k\geq 0}\frac{{}^{(k)}F_{L}^{(V)\text{Sch}}}{n^{k}},
\end{eqnarray}
%
%==================================%
for the vector type perturbation.
By using transverse traceless gauge conditions we can obtain $F^{(S)\text{Sch}}_{I}$, $F_{T,L}^{(S)\text{Sch}}$ and $F^{(V)\text{Sch}}_{L}$
from $F^{(S)\text{Sch}}_{IJ}$ and $F^{(V)\text{Sch}}_{I}$. Thus the equations to be solved are the equations of $F^{(S)\text{Sch}}_{IJ}$
for the scalar type perturbation and $F^{(V)\text{Sch}}_{I}$ for the vector type perturbation.
The perturbation equations for $F_{IJ}^{(S)\text{Sch}}$ are
%============<Equation>=============%
%
\begin{eqnarray}
\frac{\partial}{\partial\sR}\sR(\sR-1)\frac{\partial}{\partial\sR}{}^{(k)}F_{00}^{(S)\text{Sch}}-\frac{{}^{(k)}F_{00}^{(S)\text{Sch}}}{\sR-1}={}^{(k)}\mathcal{S}^{(S)}_{00},\\
\frac{\partial}{\partial\sR}\sR(\sR-1)\frac{\partial}{\partial\sR}{}^{(k)}F_{01}^{(S)\text{Sch}}-\frac{{}^{(k)}F_{01}^{(S)\text{Sch}}}{\sR-1}={}^{(k)}\mathcal{S}^{(S)}_{01},\\
\frac{\partial}{\partial\sR}\sR(\sR-1)\frac{\partial}{\partial\sR}{}^{(k)}F_{02}^{(S)\text{Sch}}-\frac{{}^{(k)}F_{02}^{(S)\text{Sch}}}{4\sR(\sR-1)}={}^{(k)}\mathcal{S}^{(S)}_{02},\\
\frac{\partial}{\partial\sR}\sR(\sR-1)\frac{\partial}{\partial\sR}{}^{(k)}F_{03}^{(S)\text{Sch}}-\frac{{}^{(k)}F_{03}^{(S)\text{Sch}}}{4\sR(\sR-1)}={}^{(k)}\mathcal{S}^{(S)}_{03},\\
\frac{\partial}{\partial\sR}\sR(\sR-1)\frac{\partial}{\partial\sR}{}^{(k)}F_{11}^{(S)\text{Sch}}-\frac{{}^{(k)}F_{11}^{(S)\text{Sch}}}{\sR}={}^{(k)}\mathcal{S}^{(S)}_{11},\\
\frac{\partial}{\partial\sR}\sR(\sR-1)\frac{\partial}{\partial\sR}{}^{(k)}F_{12}^{(S)\text{Sch}}-\frac{{}^{(k)}F_{12}^{(S)\text{Sch}}}{4\sR(\sR-1)}={}^{(k)}\mathcal{S}^{(S)}_{12},\\
\frac{\partial}{\partial\sR}\sR(\sR-1)\frac{\partial}{\partial\sR}{}^{(k)}F_{13}^{(S)\text{Sch}}-\frac{{}^{(k)}F_{13}^{(S)\text{Sch}}}{4\sR(\sR-1)}={}^{(k)}\mathcal{S}^{(S)}_{13},\\
\frac{\partial}{\partial\sR}\sR(\sR-1)\frac{\partial}{\partial\sR}{}^{(k)}F_{22}^{(S)\text{Sch}}={}^{(k)}\mathcal{S}^{(S)}_{22},\\
\frac{\partial}{\partial\sR}\sR(\sR-1)\frac{\partial}{\partial\sR}{}^{(k)}F_{23}^{(S)\text{Sch}}={}^{(k)}\mathcal{S}^{(S)}_{23},\\
\frac{\partial}{\partial\sR}\sR(\sR-1)\frac{\partial}{\partial\sR}{}^{(k)}F_{33}^{(S)\text{Sch}}={}^{(k)}\mathcal{S}^{(S)}_{33}.
\end{eqnarray}
%
%==================================%
The $k-$th order source term ${}^{(k)}\mathcal{S}^{(S)}_{ab}$ is coming from the lower order solutions and the source term does not have 
the decoupling property. 
The perturbation variables, ${}^{(k)}F^{(S)\text{Sch}}_{IJ}$, depend on $\sR$ and $\theta$
\footnote{
The Schwarzschild black hole has a spherical symmetry. So the perturbation equation can be reduced to the ordinarily differential equation
in principle. Here, for the usefulness in the following discussion, we use the harmonics on $S^{D-4}$ and the perturbation equation
becomes the apparent partial differential equation.
}. However the perturbation equation becomes ordinarily 
differential equation with respect to $\sR$ at large $D$. Furthermore each $k$-th order variables ${}^{(k)}F_{IJ}^{(S)\text{Sch}}$ are decoupling and 
we can solve the solution by the straightforward integration procedure at each order. 
The vector type perturbation has the same property. The perturbation equation for $F^{(V)\text{Sch}}_{I}$ is
%============<Equation>=============%
%
\begin{eqnarray}
\frac{\partial}{\partial\sR}\sR(\sR-1)\frac{\partial}{\partial\sR}{}^{(k)}F^{(V)\text{Sch}}_{0}-\frac{{}^{(k)}F^{(V)\text{Sch}}_{0}}{4\sR(\sR-1)}={}^{(k)}\mathcal{S}^{(V)}_{0}, \\
\frac{\partial}{\partial\sR}\sR(\sR-1)\frac{\partial}{\partial\sR}{}^{(k)}F^{(V)\text{Sch}}_{1}-\frac{{}^{(k)}F^{(V)\text{Sch}}_{1}}{4\sR(\sR-1)}={}^{(k)}\mathcal{S}^{(V)}_{1}, \\
\frac{\partial}{\partial\sR}\sR(\sR-1)\frac{\partial}{\partial\sR}{}^{(k)}F^{(V)\text{Sch}}_{2}={}^{(k)}\mathcal{S}^{(V)}_{2}, \\
\frac{\partial}{\partial\sR}\sR(\sR-1)\frac{\partial}{\partial\sR}{}^{(k)}F^{(V)\text{Sch}}_{3}={}^{(k)}\mathcal{S}^{(V)}_{3}.
\end{eqnarray}
%
%==================================%
The leading order solution satisfying the regularity condition on the horizon and at infinity is given by
%============<Equation>=============%
%
\begin{eqnarray}
{}^{(0)}F_{00}^{(S)\text{Sch}}=\frac{A(\theta)}{\sR-1},~{}^{(0)}F_{01}^{(S)\text{Sch}}=\frac{A(\theta)}{\sR-1},~
{}^{(0)}F_{02}^{(S)\text{Sch}}=\frac{B(\theta)}{\sqrt{\sR(\sR-1)}}, \\
{}^{(0)}F_{03}^{(S)\text{Sch}}=\frac{C(\theta)}{\sqrt{\sR(\sR-1)}},~{}^{(0)}F_{11}^{(S)\text{Sch}}=\frac{D(\theta)}{\sR},~
{}^{(0)}F_{12}^{(S)\text{Sch}}=\frac{B(\theta)}{\sqrt{\sR(\sR-1)}}, \\
{}^{(0)}F_{13}^{(S)\text{Sch}}=\frac{C(\theta)}{\sqrt{\sR(\sR-1)}},~{}^{(0)}F_{22}^{(S)\text{Sch}}=0,~
{}^{(0)}F_{23}^{(S)\text{Sch}}=0,~{}^{(0)}F_{33}^{(S)\text{Sch}}=0,
\end{eqnarray}
%
%==================================%
for the scalar type perturbation and
%============<Equation>=============%
%
\begin{eqnarray}
{}^{(0)}F_{0}^{(V)\text{Sch}}=\frac{W(\theta)}{\sqrt{\sR(\sR-1)}},~
{}^{(0)}F_{1}^{(V)\text{Sch}}=\frac{W(\theta)}{\sqrt{\sR(\sR-1)}},
{}^{(0)}F_{2}^{(V)\text{Sch}}={}^{(0)}F_{3}^{(V)\text{Sch}}=0,
\end{eqnarray}
%
%==================================%
for the vector type perturbation.
The leading order perturbation solution contains undetermined integration functions $A(\theta)$, $B(\theta)$, $C(\theta)$, $D(\theta)$
and $W(\theta)$ for each type perturbation. If we go to the higher order of $k$ we can obtain the non-trivial relation between integration 
functions. Especially the solution at $k=1$ and $k=2$ order gives quasinormal mode frequency of the Schwarzschild black hole
for the vector and scalar type perturbation respectively. The detail analysis and results of this will be given in section \ref{4} and Appendix \ref{vec}.

\subsection{Stationary solution}

We apply the boost transformation to the perturbation solution of the Schwarzschild black hole.
At first, before taking large $D$ limit, we give the general procedure for the perturbation of the stationary black hole solution
given by
%============<Equation>=============%
%
\begin{eqnarray}
&&
ds^{2} = -A(\rho,z)^{2}dt^{2} +B(\rho.z)^{2}d\rho^{2} +F(\rho,z)^{2}(d\phi -w(\rho,z)dt)^{2} \notag \\
&&~~~~~~~~~~~~~~~~~~~~~~~~~~~~~~~~~~~~~~~~~~~~
+G(\rho,z)^{2}dz^{2} +H(\rho,z)^{2}d\Omega^{2}_{D-4}.
\end{eqnarray}
%
%==================================%
We consider the scalar and vector type perturbation on $S^{D-4}$.
To give perturbation variables we use the following vielbein 
%============<Equation>=============%
%
\begin{eqnarray}
e^{(0)} = A(\rho,z) dt,~~
e^{(1)} = B(\rho,z) d\rho,~~
e^{(2)} = F(\rho,z) (d\phi -w(\rho,z)dt)
\label{vMP}
\end{eqnarray}
%
%==================================%
and
%============<Equation>=============%
%
\begin{eqnarray}
e^{(3)} = G(\rho,z) dz,~~
e^{(A)} = H(\rho,z) \hat{e}^{(A)},
\end{eqnarray}
%
%==================================%
where $\hat{e}^{(A)}$ is the vielbein on the unit sphere $S^{D-4}$ again. Then the perturbed metric $h_{\mu\nu}$ 
is given by the components with respect to these vielbeins as
%============<Equation>=============%
%
\begin{eqnarray}
&&
h^{(S)}_{\mu\nu}dx^{\mu}dx^{\nu} = e^{-i\omega t}e^{i m\phi}\Bigl[ f^{(S)}_{IJ} \mathbb{Y}^{(S)}_{j}e^{(I)}e^{(J)} 
+ f^{(S)}_{I}\sD_{A}\mathbb{Y}^{(S)}_{j}e^{(I)}e^{(A)}  \notag \\
&&~~~~~~~~~~~~~~~~~~~~~~~~~~~~
+H(\rho,z)^{2}(H^{(S)}_{T}\mathbb{Y}^{(S)}_{j}q_{AB} + H^{(S)}_{L}\mathbb{Y}^{(S)j}_{AB})dx^{A}dx^{B} \Bigr],
\end{eqnarray}
%
%==================================%
for scalar type perturbation and
%============<Equation>=============%
%
\begin{eqnarray}
h^{(V)}_{\mu\nu}dx^{\mu}dx^{\nu}= e^{-i\omega t}e^{i m\phi}\Bigl[
f^{(V)}_{I}\mathbb{Y}^{(V)j}_{A}e^{(I)}e^{(A)}+H^{(V)}_{L}\mathbb{Y}^{(V)j}_{AB}e^{(A)}e^{(B)}
\Bigr],
\end{eqnarray}
%
%==================================%
for the vector type perturbation. 
Then we perform the boost transformation on these perturbed metrics of the leading order solution obtained in section \ref{2}. The boost transformation gives the relation between vielbeins as
%============<Equation>=============%
%
\begin{eqnarray}
e^{(0)}_{\text{Sch}} \rightarrow \frac{2-r_{0}(z)^{2}(1-\sinh^{2}{\rho})}{r_{0}(z)^{2}\cosh{\rho}\sqrt{1+r_{0}(z)^{2}\sinh^{2}{\rho}}}e^{(0)}
-\sinh{\rho}\sqrt{\frac{1-r_{0}(z)^{2}}{1+r_{0}(z)^{2}\sinh^{2}{\rho}}}e^{(2)},\\
e^{(2)}_{\text{Sch}} \rightarrow \frac{\cosh{\rho}}{\sqrt{1+r_{0}(z)^{2}\sinh^{2}{\rho}}}e^{(2)}
-\frac{2+r_{0}(z)^{2}\sinh^{2}{\rho}}{r_{0}\sinh{\rho}}\sqrt{\frac{1-r_{0}(z)^{2}}{1+r_{0}(z)\sinh^{2}{\rho}}}e^{(0)},
\end{eqnarray}
%
%==================================%
where we used the vielbein of the Schwarzschild black hole (\ref{vSch}) and of the stationary solution (\ref{vMP}) for the 
leading order metric (\ref{LOsol})
\footnote{
We consider the asymptotically flat black hole $\sV_{0}(z)=1$.
}.
This relation between vielbeins gives also the relation between the perturbation variables of the Schwarzschild black hole 
and general stationary solutions. Since we know the decoupled perturbation variables on the Schwarzschild black hole, we can obtain
the following decoupled perturbation variables, $F^{(S)}_{IJ}$, $F^{(S)}_{I}$, $F^{(S)}_{T,L}$, $F^{(V)}_{I}$ and $F^{(V)}_{L}$ 
of the general stationary solution at large $D$ as
%============<Equation>=============%
%
\begin{eqnarray}
&&
F^{(S)}_{00}=\frac{\cosh^{2}{\rho}}{1+r_{0}(z)^{2}\sinh^{2}{\rho}}f^{(S)}_{00} 
-\frac{2\sinh{\rho}\cosh{\rho}\sqrt{1-r_{0}(z)^{2}}}{1+r_{0}(z)^{2}\sinh^{2}{\rho}}f^{(S)}_{02}\notag \\
&&~~~~~~~~~~~~~~~~~~~~~~~~~~~~~~~
+\frac{(1-r_{0}(z)^{2})\sinh^{2}{\rho}}{1+r_{0}(z)^{2}\sinh^{2}{\rho}}f^{(S)}_{22}+f^{(S)}_{11},
\end{eqnarray}
%
%==================================%
%============<Equation>=============%
%
\begin{eqnarray}
&&
F^{(S)}_{11}=\frac{\cosh^{2}{\rho}}{1+r_{0}(z)^{2}\sinh^{2}{\rho}}f^{(S)}_{00} 
-\frac{2\sinh{\rho}\cosh{\rho}\sqrt{1-r_{0}(z)^{2}}}{1+r_{0}(z)^{2}\sinh^{2}{\rho}}f^{(S)}_{02}\notag \\
&&~~~~~~~~~~~~~~~~~~~~~~~~~~~~~~~
+\frac{(1-r_{0}(z)^{2})\sinh^{2}{\rho}}{1+r_{0}(z)^{2}\sinh^{2}{\rho}}f^{(S)}_{22}-f^{(S)}_{11},
\end{eqnarray}
%
%==================================%
%============<Equation>=============%
%
\begin{eqnarray}
&&
F^{(S)}_{02}=\frac{\cosh^{2}{\rho}-(1-r_{0}(z)^{2})\sinh^{2}{\rho}}{1+r_{0}(z)^{2}\sinh^{2}{\rho}}f^{(S)}_{02} \notag \\
&&~~~~~~~~~~~~~~~~~~~~
-\frac{\sqrt{1-r_{0}(z)^{2}}\sinh{\rho}\cosh{\rho}}{1+r_{0}(z)^{2}\sinh^{2}{\rho}}(f^{(S)}_{00}+f^{(S)}_{02}),
\end{eqnarray}
%
%==================================%
%============<Equation>=============%
%
\begin{eqnarray}
&&
F^{(S)}_{22}=\frac{\cosh^{2}{\rho}}{1+r_{0}(z)^{2}\sinh^{2}{\rho}}f^{(S)}_{22} 
-\frac{2\sinh{\rho}\cosh{\rho}\sqrt{1-r_{0}(z)^{2}}}{1+r_{0}(z)^{2}\sinh^{2}{\rho}}f^{(S)}_{02}\notag \\
&&~~~~~~~~~~~~~~~~~~~~~~~~~~~~~~~
+\frac{(1-r_{0}(z)^{2})\sinh^{2}{\rho}}{1+r_{0}(z)^{2}\sinh^{2}{\rho}}f^{(S)}_{00},
\end{eqnarray}
%
%==================================%
%============<Equation>=============%
%
\begin{eqnarray}
F^{(S)}_{01}=\frac{\cosh{\rho}}{\sqrt{1+r_{0}(z)^{2}\sinh^{2}{\rho}}}f^{(S)}_{01}
-\sinh{\rho}\sqrt{\frac{1-r_{0}(z)^{2}}{1+r_{0}(z)^{2}\sinh^{2}{\rho}}}f^{(S)}_{12},
\end{eqnarray}
%
%==================================%
%============<Equation>=============%
%
\begin{eqnarray}
F^{(S)}_{12}=\frac{\cosh{\rho}}{\sqrt{1+r_{0}(z)^{2}\sinh^{2}{\rho}}}f^{(S)}_{12}
-\sinh{\rho}\sqrt{\frac{1-r_{0}(z)^{2}}{1+r_{0}(z)^{2}\sinh^{2}{\rho}}}f^{(S)}_{01},
\end{eqnarray}
%
%==================================%
%============<Equation>=============%
%
\begin{eqnarray}
F^{(S)}_{03}=\frac{\cosh{\rho}}{\sqrt{1+r_{0}(z)^{2}\sinh^{2}{\rho}}}f^{(S)}_{03}
-\sinh{\rho}\sqrt{\frac{1-r_{0}(z)^{2}}{1+r_{0}(z)^{2}\sinh^{2}{\rho}}}f^{(S)}_{23},
\end{eqnarray}
%
%==================================%
%============<Equation>=============%
%
\begin{eqnarray}
F^{(S)}_{23}=\frac{\cosh{\rho}}{\sqrt{1+r_{0}(z)^{2}\sinh^{2}{\rho}}}f^{(S)}_{23}
-\sinh{\rho}\sqrt{\frac{1-r_{0}(z)^{2}}{1+r_{0}(z)^{2}\sinh^{2}{\rho}}}f^{(S)}_{03},
\end{eqnarray}
%
%==================================%
%============<Equation>=============%
%
\begin{eqnarray}
F^{(S)}_{0}=\frac{\cosh{\rho}}{\sqrt{1+r_{0}(z)^{2}\sinh^{2}{\rho}}}f^{(S)}_{0}
-\sinh{\rho}\sqrt{\frac{1-r_{0}(z)^{2}}{1+r_{0}(z)^{2}\sinh^{2}{\rho}}}f^{(S)}_{2},
\end{eqnarray}
%
%==================================%
%============<Equation>=============%
%
\begin{eqnarray}
F^{(S)}_{2}=\frac{\cosh{\rho}}{\sqrt{1+r_{0}(z)^{2}\sinh^{2}{\rho}}}f^{(S)}_{2}
-\sinh{\rho}\sqrt{\frac{1-r_{0}(z)^{2}}{1+r_{0}(z)^{2}\sinh^{2}{\rho}}}f^{(S)}_{0},
\end{eqnarray}
%
%==================================%
and
%============<Equation>=============%
%
\begin{eqnarray}
F^{(S)}_{13}=f^{(S)}_{13},~~F^{(S)}_{33}=f^{(S)}_{33},~~F^{(S)}_{1}=f^{(S)}_{1},~~F^{(S)}_{3}=f^{(S)}_{3},~~F^{(S)}_{T,L}=H^{(S)}_{T,L},
\end{eqnarray}
%
%==================================%
for the scalar type perturbation and
%============<Equation>=============%
%
\begin{eqnarray}
F^{(V)}_{0}=\frac{\cosh{\rho}}{\sqrt{1+r_{0}(z)^{2}\sinh^{2}{\rho}}}f^{(V)}_{0}
-\sinh{\rho}\sqrt{\frac{1-r_{0}(z)^{2}}{1+r_{0}(z)^{2}\sinh^{2}{\rho}}}f^{(V)}_{2},
\end{eqnarray}
%
%==================================%
%============<Equation>=============%
%
\begin{eqnarray}
F^{(V)}_{2}=\frac{\cosh{\rho}}{\sqrt{1+r_{0}(z)^{2}\sinh^{2}{\rho}}}f^{(V)}_{2}
-\sinh{\rho}\sqrt{\frac{1-r_{0}(z)^{2}}{1+r_{0}(z)^{2}\sinh^{2}{\rho}}}f^{(V)}_{0},
\end{eqnarray}
%
%==================================%
and
%============<Equation>=============%
%
\begin{eqnarray}
~F^{(V)}_{1}=f^{(V)}_{1},~~F^{(V)}_{3}=f^{(V)}_{3},~~F^{(V)}_{L}=H^{(V)}_{L},
\end{eqnarray}
%
%==================================%
for the vector type perturbation. Each perturbation variables are expanded at large $D$ as 
%============<Equation>=============%
%
\begin{eqnarray}
F^{(S)}_{IJ}=\sum_{k\geq 0}\frac{{}^{(k)}F^{(S)}_{IJ}}{n^{k}},~
F^{(S)}_{I}=\sum_{k\geq 0}\frac{{}^{(k)}F^{(S)}_{I}}{n^{k}},~
F^{(S)}_{T,L}=\sum_{k\geq 0}\frac{{}^{(k)}F_{T,L}^{(S)}}{n^{k}},
\end{eqnarray}
%
%==================================%
and
%============<Equation>=============%
%
\begin{eqnarray}
F^{(V)}_{I}=\sum_{k\geq 0}\frac{{}^{(k)}F^{(V)}_{I}}{n^{k}},~
F^{(V)}_{T,L}=\sum_{k\geq 0}\frac{{}^{(k)}F_{L}^{(V)}}{n^{k}}.
\end{eqnarray}
%
%==================================%
Then the perturbation equation for these decoupling variables can be obtained from the Lichnerowicz equation under the transverse traceless gauge condition. Now we have 
obtained the leading order metric of the stationary black hole in the $1/D$ expansion in section \ref{2}. Then the perturbation 
equation for $k=0$ order is also obtained by using the leading order metric. The perturbation equation of the decoupling variables can 
be found to satisfy same equations with one of the Schwarzschild black hole at the leading order by changing the radial coordinate from
$\rho$ to $\sR$ by
\footnote{
This relation can be understood by the horizon position. The Schwarzschild black hole has the horizon at $\sR=1$ and the
general stationary black hole horizon position is set to be $\rho=0$. 
}
%============<Equation>=============%
%
\begin{eqnarray}
\sR=\cosh^{2}{\rho}.
\end{eqnarray}
%
%==================================%
At higher orders the perturbation equation of the stationary solution becomes different from the boosted one of the Schwarzschild black hole in source terms. This is because the boost relation is valid only in the leading order metric, and higher order
structure gives the essential effect coming from the difference from the Schwarzschild black holes. However the homogeneous part of the equation
at each order is always same with one of the Schwarzschild black hole. So, in principle, the perturbation equation is a set of decoupled ordinarily 
differential equations and we can solve it by a straightforward integration method with the Green's function. 

The regular leading order solution of the perturbation equation for the stationary black hole is
%============<Equation>=============%
%
\begin{eqnarray}
{}^{(0)}F_{00}^{(S)}=\frac{A(\theta)}{\sR-1},~{}^{(0)}F_{01}^{(S)}=\frac{A(\theta)}{\sR-1},~
{}^{(0)}F_{02}^{(S)}=\frac{B(\theta)}{\sqrt{\sR(\sR-1)}}, \\
{}^{(0)}F_{03}^{(S)}=\frac{C(\theta)}{\sqrt{\sR(\sR-1)}},~{}^{(0)}F_{11}^{(S)}=\frac{D(\theta)}{\sR},~
{}^{(0)}F_{12}^{(S)}=\frac{B(\theta)}{\sqrt{\sR(\sR-1)}}, \\
{}^{(0)}F_{13}^{(S)}=\frac{C(\theta)}{\sqrt{\sR(\sR-1)}},~{}^{(0)}F_{22}^{(S)}=0,~
{}^{(0)}F_{23}^{(S)}=0,~{}^{(0)}F_{33}^{(S)}=0,
\end{eqnarray}
%
%==================================%
for the scalar type perturbation and
%============<Equation>=============%
%
\begin{eqnarray}
{}^{(0)}F_{0}^{(V)}=\frac{W(\theta)}{\sR},~
{}^{(0)}F_{1}^{(V)}=\frac{W(\theta)}{\sR},
{}^{(0)}F_{2}^{(V)}={}^{(0)}F_{3}^{(V)}=0,
\end{eqnarray}
%
%==================================%
for the vector type perturbation. 
In the derivation of the perturbation equations and solutions we need only the leading order 
solution of general stationary solutions given in section \ref{2}. To go to more higher order perturbation equations, 
the higher order solutions of stationary solutions are required. In the next section we consider 
the higher order perturbation equation of the Myers-Perry black hole, which is a known exact solution.

Note that the quasinormal mode frequency of the decoupled mode perturbation is obtained by solving higher order perturbation 
equation in $1/D$. The higher order structure of the perturbation equation reflects the higher order structure of the background metric, 
which does not have the boost property. Hence the frequency does not have the boost property either. This implies that the quasinormal mode frequency 
of the Myers-Perry black hole cannot be obtained by the boost transformation from one of the Schwarzschild black hole.  
This non-boost property allows the existence of the instability mode in the decoupled sector of the perturbation of the 
Myers-Perry black hole. In contrast the quasinormal mode of the non-decoupled sector perturbation is obtained in the leading 
order structure and we expect that such mode has the boost property as the universal feature \cite{Emparan:2014cia}.  

\section{QNMs of Myers-Perry black hole} \label{4}

We pursue the decoupling property of perturbation variables at large $D$ expansion shown in the previous section to more higher 
order structure to obtain the quasinormal mode (QNM) frequency of the $D=n+3$ dimensional singly rotating Myers-Perry black hole. 
This was calculated numerically 
in \cite{Dias:2014eua} for non-axisymmetric perturbations. Here we give the analytic formula of the QNM frequency both for the axisymmetric 
and non-axisymmetric perturbations. The metric of the Myers-Perry black hole is given in
the Appendix \ref{A} and it is written by the following parameter setting
%============<Equation>=============%
%
\begin{eqnarray}
\sV_{0}(z)=1,~~
\hat{\kappa}=\frac{1}{2},~~
\Omega_{H} =\frac{a}{1+a^{2}},~~
R(\theta)=\cos{\theta},~~
r_{0}(\theta)^{-2}=\frac{1+a^{2}}{1+a^{2}\cos^{2}{\theta}},
\end{eqnarray}
%
%==================================%
in the leading order metric obtained in section \ref{2}.
Our detail analysis in this section concentrates on the scalar type perturbation since the analysis of the vector 
type perturbation is performed in the same manner. The only result of the QNM frequency for the vector type perturbation 
will be given later. The scalar type perturbation variables are defined in the same way with one in section \ref{3} as
%============<Equation>=============%
%
\begin{eqnarray}
&&
h_{\mu\nu}dx^{\mu}dx^{\nu} = e^{-i\omega t}e^{i m\phi}\Bigl[ f_{ab} \mathbb{Y}^{(S)}_{j}e^{(a)}e^{(b)} 
+ f_{a}\sD_{A}\mathbb{Y}^{(S)}_{j}e^{(a)}e^{(A)}  \notag \\
&&~~~~~~~~~~~~~~~~~~~~~~~~~~~~~
+r^{2}\cos^{2}{\theta}(H_{T}\mathbb{Y}^{(S)}_{j}q_{AB} + H_{L}\mathbb{Y}^{(S)j}_{AB})dx^{A}dx^{B} \Bigr],
\end{eqnarray}
%
%==================================%
where the vielbein was defined in section \ref{3}
\footnote{
We omit the index ${}^{(S)}$ representing the scalar type perturbation in section \ref{3}.
}. So the perturbation is parameterized by $\omega$, $m$ and $j$.
$j$ describes the deformation parameter on $S^{D-4}$. Thus $j=0$ correspond to the perturbation without the deformation on $S^{D-4}$.
Then the decoupled perturbation variables can be constructed
and expanded at large $D$ as
%============<Equation>=============%
%
\begin{eqnarray}
F_{IJ}=\sum_{k\geq 0}\frac{F^{(k)}_{IJ}}{n^{k}},~~
F_{a}=\sum_{k\geq 0}\frac{F^{(k)}_{a}}{n^{k}},~~
F_{T,L}=\sum_{k\geq 0}\frac{F_{T,L}^{(k)}}{n^{k}},
\end{eqnarray}
%
%==================================%
where $D=n+3$. The boundary condition for the perturbation on the horizon is the ingoing 
boundary condition \cite{Dias:2014eua} as
%============<Equation>=============%
%
\begin{eqnarray}
h_{\mu\nu}(r,\theta) = \hat{h}_{\mu\nu}(r,\theta)(r-r_{+})^{-i(\omega-m\Omega_{H})/2\kappa},
\label{QNMBCHo}
\end{eqnarray}
%
%==================================%
where $r_{+}$ is the horizon position and $\kappa$ is a surface gravity. The component of $\hat{h}_{\mu\nu}$ in the Eddington-Finkelstein ingoing coordinate
should be regular on the horizon. Here we omit the $t$, $\phi$ and $x^{A}$ dependence just for the simplicity
\footnote{
$x^{A}$ is a coordinate on $S^{D-4}$.
}. The boundary
condition at infinity $\sR\gg 1$ is the decoupled mode condition \cite{Emparan:2014aba} as
%============<Equation>=============%
%
\begin{eqnarray}
h_{\mu\nu}=O(\sR^{-1}).
\label{QNMBCIn}
\end{eqnarray}
%
%==================================% 
This condition is equivalent to the outgoing wave boundary condition at infinity for the decoupled mode defined by $\omega r_{0}=O(D^{0})$.

In the following we give the brief summary of perturbation solutions of the scalar type perturbation up to $k=2$ order. At $k=2$ we obtain QNM frequency
for the first time. 

\paragraph{Leading order solution $(k=0)$}
The leading order solution satisfying the boundary conditions is
%============<Equation>=============%
%
\begin{eqnarray}
F_{00}^{(0)}=\frac{A_{(0)}(\theta)}{\sR-1},~F_{01}^{(0)}=\frac{A_{(0)}(\theta)}{\sR-1},~
F_{02}^{(0)}=\frac{B_{(0)}(\theta)}{\sqrt{\sR(\sR-1)}}, \notag \\
F_{03}^{(0)}=\frac{C_{(0)}(\theta)}{\sqrt{\sR(\sR-1)}},~F_{11}^{(0)}=\frac{D_{(0)}(\theta)}{\sR},~
F_{12}^{(0)}=\frac{B_{(0)}(\theta)}{\sqrt{\sR(\sR-1)}},  \\
F_{13}^{(0)}=\frac{C_{(0)}(\theta)}{\sqrt{\sR(\sR-1)}},~F_{22}^{(0)}=0,~
F_{23}^{(0)}=0,~F_{33}^{(0)}=0. \notag
\end{eqnarray}
%
%==================================%
At leading order we have four independent integration functions, which cannot be determined by the boundary condition.
The solution for $F^{(0)}_{I}$ and $F^{(0)}_{T,L}$ can be obtained by the transverse traceless gauge condition, and we omit them 
since it is not so important.  

\paragraph{Next to leading order solution $(k=1)$}

The next to leading order solution of $F_{IJ}$ is represented by
%============<Equation>=============%
%
\begin{eqnarray}
F_{00}^{(1)}=\frac{A_{(1)}(\theta)}{\sR-1}+\hat{F}_{00}^{(1)},~
F_{01}^{(1)}=\frac{A_{(1)}(\theta)}{\sR-1}+\hat{F}_{01}^{(1)},~ \notag \\
F_{02}^{(1)}=\frac{B_{(1)}(\theta)}{\sqrt{\sR(\sR-1)}}+\hat{F}_{02}^{(1)},~
F_{03}^{(1)}=\frac{C_{(1)}(\theta)}{\sqrt{\sR(\sR-1)}}+\hat{F}_{03}^{(1)},~ \notag \\
F_{11}^{(1)}=\frac{D_{(1)}(\theta)}{\sR}+\hat{F}_{11}^{(1)},~
F_{12}^{(1)}=\frac{B_{(1)}(\theta)}{\sqrt{\sR(\sR-1)}}+\hat{F}_{12}^{(1)}, \notag \\
F_{13}^{(1)}=\frac{C_{(1)}(\theta)}{\sqrt{\sR(\sR-1)}}+\hat{F}_{13}^{(1)},~
F_{22}^{(1)}=\hat{F}_{22}^{(1)},~
F_{23}^{(1)}=0,~F_{33}^{(1)}=\hat{F}_{33}^{(1)},
\end{eqnarray}
%
%==================================%
where the solution with the hat comes from the integration of the source term composed 
by the leading order solution. At this order we have new four integration functions
$A_{(1)}(\theta)$, $B_{(1)}(\theta)$, $C_{(1)}(\theta)$ and $D_{(1)}(\theta)$. 
To satisfy boundary conditions  we obtain some non-trivial relations for $A_{(0)}(\theta)$, $B_{(0)}(\theta)$, 
$C_{(0)}(\theta)$ and $D_{(0)}(\theta)$. Conditions we find are
%============<Equation>=============%
%
\begin{eqnarray}
B_{(0)}(\theta)=0,~~C_{(0)}(\theta)=0,
\end{eqnarray}
%
%==================================% 
and
%============<Equation>=============%
%
\begin{eqnarray}
&&
\cos{\theta}(1+a^{2}\cos^{2}{\theta})\left( 
(1+a^{2})\cos{\theta}A_{(0)}(\theta) -\sin{\theta}D_{(0)}'(\theta)
\right) \notag \\
&&~~~~
+\bigl[
j +(-1+iam +a^{2}(-3+2j -i\omega)-i\omega)\cos^{2}{\theta}
\notag \\
&&~~~~~~~~~~~~~~~~
+a^{2}(1+iam+a^{2}(j -1-i\omega)-i\omega)\cos^{4}{\theta}
\bigr]D_{(0)}(\theta)=0.
\end{eqnarray}
%
%==================================%
Using this condition we can eliminate $A_{(0)}(\theta)$. To obtain QNM frequency we should specify 
the function $D_{(0)}(\theta)$, and we can do it at next to next to leading order.

\paragraph{Next-to-next-to leading order solution $(k=2)$}

At this order we can obtain the solutions in the same way with the next to leading order. 
We find additional non-trivial condition for integration functions. By eliminating $A_{(0)}(\theta)$, 
the additional conditions are written as
%============<Equation>=============%
%
\begin{eqnarray}
&&
c_{1}(\theta)C_{(1)}''(\theta)+c_{2}(\theta)C_{(1)}'(\theta)+c_{3}(\theta)C_{(1)}(\theta) \notag \\
&&~~~~~~~~~~~~~~
+c_{4}(\theta)D_{(0)}''(\theta)+c_{5}(\theta)D_{(0)}'(\theta)+c_{6}(\theta)D_{(0)}(\theta)=0,
\label{eq1}
\end{eqnarray}
%
%==================================%
and
%============<Equation>=============%
%
\begin{eqnarray}
&&
d_{1}(\theta)D_{(0)}'''(\theta)+d_{2}(\theta)D_{(0)}''(\theta)+d_{3}(\theta)D_{(0)}'(\theta) \notag \\
&&~~~~~~~~~~~~~~
+d_{4}(\theta)D_{(0)}(\theta)+d_{5}(\theta)C_{(1)}'(\theta)+d_{6}(\theta)C_{(1)}(\theta)=0.
\label{eq2}
\end{eqnarray}
%
%==================================%
Each functions $c_{i}(\theta)$ and $d_{i}(\theta)$ have messy forms so we do not show them here. 
They are given in the Appendix \ref{cd}. In general it seems that we 
should solve these differential equations to obtain the QNM frequency. However, we find that the QNM frequency 
can be obtained only from the boundary conditions
\footnote{
Actually we can confirm that eqs. (\ref{eq1}) and (\ref{eq2}) can be solved regularly under the QNM frequency obtained below. 
Thus there may be some non-trivial structure in the solutions of eqs. (\ref{eq1}) and (\ref{eq2}) and the boundary conditions. 
}
. At $\theta=\pi/2$ the boundary condition is read from eqs (\ref{eq1})
and (\ref{eq2}) as
%============<Equation>=============%
%
\begin{eqnarray}
C_{(1)}(\theta)|_{\theta=\pi/2} = \cos^{j-3}{\theta},~~
D_{(0)}(\theta)|_{\theta=\pi/2} = \cos^{j}{\theta}.
\end{eqnarray}
%
%==================================%
On the other hand the behavior at $\theta=0$ is
%============<Equation>=============%
%
\begin{eqnarray}
C_{(1)}(\theta)|_{\theta=0} = \sin^{\delta^{S}_{\omega}-1}{\theta},~~
D_{(0)}(\theta)|_{\theta=0} = \sin^{\delta^{S}_{\omega}}{\theta},
\end{eqnarray}
%
%==================================%
where $\delta^{S}_{\omega}$ depends on $a$, $m$, $j$ and $\omega$.
To see which value $\delta^{S}_{\omega}$ should take, we observe the spherical harmonics, $\mathbb{S}^{\ell}(\theta)e^{im\phi}\mathbb{Y}^{(S)}_{j}$, 
on $S^{D-2}$,
defined by
%============<Equation>=============%
%
\begin{eqnarray}
&&
\bigl[ \Delta_{S^{D-2}} +\ell(\ell+D-3) \bigr] \mathbb{S}^{\ell}e^{im\phi}\mathbb{Y}^{(S)}_{j} \notag \\
&&
=\Biggl[ \frac{1}{\sin{\theta}\cos^{D-4}{\theta}}\frac{d}{d\theta}\left(
\sin{\theta}\cos^{D-4}{\theta}\frac{d}{d\theta} \mathbb{S}^{\ell}
\right) \notag \\
&&~~~~~~~~~~
-\frac{m^{2}}{\sin^{2}{\theta}}\mathbb{S}^{\ell}-\frac{j(j+D-5)}{\cos^{2}{\theta}}\mathbb{S}^{\ell}
+\ell(\ell+D-3)\mathbb{S}^{\ell} \Biggr]e^{im\phi}\mathbb{Y}^{(S)}_{j} =0.
\label{SPHeq}
\end{eqnarray}
%
%==================================%
The solution of this equation can be written by the hypergeometric function as
%============<Equation>=============%
%
\begin{eqnarray}
\mathbb{S}^{\ell} = \cos^{j}{\theta}~\sin^{|m|}{\theta}~{}_{2}F_{1}(-k_{S},k_{S}+|m|+j+n/2,|m|+1;\sin^{2}{\theta}),
\end{eqnarray}
%
%==================================%
where $k_{S}$ is a non-negative integer satisfying
%============<Equation>=============%
%
\begin{eqnarray}
\ell = j+|m|+2k_{S}. \label{kdef}
\end{eqnarray}
%
%==================================%
At large $D$ this solution becomes
%============<Equation>=============%
%
\begin{eqnarray}
\mathbb{S}^{\ell} &=& \sin^{\ell-j}{\theta}\cos^{j}{\theta}+O(1/D) \notag \\
&=& \sin^{|m|+2k_{S}}{\theta}\cos^{j}{\theta}+O(1/D).  \label{SPHsol}
\end{eqnarray}
%
%==================================%
$k_{S}$ describes the overtone number of $\mathbb{S}^{\ell}$ along $\theta$ direction, and this overtone number can be 
observed from the behavior of $\mathbb{S}^{\ell}$ at $\theta=0$ at large $D$ limit as seen eq. (\ref{SPHsol}).  
From this observation on the spherical harmonics we impose the following condition on $\delta^{S}_{\omega}$ 
as the harmonics condition  
%============<Equation>=============%
%
\begin{eqnarray}
\delta^{S}_{\omega} &=& \ell-j \notag \\
&=& 2k_{s}+|m|, \label{QNMcond}
\end{eqnarray}
%
%==================================%
where $\ell$ is parametrized as eq. (\ref{kdef}) by non-negative integer $k_{S}$. Then eqs. (\ref{eq1}) and (\ref{eq2}) 
under the condition (\ref{QNMcond}) give one non-trivial algebraic condition on $\omega$. This non-trivial condition on 
$\omega$ can be regarded as the QNM condition. In the following we solve the condition eq. (\ref{QNMcond}) for $\omega$ 
in some cases. The explicit form of the QNM condition (\ref{QNMcond}) is given in the Appendix \ref{C}.

Note that the condition (\ref{QNMcond}) is equivalent to the regularity condition to derive eq. (\ref{acond}) 
in section \ref{2}. The condition (\ref{QNMcond}) is suggested by the analysis for spherical harmonics at large $D$. 
However, as we can see in Appendix \ref{ASP}, the analysis of the spheroidal harmonics also gives same condition (\ref{QNMcond}). 
Thus we expect that the condition (\ref{QNMcond}) would hold also in the gravitational perturbation of the Myers-Perry black hole.

\subsection{QNM frequency}

We show some explicit results of the QNM frequency of the scalar and vector type perturbations by solving the QNM 
condition (\ref{QNMcond}). The instability exists only in the scalar type perturbation and the vector type perturbation 
is always stable for the decoupled sector perturbation
\footnote{
The instability mode is conjectured to exist only in the decoupled sector (saturated mode) \cite{Dias:2014eua}. 
The tensor type perturbation of the singly 
rotating Myers-Perry black hole does not have the decoupled sector and it was shown to be stable \cite{Kodama:2009bf}. 
}. We set the horizon radius to unity as
%============<Equation>=============%
%
\begin{eqnarray}
r_{0}=1,
\end{eqnarray}
%
%==================================%
by fixing the unit. 

\subsubsection{Scalar type perturbation}

The QNM frequency is given by the solution of the algebraic equation (\ref{QNMcond}), which can be solved analytically. 
We solve eq. (\ref{QNMcond}) for some cases below. Since the $j\neq 0$ modes do not show any instability, we consider only 
$j=0$ modes. The perturbation with $j\neq 0$ describes the deformation of $S^{D-4}$ of the Myers-Perry black hole. So we might 
be able to say that the axisymmetric instability occurs only in the "S-wave" sector in the perturbation
similar to the Gregory-Laflamme instability of the black brane \cite{Gregory:1993vy,Kudoh:2006bp}. 

Then, for this mode, the quasinormal mode condition (\ref{QNMcond}) can be written in a relatively simple form as
%============<Equation>=============%
%
\begin{eqnarray}
&&
(1+a^{2})^{3}\omega^{3}-(1+a^{2})^{2}(4i-3\ell+3am)\omega^{2} \notag \\
&&~~~~
+(1+a^{2})(-4+7\ell -3\ell^{2}-6iam(\ell-1) +a^{2}-4+3\ell+3m^{2})\omega \notag \\
&&~~~~
+(-i\ell^{3}+\ell^{2}(3i+ia^{2}+3am)-\ell(2i+5am+a^{3}m-ia^{2}(-2+3m^{2})) \notag \\
&&~~~~~~~~~~~~~~~~~~
+am(2-2iam-a^{2}(m^{2}-2)))=0.
\label{QNMcondj0}
\end{eqnarray}
%
%==================================% 
At $a=0$ the scalar type perturbation has the physical degree of freedom for $\ell\geq 2$ \cite{Kodama:2003kk} and, hence, 
we study the modes of $\ell\geq 2$. 

\paragraph{Schwarzschild black hole}
At $a=0$ where the solution is the Schwarzschild black hole we can solve eq. (\ref{QNMcondj0}) explicitly by
%============<Equation>=============%
%
\begin{eqnarray}
\omega=\pm\sqrt{\ell-1}-i(\ell-1),~~
\omega=-i \ell. \label{SCHQNM}
\end{eqnarray}
%
%==================================%
These modes correspond to the decoupled scalar and vector type mode of Schwarzschild black hole on $S^{D-2}$ \cite{Emparan:2014aba}.

\paragraph{Axisymmetric perturbation} 

\begin{figure}[t]
 \begin{center}
  \includegraphics[width=65mm,angle=0]{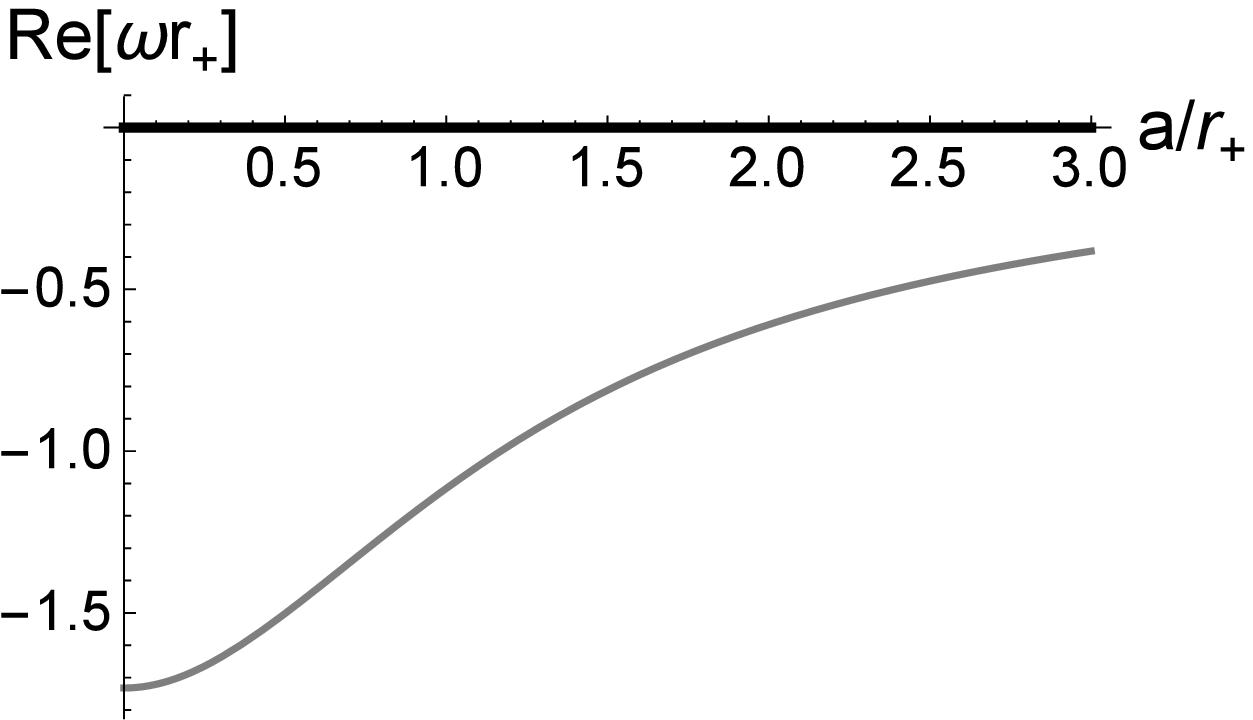}
  \hspace{5mm}
  \includegraphics[width=65mm,angle=0]{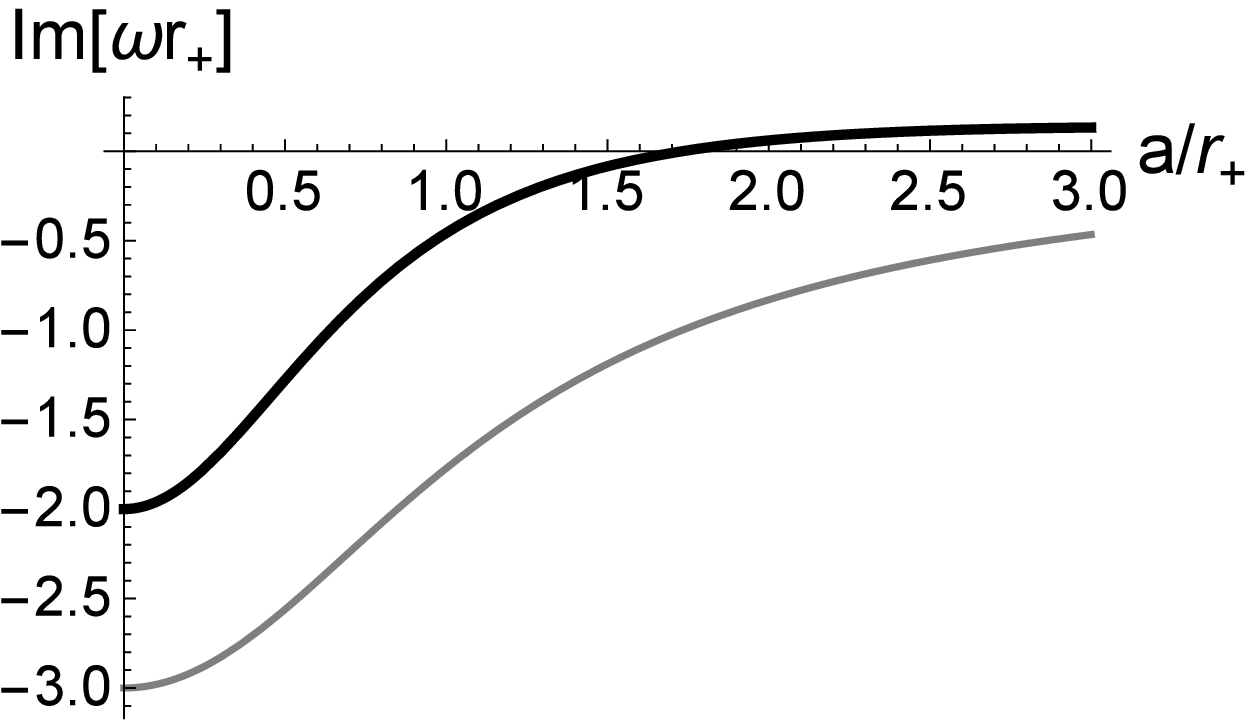}
 \end{center}
 \vspace{-5mm}
 \caption{The QNM frequency for $(\ell,m,j)=(4,0,0)$. The black thick line is the pure imaginary mode and 
it shows the instability at $a^{2}>3$. The gray thick line is the stable mode.  }
 \label{fig1}
\end{figure}

Next we consider the axisymmetric perturbation $m=0$ with $a\neq 0$. In this perturbation we find that there are 
stationary perturbations $\omega=0$ at $a_{c}$ where
%============<Equation>=============%
%
\begin{eqnarray}
a^{2}_{c}&=&\ell-1 \notag \\
&=&2k_{S}-1,
\label{accond}
\end{eqnarray}
%
%==================================%
and we used eq. (\ref{kdef}) with $j=m=0$. The assumption of $\ell\geq 2$ implies $k_{S}\geq 1$ for $j=m=0$, and this result reproduces the result 
(\ref{acond}) derived from the effective theory obtained in section \ref{2}. 
In figure \ref{fig1} we give the plot of the QNM frequency with $(\ell,m,j)=(4,0,0)$. The black thick line, purely imaginary frequency mode, 
shows the instability mode
and it becomes the marginally stable mode at $a^{2}_{c}=3$. 
%============<Figure>=============%
%
%
%===================================%
For the axisymmetric perturbation with $\omega=0$ and $j=0$ we can solve eqs. (\ref{eq1}) and (\ref{eq2})
explicitly by
%============<Equation>=============%
%
\begin{eqnarray}
C_{0}(\theta)=0,~~
D_{0}(\theta)=\tilde{D}_{0}\frac{\sin^{1+a^{2}}{\theta}}{1+a^{2}\cos^{2}{\theta}},
\end{eqnarray}
%
%==================================%
where $\tilde{D}_{0}$ is a constant. Then we can see that this stationary solution is equivalent to the perturbation
solution (\ref{Psol}) obtained in the effective theory. This equivalence suggests the validity of our QNM condition (\ref{QNMcond}).

\paragraph{Non-axisymmetric perturbation} 

The non-axisymmetric mode perturbation also shows the instability at $a^{2}>a^{2}_{c}$ 
where $a_{c}$ is given in eq. (\ref{accond}) as
%============<Equation>=============%
%
\begin{eqnarray}
a^{2}_{c}&=&\ell-1 \notag \\
&=&|m|+2k_{S}-1.
\end{eqnarray}
%
%==================================%
At the critical rotation 
the instability mode satisfies the superradiance condition as
%============<Equation>=============%
%
\begin{eqnarray}
\omega|_{a=a_{c}} = \frac{a_{c}m}{1+a^{2}_{c}}=m\Omega_{H}(a_{c}).
\end{eqnarray}
%
%==================================%
Hence the dynamical instability mode shows the superradiance instability at the same time.  
This coincidence of the onsets of dynamical and superradiant instability has been also observed
in the QNMs of the Myers-Perry black hole with equal spin \cite{Emparan:2014jca} up to $O(1/D)$ 
correction. The numerical results \cite{Hartnett:2013fba} also suggest the coincidence for the equal spin case. 
On the other hand the numerical analysis of QNMs of the singly rotating Myers-Perry black hole does not
show such coincidence \cite{Dias:2014eua,Shibata:2010wz}: the superradiant instability appears at slower rotation than the critical 
rotation of the dynamical instability. Thus our coincidence is just the property only at large $D$
limit. If one considers $1/D$ corrections of QNMs of singly rotating Myers-Perry black hole, 
there would be a difference between the onset of the superradiant and dynamical instability.  

\begin{figure}[t]
 \begin{center}
  \includegraphics[width=65mm,angle=0]{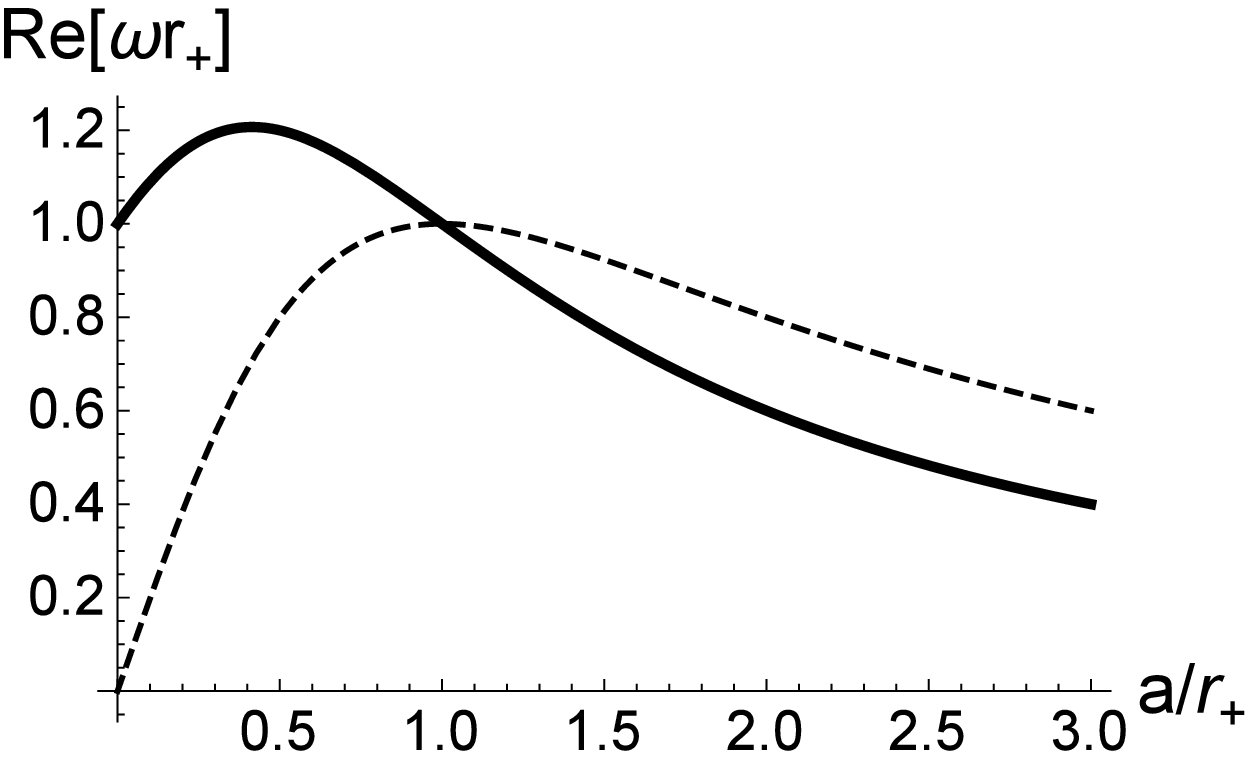}
  \hspace{5mm}
  \includegraphics[width=65mm,angle=0]{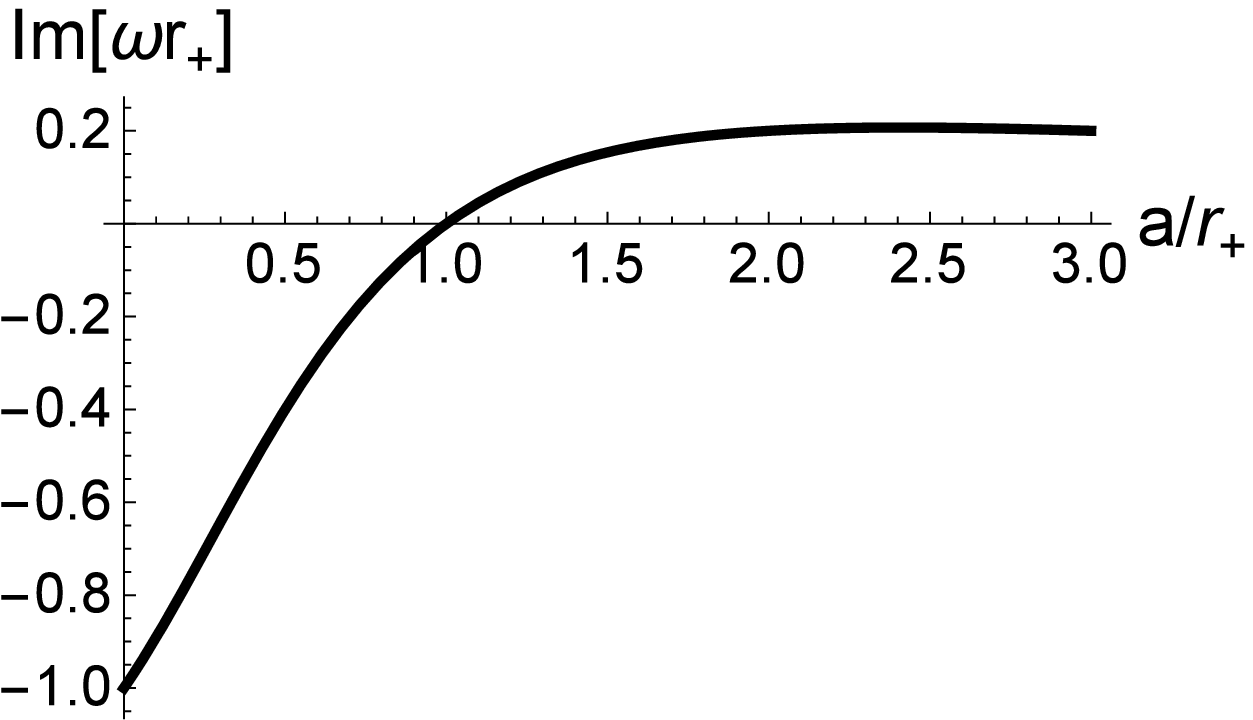}
 \end{center}
 \vspace{-5mm}
 \caption{The plot of the bar mode QNM frequency of $(\ell,m,j)=(2,2,0)$ is shown. The black dashed line is the superradiance 
condition. The onset of the dynamically unstable mode is same as the one of the superradiance instability.}
 \label{fig2}
\end{figure}

For the bar mode defined by $m=\ell$, we can solve eq. (\ref{QNMcondj0}) explicitly by
%============<Equation>=============%
%
\begin{eqnarray}
\omega_{\pm}=\frac{\pm\sqrt{m-1}+a(m-1)}{1+a^{2}}-i\frac{m-1\mp a\sqrt{m-1}}{1+a^{2}},~~
\omega_{(0)} = \frac{a m}{1+a^{2}} -i \frac{m}{1+a^{2}}.
\end{eqnarray}
%
%==================================%
$\omega_{+}$ shows the instability at $a>a_{c}$ and saturates the superradiance condition at $a=a_{c}$. In figure \ref{fig2}
we show the plot of the bar mode QNM frequency for $m=2$. 
%============<Figure>=============%
%
%
%===================================%
The QNM frequency for the bar mode was obtained numerically in \cite{Dias:2014eua}. The behavior of the frequency shows 
good agreements with our analytic results.

\begin{figure}[t]
 \begin{center}
  \includegraphics[width=65mm,angle=0]{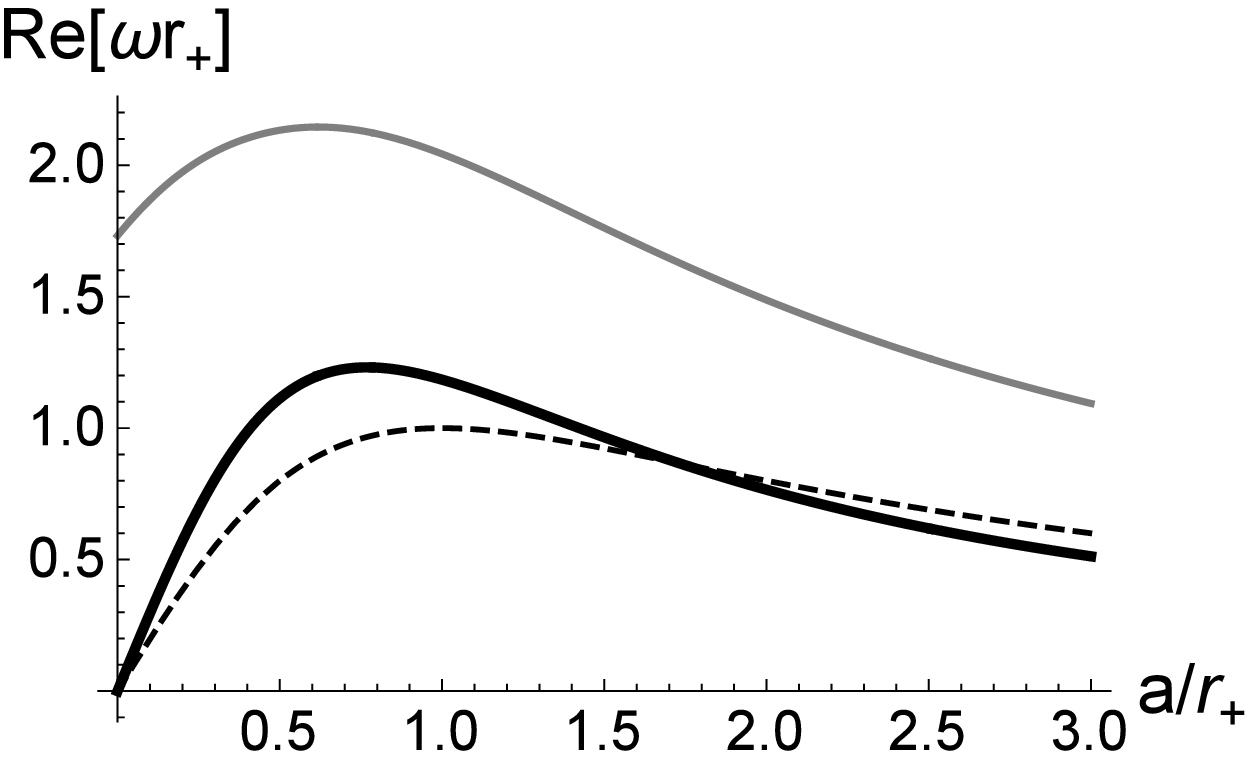}
  \hspace{5mm}
  \includegraphics[width=65mm,angle=0]{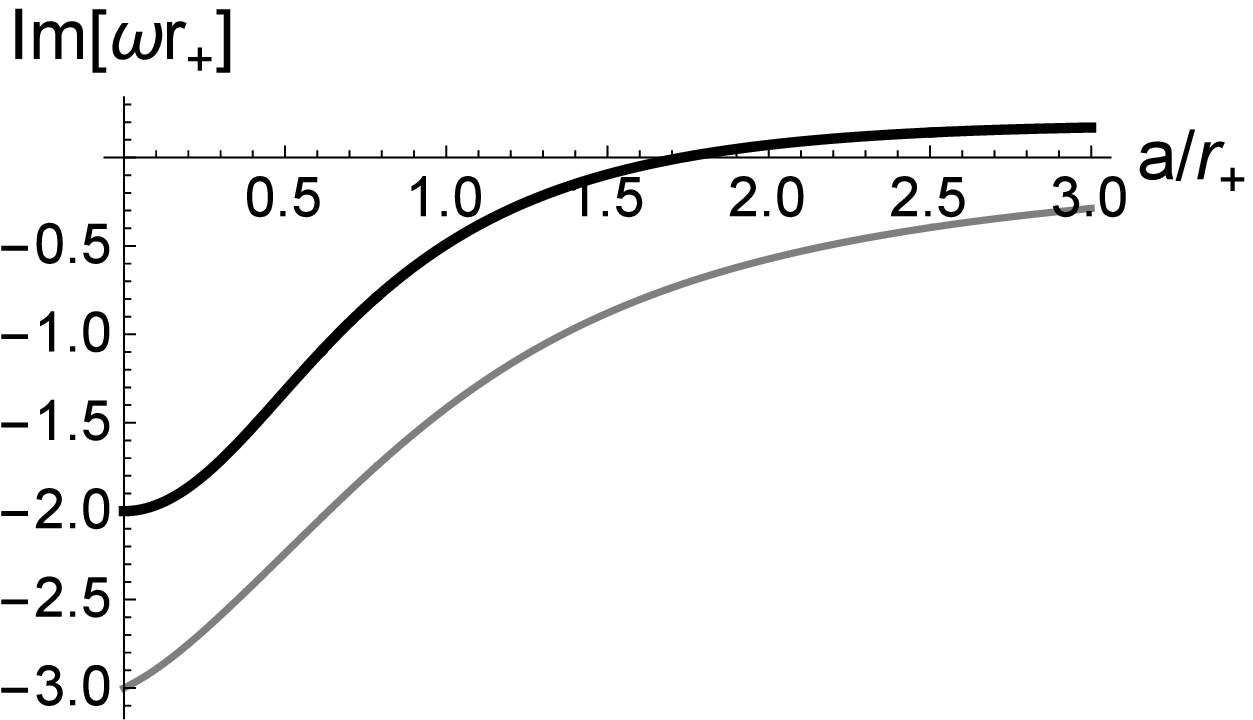}
 \end{center}
 \vspace{-5mm}
 \caption{The plot of the QNM frequency of $(\ell,m,j)=(4,2,0)$ is shown. The black dashed line is the superradiance condition.
The gray thick line is the stable mode $\omega_{-}$, while the black thick line shows the unstable mode $\omega_{+}$. The dynamical
instability appears exactly when the superradiance condition is satisfied. }
 \label{fig3}
\end{figure}

The non-axisymmetric perturbation with $\ell>m$ also shows the instability. But its origin is different from one 
of the bar mode. The instability mode of the bar mode perturbation comes from the scalar type perturbation 
of the Schwarzschild black hole at $a=0$ where the frequency is complex (see eq. (\ref{SCHQNM})). On the other hand non-axisymmetric mode with $\ell>m$ has an instability in the mode
coming from the vector type perturbation of the Schwarzschild black hole, whose frequency is a pure imaginary as (\ref{SCHQNM}). This is same situation with the 
perturbation of the Myers-Perry black hole with equal spin \cite{Emparan:2014jca}. In figure \ref{fig3} we give the plot
of QNM frequency for $\ell=4$, $m=2$. As we can see, the mode showing instability becomes purely imaginary at $a=0$, 
which corresponds to the vector type perturbation of the Schwarzschild black hole. 
%============<Figure>=============%
%
%
%===================================%

\subsubsection{Vector type perturbation}

The QNM frequency of the vector type perturbation can be given as
%============<Equation>=============%
%
\begin{eqnarray}
\omega^{V} = \omega^{V}_{\text{LO}} +\frac{\omega^{V}_{\text{NLO}}}{n}+O(1/n^{2}),
\end{eqnarray}
%
%==================================%
where
%============<Equation>=============%
%
\begin{eqnarray}
\omega^{V}_{\text{LO}} = \frac{a m}{1+a^{2}} -i\left( j-1 +\frac{\ell-j}{1+a^{2}}\right),
\label{vQNM}
\end{eqnarray}
%
%==================================%
and
%============<Equation>=============%
%
\begin{eqnarray}
&&
\omega^{V}_{\text{NLO}} =\frac{2 a m}{1+a^{2}}\Bigl(
\ell -1 +a^{2}(j-1) +\log{(1+a^{2})}
\Bigr) \notag \\
&&~~~~~~~~
-i\Biggl(
(j-1)^{2} +\frac{(1+a^{2})(\ell-j)(\ell+j-2)-2a^{2}m^{2}}{(1+a^{2})^{2}} \notag \\
&&~~~~~~~~~~~~~~~~~~~~~
+\left(
j-1 -\frac{\ell-j}{1+a^{2}}+\frac{2(\ell-j)}{(1+a^{2})^{2}}
\right)\log{(1+a^{2})}
\Biggr).
\end{eqnarray}
%
%==================================%
We have solved the perturbation equation up to $k=2$ order and obtained the QNM frequency up to $1/D$ correction.
This QNM frequency is obtained by imposing the same harmonics condition with the scalar type perturbation on the behavior at $\theta=0$
motivated by eq. (\ref{SPHsol}). 
The QNM frequency (\ref{vQNM}) reduces to the vector type perturbation on $S^{D-2}$ of the Schwarzschild black hole \cite{Emparan:2014aba}
at $a=0$ as
%============<Equation>=============%
%
\begin{eqnarray}
\omega = -i(\ell-1)\left( 1+\frac{\ell-1}{n}+O(1/n^{2}) \right).
\end{eqnarray}
%
%==================================%

\section{Summary} \label{5}

The large $D$ expansion method has been found to be useful to solve the gravitational problem as shown in \cite{Asnin:2007rw,Emparan:2013moa,Emparan:2013xia,Emparan:2013oza,Emparan:2014cia,Emparan:2014jca,Emparan:2014aba,Emparan:2015rva,Emparan:2015hwa,Bhattacharyya:2015dva}. In this paper we have performed
the next step for the further developing of the large $D$ expansion method by constructing the effective theory of the large $D$ stationary black hole
in asymptotically flat or AdS spacetime. Considering the additional degree of freedom of the horizon, {\it rotation radius of the black hole}, the effective 
theory becomes non-trivial and allows the existence of various solutions compared to the static case in the asymptotically flat spacetime which 
has only one unique solution, Schwarzschild black hole \cite{Emparan:2015hwa}. This new degree of freedom appears as the Lorentz boost of the black hole along the rotational direction, 
and we can rewrite the solution of the effective equation in very simple and illuminative expression by using this boost property. 

As applications of the effective theory we have considered the ellipsoidal embeddings and obtained the Myers-Perry black hole and bumpy black hole 
solutions in perturbative manner. Then we have succeeded to derive the threshold angular momentum of the instability of the Myers-Perry black hole. 
The bumpy black hole construction and perturbation analysis of the Myers-Perry black hole had been investigated numerically. These numerical 
studies of the solution and perturbations need very advanced and sophisticated technique. However the effective theory obtained in this 
paper is rather simple equation and we can obtain the solution easily. 

As another application we have obtained the quasinormal mode frequencies of the singly rotating Myers-Perry black hole. Our effective theory is only
for the stationary black hole. But, if we consider the dynamics with $\omega r_{0}=O(D^{0})$, the time-dependence becomes sub-dominant compared with 
the radial dynamics. As a result we could solve the perturbation equation as the ordinarily differential equation with respect to the radial direction.
Furthermore the leading order solution of our effective theory has a very important property, the boost property. All stationary black hole solution 
is represented as the boost transformation of the Schwarzschild black hole. Thus various things which Schwarzschild black hole possess are common
with stationary black holes. Using this useful feature we have solved the perturbation equation of the singly rotating Myers-Perry black hole explicitly
and obtained the quasinormal mode condition, which is just an algebraic equation, analytically. Although originally the perturbation equation is the partial
differential equation system, the large radial gradient and boost property of the large $D$ Myers-Perry black hole reduce the equation to the analytically 
solvable ordinarily differential equation. 

We can consider many directions of the extension of our work. In this paper we consider the solution with only one angular momentum in vacuum. 
Thus it is interesting to include more angular momentum, matter fields such as a gauge field or some compact dimensions into the effective theory of 
stationary black holes. These inclusions would give further additional degree of freedom for the deformation of the horizon. So its dynamics becomes much
richer. Another direction of the extension is the more detail investigation of the effective theory of stationary black hole by searching other solution 
such as black ring solutions and going to more higher order nonlinear solutions. By constructing such solution we can draw the phase diagram of black holes 
analytically and compare with numerical results. Actually the higher order investigation of the effective theory for the static solution \cite{Emparan:2015hwa}
allows very interesting phenomena such as the existence of the critical dimension of the non-uniform black string \cite{RTnubs}. 

%These constructions and investigations had been considered as almost impossible task in analytic way. However the large $D$ expansion makes it possible and %gives new interesting aspect of the black hole physics.

\section*{Acknowledgments}

The authors are very grateful to Roberto Emparan and Hideo Kodama for valuable comments on the draft and useful discussions.   
KT was supported by JSPS Grant-in-Aid for Scientific Research No.26-3387.

\appendix

\section{Singly rotating AdS Myers-Perry black hole} \label{A}

In this appendix we give the metric of the AdS Myers-Perry black hole and its large $D$ limit in our notation. 
The $D=n+3$ dimensional singly rotating AdS Myers-Perry black hole is given by \cite{Hawking:1998kw,Gibbons:2004uw}
\footnote{
We change the $\phi$ coordinate by
%============<Equation>=============%
%
\begin{eqnarray}
\phi \rightarrow \phi -\frac{a}{L^{2}}dt \notag
\end{eqnarray}
%
%==================================%
from \cite{Hawking:1998kw,Gibbons:2004uw} to satisfy the boundary condition (\ref{BCIn}).
}
%============<Equation>=============%
%
\begin{eqnarray}
&&
ds^{2} = -A(r,\theta)^{2} dt^{2} + B(r,\theta)^{2}d\hat{\sR}^{2} +F(r,\theta)^{2}(d\phi-w(r,\theta)dt)^{2} \notag \\
&&~~~~~~~~~~~~~~~~~~~~~~~~~~~~~~~~~~~~~~~~~
+G(r,\theta)^{2}d\theta^{2} +r^{2}\cos^{2}{\theta}d\Omega^{2}_{D-4},
\end{eqnarray}
%
%==================================%
where
%============<Equation>=============%
%
\begin{eqnarray}
A(r,\theta)^{2}= \frac{\Delta_{\hat{\sR}}\Delta_{\theta}(r^{2}+L^{2})(r^{2}+a^{2}\cos^{2}{\theta})}
{L^{2}\Xi\Delta_{\hat{\sR}}(r^{2}+a^{2}\cos^{2}{\theta})+r^{2}L^{2}\Delta_{\theta}(r^{2}+a^{2})},
\end{eqnarray}
%
%==================================%
%============<Equation>=============%
%
\begin{eqnarray}
B(r,\theta)^{2} = \frac{r^{2}}{n^{2}\hat{\sR}^{2}}\frac{\hat{\sR}(r^{2}+a^{2}\cos^{2}{\theta})}{\Delta_{\hat{\sR}}},
\end{eqnarray}
%
%==================================%
%============<Equation>=============%
%
\begin{eqnarray}
F(r,\theta)^{2} = \sin^{2}{\theta}\left( \frac{r^{2}+a^{2}}{\Xi} 
+\frac{a^{2}r^{2}\sin^{2}{\theta}}{\Xi^{2}(r^{2}+a^{2}\cos^{2}{\theta})\hat{\sR}} \right),
\end{eqnarray}
%
%==================================%
%============<Equation>=============%
%
\begin{eqnarray}
w(r,\theta) = \frac{ar^{2}\Delta_{\theta}}
{L^{2}\Xi\Delta_{\hat{\sR}}(r^{2}+a^{2}\cos^{2}{\theta}) +r^{2}L^{2}\Delta_{\theta}(r^{2}+a^{2})},
\end{eqnarray}
%
%==================================%
and
%============<Equation>=============%
%
\begin{eqnarray}
G(r,\theta)^{2} = \frac{r^{2}+a^{2}\cos^{2}{\theta}}{\Delta_{\theta}}.
\end{eqnarray}
%
%==================================%
Here we defined
%============<Equation>=============%
%
\begin{eqnarray}
\hat{\sR} =\left(\frac{r}{r_{0}}\right)^{n},~~
\Xi=1-\frac{a^{2}}{L^{2}},
\label{Rdef}
\end{eqnarray}
%
%==================================%
and following functions
%============<Equation>=============%
%
\begin{eqnarray}
\Delta_{\hat{\sR}}=r^{2}-(r^{2}+a^{2})\left( 1+ \frac{r^{2}}{L^{2}} \right)\hat{\sR},~~
\Delta_{\theta} =1-\frac{a^{2}}{L^{2}}\cos^{2}{\theta}.
\end{eqnarray}
%
%==================================%
The horizon position, $r=r_{+}$, is defined by the positive real root of $\Delta_{\hat{\sR}}$ by
%============<Equation>=============%
%
\begin{eqnarray}
\Delta_{\hat{\sR}}(r_{+})=0.
\end{eqnarray}
%
%==================================% 
$L$ is the AdS curvature and the metric is the solution of the Einstein equation with a cosmological constant
%============<Equation>=============%
%
\begin{eqnarray}
\Lambda= -\frac{(D-1)(D-2)}{L^{2}}.
\end{eqnarray}
%
%==================================%
At the large $D$ limit the metric is simplified to
\footnote{
In the following we set $r_{0}=1$.
}
%============<Equation>=============%
%
\begin{eqnarray}
&&
ds^{2} = \frac{d\sR^{2}}{n^{2}}\frac{L^{2}(1+a^{2}\cos^{2}{\theta})}{(1+a^{2})(1+L^{2})\sR(\sR-1)} \notag \\
&&~~~~~~~~
-\left( 1 -\frac{\cosh^{2}{\sigma(\theta)}}{\sR} \right)d\hat{t}^{2}
+\left( 1 +\frac{\sinh^{2}{\sigma(\theta)}}{\sR} \right)d\hat{\phi}^{2}
\notag \\
&&~~~~~~~~
-\frac{2\sinh{\sigma(\theta)}\cosh{\sigma(\theta)}}{\sR}d\hat{t}~d\hat{\phi} 
+ \frac{1+a^{2}\cos^{2}{\theta}}{\Delta_{\theta}}d\theta^{2} +\cos^{2}{\theta}d\Omega^{2}_{D-4},
\label{MPmet}
\end{eqnarray}
%
%==================================%
where we introduced
%============<Equation>=============%
%
\begin{eqnarray}
\sR=\frac{(1+a^{2})(1+L^{2})}{L^{2}}\hat{\sR},~~
\cosh^{2}{\sigma(\theta)} = \frac{(1+a^{2})\Delta_{\theta}}{(1+a^{2}\cos^{2}{\theta})\Xi}.
\end{eqnarray}
%
%==================================%
$d\hat{t}$ and $d\hat{\phi}$ are defined by
%============<Equation>=============%
%
\begin{eqnarray}
d\hat{t}^{2}=\frac{(1+L^{2})\Delta_{\theta}}{L^{2}\Xi}dt^{2},~~
d\hat{\phi}^{2}=\frac{1+a^{2}}{\Xi}\sin^{2}{\theta}d\phi^{2}.
\end{eqnarray}
%
%==================================%
At the large $D$ limit the horizon is located on 
%============<Equation>=============%
%
\begin{eqnarray}
\sR\Big|_{r=r_{+}}=1+O(1/n).
\end{eqnarray}
%
%==================================%
This solution can be reproduced by the leading order solution in section \ref{2} by 
%============<Equation>=============%
%
\begin{eqnarray}
\sV_{0}(\theta)^{2}=\frac{(1+L^{2})\Delta_{\theta}}{L^{2}\Xi},~~
\cR_{\tH}(z)=\cos{\theta},
\end{eqnarray}
%
%==================================%
and
%============<Equation>=============%
%
\begin{eqnarray}
\frac{dz}{d\theta} =\sqrt{\frac{1+a^{2}\cos^{2}{\theta}}{\Delta_{\theta}}},~~
\sR=\cosh^{2}\rho.
\end{eqnarray}
%
%==================================%
The surface gravity and horizon angular velocity at the leading order of large $D$ limit are
%============<Equation>=============%
%
\begin{eqnarray}
\hat{\kappa} = \frac{1}{2}\frac{1+L^{2}}{L^{2}},~~
\Omega_{H} =\frac{a}{1+a^{2}}\frac{1+L^{2}}{L^{2}}.
\label{AdSTW}
\end{eqnarray}
%
%==================================%
The metric (\ref{MPmet}) becomes the large $D$ metric of the Schwarzschild black hole \cite{Emparan:2013xia}
%============<Equation>=============%
%
\begin{eqnarray}
ds^{2} = \frac{4}{n^{2}}d\rho^{2}-\tanh^{2}{\rho}~dt^{2}+dz^{2}+\sin^{2}{\theta}d\phi^{2}+\cos^{2}{\theta}d\Omega^{2}_{D-4},
\end{eqnarray}
%
%==================================% 
when $a=0$ and $1/L=0$. We changed the coordinate by $\sR=\cosh^{2}{\rho}$.

\paragraph{Structure in $1/D$ correction}
Let us consider the $1/D$ correction in $g_{tt}$ of the AdS Myers-Perry black hole at $a=0$. Using 
%============<Equation>=============%
%
\begin{eqnarray}
r=1+\frac{\log{\hat{\sR}}}{n}+O(1/n^{2})
\end{eqnarray}
%
%==================================% 
we find that $g_{tt}$ becomes 
%============<Equation>=============%
%
\begin{eqnarray}
g_{tt}&=&-A(r,\theta)^{2} \notag \\
&=& 1+\frac{1}{L^{2}}-\frac{1}{\hat{\sR}}+\frac{2\log{\hat{\sR}}}{nL^{2}}+O(1/n^{2}).
\end{eqnarray}
%
%==================================%
At large $\sR$ $g_{tt}$ has a linear term in $\rho$ since the relation between $\rho$ and $\hat{\sR}$ is
%============<Equation>=============%
%
\begin{eqnarray}
\hat{\sR}=\frac{L^{2}}{1+L^{2}}\cosh^{2}{\rho}.
\end{eqnarray}
%
%==================================%
Thus, in general, the higher order corrections in $A(\rho,\theta)$ cannot be assumed to be damping exponentially at the 
overlap region $n\gg \rho\gg 1$ in the presence of the cosmological constant as mentioned in section \ref{2}.

\subsection{Ellipsoidal embedding}

The AdS-Myers-Perry black hole is described as the solution of the effective equation (\ref{EFT}) for the ellipsoidal 
embedding in the AdS background. The ellipsoidal coordinate in AdS is given by
\footnote{
The coordinate transformation from the usual AdS coordinate to the ellipsoidal coordinate is \cite{Hawking:1998kw}
%============<Equation>=============%
%
\begin{eqnarray}
r^{2}\sin^{2}{\theta} \rightarrow \Xi^{-1}(r^{2}+a^{2})\sin^{2}{\theta},~~
r^{2}\cos^{2}{\theta} \rightarrow r^{2}\cos^{2}{\theta}. \notag
\end{eqnarray}
%
%==================================%
}
%============<Equation>=============%
%
\begin{eqnarray}
&&
ds^{2}=-\frac{\Delta_{\theta}(L^{2}+r^{2})}{L^{2}\Xi}dt^{2}+\frac{L^{2}(r^{2}+a^{2}\cos^{2}{\theta})}{(r^{2}+a^{2})(L^{2}+r^{2})}dr^{2} \notag \\
&&~~~~~~~~~~
+\frac{r^{2}+a^{2}\cos^{2}{\theta}}{\Delta_{\theta}}d\theta^{2}+\frac{(r^{2}+a^{2})\sin^{2}{\theta}}{\Xi}d\phi^{2}+r^{2}\cos^{2}{\theta}d\Omega^{2}_{D-4}.
\end{eqnarray}
%
%==================================%
The ellipsoidal embedding is defined by 
%============<Equation>=============%
%
\begin{eqnarray}
r=r(\theta)=
\frac{\cR_{\tH}(\theta)}{\cos{\theta}},
\end{eqnarray}
%
%==================================% 
and the leading order metric of the AdS Myers-Perry black hole in the $1/D$ expansion is reproduced by $r(\theta)=1$ when we fix the 
horizon position by $r_{+}=1$ with the surface gravity and horizon angular velocity given in eq. (\ref{AdSTW}). We consider the perturbation around this 
AdS Myers-Perry black hole embedding. As done in section \ref{2} the obtained perturbed solution would give the bumpy black hole 
solution in AdS in the perturbative manner. At first we fix the surface gravity to 
%============<Equation>=============%
%
\begin{eqnarray}
\hat{\kappa}=\frac{1+L^{2}}{2L^{2}},
\end{eqnarray}
%
%==================================%
by using the time coordinate normalization in near zone. The perturbed embedding is 
%============<Equation>=============%
%
\begin{eqnarray}
\cR_{\tH}(\theta)=\cos{\theta}\Bigl[1 + \epsilon \hat{\cR}(\theta)+O\left(\epsilon^{2}\right)\Bigr], 
\end{eqnarray}
%
%==================================%
where
%============<Equation>=============%
%
\begin{eqnarray}
\epsilon=\Omega_{H}-\frac{a}{1+a^{2}}\frac{1+L^{2}}{L^{2}}.
\end{eqnarray}
%
%==================================%
Then, perturbing eq. (\ref{EFT}) with respect to $\epsilon$, we find the perturbative solution
%============<Equation>=============%
%
\begin{eqnarray}
\hat{\cR}(\theta)= \frac{a(1+a^{2})^{2}\sin^{2}{\theta}}{\Xi(1-a^{2})(1+a^{2}\cos^{2}{\theta})}
+A\frac{(\sin{\theta})^{\delta_{\text{MP}}}\Delta_{\theta}^{1-\delta_{\text{MP}}/2}}{1+a^{2}\cos^{2}{\theta}},
\label{PsolAdS}
\end{eqnarray}
%
%==================================%
where
%============<Equation>=============%
%
\begin{eqnarray}
\delta_{\text{MP}}=\frac{(1+a^{2})(L^{2}-1)}{L^{2}\Xi}.
\end{eqnarray}
%
%==================================%
This perturbation solution becomes regular at $\theta=0$ when $a=a_{c}$ given by
%============<Equation>=============%
%
\begin{eqnarray}
a^{2}_{c}=\frac{(2k-1)L^{2}+1}{L^{2}+2k-1}
\label{accondAdS}
\end{eqnarray}
%
%==================================%
where $k$ is a positive integer. At $L\rightarrow\infty$ eq. (\ref{accondAdS}) reproduces the threshold angular momentum (\ref{acond}).
Hence we expect that eq. (\ref{accondAdS}) gives the threshold angular momentum of the AdS Myers-Perry black hole and branching angular 
momentum for the bumpy black hole in AdS. It is interesting to check this result by solving the perturbation equation of the AdS Myers-Perr black 
hole and obtaining the QNM frequency as done for the Myers-Perry black hole in section \ref{4}.

\section{Next-to-leading order analysis}\label{Am}

Here we briefly give results for next-to-leading order solutions. In particular we will see how the constancy condition of the surface gravity
and horizon angular velocity, eq. (\ref{kOccond}), can be obtained.  

Our $D=n+3$ dimensional metric ansatz is 
%============<Equation>=============%
%
\begin{eqnarray}
ds^{2} &=& \frac{N^{2}(\rho,z)}{n^{2}}d\rho^{2}-A(\rho,z)^{2} dt^{2} +F(\rho,z)^{2}(d\phi-W(\rho,z)dt)^{2} \notag \\
&&~~~~~~~~~~~~~~~~~~~~~~~~~~~
+G(\rho,z)^{2}dz^{2} +H(\rho,z)^{2}q_{AB}dx^{A}dx^{B}.
\end{eqnarray}
%
%==================================%
We solve the Einstein equations for metric functions and the extrinsic curvature on $\rho=\text{constant}$ surface defined by 
%============<Equation>=============%
%
\begin{eqnarray}
K^a_{~b} =\frac{n}{2N(\rho,z)}g^{ac}\partial_\rho g_{cb}.
\end{eqnarray}
%
%==================================%
The Einstein equations are decomposed on $\rho=\text{constant}$ surface as
%============<Equation>=============%
%
\begin{eqnarray}
&&-R+K^2-K^a_{~b} K^b_{~a}-\frac{(D-1)(D-2)}{L^{2}}=0,  \\
&&\nabla_{a}K^{a}_{~b}-\nabla_{b}K=0,  \label{vconapp}\\
&&\frac{n}{N}\partial_\rho K^a_{~b}=-K K^a_{~b} + R^a_{~b}+\frac{D-1}{L^{2}}\delta^{a}_{~b} 
 -\frac{1}{N}\nabla^a\nabla_{b} N. \label{Keqapp}
\end{eqnarray}
%
%==================================%
The leading order solutions which are regular on the horizon are
%============<Equation>=============%
%
\begin{eqnarray}
&&
A(\rho,z)^{2} = \frac{A_{0}(z)^{2}F_{0}(z)^{2}\tanh^{2}{\rho}}{F_{0}(z)^{2}-A_{0}(z)^{2}C_{t\phi}(z)^{2}\tanh^{2}{\rho}},\\
&&
F(\rho,z)^{2} =F_{0}(z)^{2}-A_{0}(z)C_{t\phi}(z)^{2}\tanh^{2}{\rho},\\
&&
W(\rho,z) = \frac{F_{0}(z)^{2}(1-C_{tt}(z))+A_{0}(z)^{2}C_{t\phi}(z)^{2}C_{tt}(z)\tanh^{2}{\rho}}
{K_{t\phi}(z)F_{0}(z)^{2}-A_{0}(z)^{2}C_{t\phi}(z)^{3}\tanh^{2}{\rho}},
\end{eqnarray}
%
%==================================% 
and
%============<Equation>=============%
%
\begin{eqnarray}
&&
G(\rho,z) = 1 - \frac{2r_{0}(z)^{2}}{n}\left(\frac{\cR_{\tH}''(z)}{\cR_{\tH}(z)}-\frac{1}{L^{2}}\right)\log{(\cosh{\rho})},   \\
&&
H(\rho,z) = \cR_{\tH}(z)\left( 1 +\frac{2}{n}\log{(\cosh{\rho})} \right). 
\end{eqnarray}
%
%==================================%
To satisfy the boundary conditions at $\rho\gg 1$, $C_{tt}$ and $C_{t\phi}$ are found to be
%============<Equation>=============%
%
\begin{eqnarray}
C_{tt}(z)=\frac{1}{A_{0}(z)^{2}}~~
C_{t\phi}(z)=\frac{F_{0}(z)\sqrt{1-A_{0}(z)^{2}}}{A_{0}(z)}.
\label{AFcondapp}
\end{eqnarray}
%
%==================================%
Then we can see that the leading order solutions satisfy the boundary condition at $\rho\gg 1$ as
%============<Equation>=============%
%
\begin{eqnarray}
A(\rho,z)=1+O(e^{-\rho}),~~W(\rho,z)=O(e^{-\rho}). 
\end{eqnarray}
%
%==================================%
Furthermore the functions in this leading order solution should satisfy one equation coming from the vector constraint (\ref{vconapp})
given by
%============<Equation>=============%
%
\begin{eqnarray}
\frac{d}{dz}\log{\left( \frac{A_{0}(z)C_{tt}(z)}{r_{0}(z)} \right)}-(1-C_{tt}(z))\frac{d}{dz}\log{\left(\frac{C_{tt}(z)}{C_{t\phi}(z)}\right)}
=0. \label{vconstLOapp}
\end{eqnarray}
%
%==================================%
This equation is rewritten by using eq. (\ref{AFcondapp}) as
%============<Equation>=============%
%
\begin{eqnarray}
\frac{d}{dz}\log{\left( A_{0}r_{0}(z) \right)} 
+\frac{1-A_{0}(z)^{2}}{A_{0}(z)^{2}}\frac{d}{dz}
\log{\left(
A_{0}(z)F_{0}(z)\sqrt{1-A_{0}(z)^{2}}
\right)}=0.
\label{AFeqapp}
\end{eqnarray}
%
%==================================%
We consider the $1/D$ corrections to the leading order solutions. Especially, by solving eq. (\ref{Keqapp}), we found that 
the $1/D$ correction to $K^{z}{}_{z}$, say ${}^{(1)}K^{z}{}_{z}$, has the following form
%============<Equation>=============%
%
\begin{eqnarray}
{}^{(1)}K^{z}{}_{z} =\frac{C^{(1)}_{zz}}{\sinh{\rho}} + \delta K^{z}{}_{z}(\rho,z), 
\end{eqnarray}
%
%==================================%  
where $C^{(1)}_{zz}$ is an integration function with respect to $\rho$-integration. $\delta K^{z}{}_{z}(\rho,z)$ is
a solution coming from the integration of the source term at next-to-leading order. $C^{(1)}_{zz}$ is used to eliminate
the divergence of $O(\rho^{-1})$ in $\delta K^{z}{}_{z}(\rho,z)$ at the horizon $\rho=0$. Even though the $O(\rho^{-1})$
divergence is eliminated by the integration function, ${}^{(1)}K^{z}{}_{z}$ cannot satisfy the regularity condition at the horizon. Actually 
$\delta K^{z}{}_{z}(\rho,z)$ has the following behavior after eliminating $O(\rho^{-1})$ divergence at the horizon
%============<Equation>=============%
%
\begin{eqnarray}
&&
\delta K^{z}{}_{z}(\rho,z) = 
-\frac{r_{0}(z)\log{\rho}}{(1-A_{0}(z)^{2})A_{0}(z)^{2}F_{0}(z)^{2}\rho} \notag \\
&&~~~~~~~~~~\times
\Bigl[
F_{0}(z)A_{0}'(z)+(1-A_{0}^{2})A_{0}(z)F_{0}'(z)
\Bigr]^{2} 
+O(1).
\end{eqnarray}
%
%==================================%
To satisfy the regularity condition on the horizon we should eliminate also this $O(\rho^{-1}\log{\rho})$ divergence. 
Then $A_{0}(z)$ and $F_{0}(z)$ should satisfy additional condition
%============<Equation>=============%
%
\begin{eqnarray}
F_{0}(z)A_{0}'(z)+(1-A_{0}^{2})A_{0}(z)F_{0}'(z)=0.
\label{Feqapp}
\end{eqnarray}
%
%==================================%
This condition can be solved by
%============<Equation>=============%
%
\begin{eqnarray}
F_{0}(z)=\frac{\sqrt{1-A_{0}(z)^{2}}}{\Omega_{\tH}A_{0(z)}},
\label{Fsolapp}
\end{eqnarray}
%
%==================================%
where we introduce an integration constant $\Omega_{\tH}$. Using this solution (\ref{Fsolapp}), eq. (\ref{AFeqapp})
is reduced to the equation only for $A_{0}(z)$, and it can be solved by
%============<Equation>=============%
%
\begin{eqnarray}
A_{0}(z)=2\hat{\kappa}r_{0}(z),
\label{Asolapp}
\end{eqnarray}
%
%==================================%
where $\hat{\kappa}$ is an integration constant. 

Let us summarize above results. We obtained additional condition (\ref{Feqapp}) from the boundary condition on $O(1/D)$ corrections. 
By combining this additional condition with the leading order vector constraint (\ref{AFeqapp}) we get the expressions of 
$A_{0}(z)$ and $F_{0}(z)$ in terms of $r_{0}(z)$ and integration constants as eqs. (\ref{Fsolapp}) and (\ref{Asolapp}). Using
eq. (\ref{AFcondapp}) we can see that eq. (\ref{kOccond}) is obtained by defining $\kappa=n\hat{\kappa}$.

\section{Spheroidal harmonics at large $D$}\label{ASP}

We give some analysis for the spheroidal harmonics in the $1/D$ expansion. Let us consider the massless scalar field
equation 
%============<Equation>=============%
%
\begin{eqnarray}
\Box \Psi=0,
\label{psieq}
\end{eqnarray}
%
%==================================%
in the $D=n+3$ dimensional flat spacetime in the ellipsoidal coordinate given by
%============<Equation>=============%
%
\begin{eqnarray}
&&
ds^{2} = -dt^{2}+\frac{r^{2}+a^{2}\cos^{2}{\theta}}{r^{2}+a^{2}}dr^{2}+ 
(r^{2}+a^{2}\cos^{2}{\theta})d\theta^{2}  \notag \\
&&~~~~~~~~~~~~~~~~~~~~~~~~~~~~~
+(r^{2}+a^{2})\sin^{2}{\theta}d\phi^{2} +r^{2}\cos^{2}{\theta}d\Omega^{2}_{D-4}.
\end{eqnarray}
%
%==================================%
$a$ is an oblateness parameter. The scalar field can be decomposed as 
%============<Equation>=============%
%
\begin{eqnarray}
\Psi = e^{-i\omega t}e^{im\phi}\psi(r)S(\theta) \mathbb{Y}_{j}. 
\end{eqnarray}
%
%==================================% 
$\mathbb{Y}_{j}$ is the spherical harmonics on $S^{D-4}$ with the angular momentum number $j$. 
Eq. (\ref{psieq}) becomes equations for $\psi(r)$ and $S(\theta)$ as
%============<Equation>=============%
%
\begin{eqnarray}
\Biggl[
\frac{1}{r^{n-1}}\partial_{r}r^{n-1}\partial_{r}+\frac{((r^{2}+a^{2})\omega+am)^{2}}{r^{2}+a^{2}}
-\frac{a^{2}(j(j+n-2))}{r^{2}}-\Lambda
\Biggr]\psi(r)=0,
\end{eqnarray}
%
%==================================%
and
%============<Equation>=============%
%
\begin{eqnarray}
&&
\Biggl[
\frac{1}{\sin{\theta}\cos^{n-1}{\theta}}\partial_{\theta}\sin{\theta}\cos^{n-1}{\theta}\partial_{\theta}
-\frac{(a\omega\sin^{2}{\theta}+m)^{2}}{\sin^{2}{\theta}} \notag \\
&&~~~~~~~~~~~~~~~~~~~~~~~~~~~~~~~~~~~~~~~~~~~~~
-\frac{j(j+n-2)}{\cos^{2}{\theta}}+\Lambda
\Biggr]S(\theta)=0. 
\label{Seqapp}
\end{eqnarray}
%
%==================================%
$\Lambda$ is the separation constant. The spheroidal harmonics, $S(\theta)$, is defined by the solution of eq. (\ref{Seqapp}). 
While there are numerical analysis \cite{Berti:2005gp} to find solution and eigenvalue of eq. (\ref{Seqapp}), 
but it is hard to obtain analytic solutions of eq. (\ref{Seqapp}). Here we apply the large $D$ expansion to solve eq. (\ref{Seqapp}). 
At first we assume that the parameters in eq. (\ref{Seqapp}) have following orders at large $D$
%============<Equation>=============%
%
\begin{eqnarray}
\omega=O(1),~~
a=O(1),~~
m=O(1),~~
j=O(1),~~
\Lambda=O(n).
\end{eqnarray}
%
%==================================%
The reason for $\Lambda=O(n)$ can be seen by considering $a=0$ limit. At $a=0$ eq. (\ref{Seqapp}) is reduced to the equation
for the spherical harmonics. The equation at $a=0$ can be solved by $S^{(a=0)}_{\ell}(\theta)$
%============<Equation>=============%
%
\begin{eqnarray}
S^{(a=0)}(\theta)= \cos^{j}{\theta}~\sin^{|m|}{\theta}~{}_{2}F_{1}(-k_{S},k_{S}+|m|+j+n/2,|m|+1;\sin^{2}{\theta}),
\label{Sphapp}
\end{eqnarray}
%
%==================================% 
with the separation constant
%============<Equation>=============%
%
\begin{eqnarray}
\Lambda^{(a=0)}=\ell(\ell+n).
\end{eqnarray}
%
%==================================%
$k_{S}$ is given by
%============<Equation>=============%
%
\begin{eqnarray}
k_{S}=\frac{\ell-j-|m|}{2}.
\end{eqnarray}
%
%==================================%
Then, at $a=0$, the separation constant is $O(n)$. The spheroidal harmonics becomes the spherical harmonics at $a=0$.
Thus it is natural to assume $\Lambda=O(n)$ also at $a\neq 0$. The important observation is that the spherical harmonics (\ref{Sphapp})
has the following large $D$ behaviors
%============<Equation>=============%
%
\begin{eqnarray}
S^{(a=0)}_{\ell}(\theta) = \sin^{\ell-j}{\theta}\cos{^{j}}{\theta} + O(1/n). 
\label{Sphapp2}
\end{eqnarray}
%
%==================================%
Usually the spherical harmonics $S^{(a=0)}_{\ell}(\theta)$ is understood as the function which has $\ell$ nodes. This 
$\ell$ is the overtone number along $\theta$ direction. At large $D$ limit the situation is changed, and any nodes
would disappear according to the large $D$ limit solution (\ref{Sphapp2}). Instead the quantum number $\ell$ can be
read from the behavior only around $\theta=0$ as the degree of vanishing as 
%============<Equation>=============%
%
\begin{eqnarray}
S^{(a=0)}_{\ell}(\theta) = \theta^{\ell}\left( 1+O(\theta, n^{-1}) \right).
\end{eqnarray}
%
%==================================%. 
Although it is not still unclear why we can obtain the quantum number $\ell$ only by the behavior around $\theta=0$, 
not by the behavior both at $\theta=0$ and $\theta=\pi/2$, such behavior is useful for the quasinormal mode analysis
of the Myers-Perry black hole.   

Next we consider the contributions by $a\neq 0$ to the spheroidal harmonics. Let us assume that the spheroidal 
harmonics $S(\theta)$
is modified by the contributions from $a$ and $\omega$ as
%============<Equation>=============%
%
\begin{eqnarray}
S(\theta) = S^{(a=0)}_{\ell}(\theta)\left( 1+ \delta S(\theta) \right). 
\end{eqnarray}
%
%==================================%
Substituting this into eq. (\ref{Seqapp}) and linearizing the equation with respect to $\delta S(\theta)$, we can obtain
the equation for $\delta S(\theta)$. The obtained equation, however, does not have $a$ and $\omega$ dependences in its equation at the leading order of the large $D$ limit.
This is because the contributions of $a$ and $\omega$ are not leading order effects in eq. (\ref{Seqapp}) in the $1/D$ expansion. 
Thus $\delta S(\theta)$ can be set to zero also for the spheroidal harmonics $S(\theta)$ at the leading order in the $1/D$ 
expansion. Hence the spheroidal harmonics has the following large $D$ form
%============<Equation>=============%
%
\begin{eqnarray}
S(\theta) &=& S^{(a=0)}_{\ell}(\theta) + O(1/n) \notag \\
&=& \sin^{\ell-j}{\theta}\cos{^{j}}{\theta}+ O(1/n),
\end{eqnarray}
%
%==================================%
and the separation constant also becomes
%============<Equation>=============%
%
\begin{eqnarray}
\Lambda = \ell n\left( 1+O(n^{-1}) \right),
\end{eqnarray}
%
%==================================%
where $\ell$ is non-negative integer. The quantum number along $\theta$ direction, $\ell$, can be obtained
again only by the behavior around $\theta=0$. If one considers the $1/D$ corrections to the spheroidal harmonics we can obtain
the $1/D$ correction to the separation constant. In the analysis of this paper we consider only the leading order results
of the quasinormal modes, so we do not pursue this analysis in more detail here.

\section{Detail analysis on the vector type perturbation} \label{vec}

In this appendix we show the detail analysis of the vector type perturbation of the Myers-Perry- black hole.
The vector type perturbation on the Myers-Perry black hole is decomposed to 
%============<Equation>=============%
%
\begin{eqnarray}
h^{(V)}_{\mu\nu}dx^{\mu}dx^{\nu}= e^{-i\omega t}e^{i m\phi}\Bigl[
f^{(V)\text{Sch}}_{I}\mathbb{Y}^{(V)j}_{A}e^{(I)}e^{(A)}+H^{(V)}_{L}\mathbb{Y}^{(V)j}_{AB}e^{(A)}e^{(B)}
\Bigr],
\end{eqnarray}
%
%==================================%
where the vielbein and the vector harmonics were defined in section \ref{3}. The decoupling perturbation variables 
are defined  by 
%============<Equation>=============%
%
\begin{eqnarray}
F^{(V)}_{0}=\frac{\cosh{\rho}}{\sqrt{1+r_{0}(z)^{2}\sinh^{2}{\rho}}}f^{(V)}_{0}
-\sinh{\rho}\sqrt{\frac{1-r_{0}(z)^{2}}{1+r_{0}(z)^{2}\sinh^{2}{\rho}}}f^{(V)}_{2},
\end{eqnarray}
%
%==================================%
%============<Equation>=============%
%
\begin{eqnarray}
F^{(V)}_{2}=\frac{\cosh{\rho}}{\sqrt{1+r_{0}(z)^{2}\sinh^{2}{\rho}}}f^{(V)}_{2}
-\sinh{\rho}\sqrt{\frac{1-r_{0}(z)^{2}}{1+r_{0}(z)^{2}\sinh^{2}{\rho}}}f^{(V)}_{0},
\end{eqnarray}
%
%==================================%
and
%============<Equation>=============%
%
\begin{eqnarray}
~F^{(V)}_{1}=f^{(V)}_{1},~~F^{(V)}_{3}=f^{(V)}_{3},~~F^{(V)}_{L}=H^{(V)}_{L}.
\end{eqnarray}
%
%==================================%
The decoupling perturbation variables are expanded by
%============<Equation>=============%
%
\begin{eqnarray}
F^{(V)}_{I}=\sum_{k\geq 0}\frac{{}^{(k)}F^{(V)}_{I}}{n^{k}},~
F^{(V)}_{T,L}=\sum_{k\geq 0}\frac{{}^{(k)}F_{L}^{(V)}}{n^{k}}.
\end{eqnarray}
%
%==================================
Changing coordinate by $\sR=\cosh^{2}{\rho}$, the perturbation equation at each order becomes 
%============<Equation>=============%
%
\begin{eqnarray}
\frac{\partial}{\partial\sR}\sR(\sR-1)\frac{\partial}{\partial\sR}{}^{(k)}F^{(V)}_{0}-\frac{{}^{(k)}F^{(V)}_{0}}{4\sR(\sR-1)}={}^{(k)}\mathcal{S}^{(V)}_{0}, \\
\frac{\partial}{\partial\sR}\sR(\sR-1)\frac{\partial}{\partial\sR}{}^{(k)}F^{(V)}_{1}-\frac{{}^{(k)}F^{(V)}_{1}}{4\sR(\sR-1)}={}^{(k)}\mathcal{S}^{(V)}_{1}, \\
\frac{\partial}{\partial\sR}\sR(\sR-1)\frac{\partial}{\partial\sR}{}^{(k)}F^{(V)}_{2}={}^{(k)}\mathcal{S}^{(V)}_{2}, \\
\frac{\partial}{\partial\sR}\sR(\sR-1)\frac{\partial}{\partial\sR}{}^{(k)}F^{(V)}_{3}={}^{(k)}\mathcal{S}^{(V)}_{3}.
\end{eqnarray}
%
%==================================%
The boundary condition is the same with one of the scalar type perturbation, the ingoing boundary condition on the horizon (\ref{QNMBCHo}) 
and decoupling mode condition (\ref{QNMBCIn}) at asymptotic infinity of near zone. 

\paragraph{Leading order $(k=0)$}
The regular solution at the leading order is
%============<Equation>=============%
%
\begin{eqnarray}
{}^{(0)}F^{V}_{0}={}^{(0)}F^{V}_{1}=\frac{W_{(0)}}{\sqrt{\sR(\sR-1)}},~~
{}^{(0)}F^{V}_{2}={}^{(0)}F^{V}_{3}=0.
\end{eqnarray}
%
%==================================%
The leading order solution has an undetermined function $W_{(0)}$ by the boundary condition.

\paragraph{Next-to leading order $(k=1)$}
The regular solution at the next to leading order is
%============<Equation>=============%
%
\begin{eqnarray}
{}^{(1)}F^{V}_{0}=\frac{W_{(1)}}{\sqrt{\sR(\sR-1)}}+{}^{(1)}\hat{F}^{V}_{0},~
{}^{(1)}F^{V}_{1}=\frac{W_{(1)}}{\sqrt{\sR(\sR-1)}}+{}^{(1)}\hat{F}^{V}_{1},~
\end{eqnarray}
%
%==================================%
and
%============<Equation>=============%
%
\begin{eqnarray}
{}^{(0)}F^{V}_{2}={}^{(1)}\hat{F}^{V}_{2},~
{}^{(0)}F^{V}_{3}={}^{(1)}\hat{F}^{V}_{3},
\end{eqnarray}
%
%==================================%
where terms with the hat come from the integrations of source terms. 
The next to leading order solution has an undetermined function $W_{(1)}$ by the boundary condition again.
At this order we obtain the non-trivial condition for $W_{(0)}$ as
%============<Equation>=============%
%
\begin{eqnarray}
&&
\cos{\theta}\sin^{2}{\theta}(1+a^{2}\cos^{2}{\theta})W'_{(0)}(\theta) \notag \\
&&~~~~~~~~
+\Bigl[
-j(1+a^{2}\cos^{2}{\theta})+i\cos^{2}{\theta}(a^{4}\cos^{2}{\theta}(\omega-i) 
-a^{3}m\cos^{2}{\theta} \notag \\
&&~~~~~~~~~~~~~~
+a^{2}(\cos^{2}{\theta}(\omega+i)+\omega-2i) +\omega-am )
\Bigr]W_{0}(\theta)=0.
\end{eqnarray}
%
%==================================%
This condition can be solved explicitly as
%============<Equation>=============%
%
\begin{eqnarray}
W_{(0)}(\theta)=\frac{\cos^{j}{\theta}\sin^{{}^{(0)}\delta_{\omega}^{V}}{\theta}}{1+a^{2}\cos^{2}{\theta}},
\end{eqnarray}
%
%==================================%
where
%============<Equation>=============%
%
\begin{eqnarray}
{}^{(0)}\delta^{V}_{\omega}=-(1+a^{2})(j-1)-iam+i(1+a^{2})\omega.
\label{wv0}
\end{eqnarray}
%
%==================================%
The harmonics condition on $W_{0}(\theta)$ is eq. (\ref{QNMcond}) as
%============<Equation>=============%
%
\begin{eqnarray}
{}^{(0)}\delta^{V}_{\omega}=\ell-j,
\label{regconV}
\end{eqnarray}
%
%==================================%
and $\ell$ is parameterized by the non-negative integer $k_{V}$ by
%============<Equation>=============%
%
\begin{eqnarray}
\ell = j+|m|+2k_{V}.
\end{eqnarray}
%
%==================================%
This harmonics condition gives the leading order QNM frequency for the vector type perturbation by
%============<Equation>=============%
%
\begin{eqnarray}
\omega^{V}_{\text{LO}}=\frac{am}{1+a^{2}}- i\left( j-1 +\frac{\ell-j}{1+a^{2}} \right).
\end{eqnarray}
%
%==================================%

\paragraph{Next-to-next-to leading order}
At this order we can obtain the solution and undetermined function in the same manner. Furthermore we find 
the non-trivial condition on $W_{(1)}(\theta)$. Although we omit the explicit form of it since it is not so illuminative, 
the solution of the non-trivial condition gives the following behavior around $\theta=0$ on the horizon $\sR=1$
%============<Equation>=============%
%
\begin{eqnarray}
F^{V}_{(0)}=\frac{\theta^{\delta^{V}_{\omega}}}{\sqrt{\sR-1}}(1+O(\theta,1/n^{2}))
\end{eqnarray}
%
%==================================%
where
%============<Equation>=============%
%
\begin{eqnarray}
\delta^{V}_{\omega}={}^{(0)}\delta^{V}_{\omega}+\frac{{}^{(1)}\delta^{V}_{\omega}}{n}.
\end{eqnarray}
%
%==================================%
${}^{(0)}\delta^{V}_{\omega}$ is given in eq. (\ref{wv0}). ${}^{(1)}\delta^{V}_{\omega}$ is
%============<Equation>=============%
%
\begin{eqnarray}
&&
{}^{(1)}\delta^{V}_{\omega}=a^{2}(-a^{2}(j-1)^{2}+2ia(j-1)m+j^{2}-2j-m^{2}+1)\notag \\
&&~~~~~~~
-a(2a(j-1)+im\log{(1+a^{2})})+\omega(i(a^{2}-1)\log{(1+a^{2})}\notag \\
&&~~~~~~~
+2ia^{2}(a^{2}(j-1)+iam+j-1)) +(1+a^{2})^{2}\omega^{2}.
\end{eqnarray}
%
%==================================%
Then the harmonics condition gives 
%============<Equation>=============%
%
\begin{eqnarray}
{}^{(0)}\delta^{V}_{\omega}+\frac{{}^{(1)}\delta^{V}_{\omega}}{n}=\ell-j.
\end{eqnarray}
%
%==================================%
This condition is solved by
%============<Equation>=============%
%
\begin{eqnarray}
\omega^{V} = \omega^{V}_{\text{LO}} +\frac{\omega^{V}_{\text{NLO}}}{n}+O(1/n^{2}),
\end{eqnarray}
%
%==================================%
where
%============<Equation>=============%
%
\begin{eqnarray}
&&
\omega^{V}_{\text{NLO}} =\frac{2 a m}{1+a^{2}}\Bigl(
\ell -1 +a^{2}(j-1) +\log{(1+a^{2})}
\Bigr) \notag \\
&&~~~~~~~~
-i\Biggl(
(j-1)^{2} +\frac{(1+a^{2})(\ell-j)(\ell+j-2)-2a^{2}m^{2}}{(1+a^{2})^{2}} \notag \\
&&~~~~~~~~~~~~~~~~~~~~~
+\left(
j-1 -\frac{\ell-j}{1+a^{2}}+\frac{2(\ell-j)}{(1+a^{2})^{2}}
\right)\log{(1+a^{2})}
\Biggr).
\end{eqnarray}
%
%==================================%

\section{Explicit forms of $c_{i}(\theta)$ and $d_{i}(\theta)$} \label{cd}

The scalar type perturbation can satisfy the boundary condition on the horizon at the next to next to the leading order if the
integration functions are solutions of following equations: 
%============<Equation>=============%
%
\begin{eqnarray}
&&
c_{1}(\theta)C_{(1)}''(\theta)+c_{2}(\theta)C_{(1)}'(\theta)+c_{3}(\theta)C_{(1)}(\theta) \notag \\
&&~~~~~~~~~~~~~~
+c_{4}(\theta)D_{(0)}''(\theta)+c_{5}(\theta)D_{(0)}'(\theta)+c_{6}(\theta)D_{(0)}(\theta)=0,
\end{eqnarray}
%
%==================================%
and
%============<Equation>=============%
%
\begin{eqnarray}
&&
d_{1}(\theta)D_{(0)}'''(\theta)+d_{2}(\theta)D_{(0)}''(\theta)+d_{3}(\theta)D_{(0)}'(\theta) \notag \\
&&~~~~~~~~~~~~~~
+d_{4}(\theta)D_{(0)}(\theta)+d_{5}(\theta)C_{(1)}'(\theta)+d_{6}(\theta)C_{(1)}(\theta)=0.
\end{eqnarray}
%
%==================================%
Each coefficients are following:
%============<Equation>=============%
%
\begin{eqnarray}
c_{1}(\theta)=\frac{\cos^{2}{\theta} \sin^{3}{\theta}\left(1+a^2 \cos{\theta}^2\right)}{2 \left(1+a^2\right)^{5/2}},
\end{eqnarray}
%
%==================================%
%============<Equation>=============%
%
\begin{eqnarray}
&&
c_{2}(\theta)=
\frac{\cos{\theta} \sin^{2}{\theta}}{2 \left(1+a^2\right)^{5/2}}
\Bigl[
2 a^2 \cos^{4}{\theta} \left(a^2 (j-i \omega )+i a m-i \omega +2\right) \notag \\
&&~~~~~~~~~
+\cos^{2}{\theta}
   \left(a^2 (4 j-2 i \omega -5)+2 i a m-2 i \omega -2\right)+2 j+1
\Bigr],
\end{eqnarray}
%
%==================================%
%============<Equation>=============%
%
\begin{eqnarray}
&&
c_{3}(\theta)=
\frac{\sin^{2}{\theta}}{2 \left(1+a^2\right)^{5/2}}
\Bigl[
 a^2 \cos^{6}{\theta} \left(a^2 (j-i \omega )+i a m-i \omega +2\right)^2 \notag \\
&&~~~~~~~
+\cos^{2}{\theta}\left(a^2 \left(3 j^2+j (-4-2 i \omega )-1\right)+2 i a j m+j (-4-2 i
   \omega )+4\right)
 \notag \\
&&~~~~~~~
+\cos^{4}{\theta} \Bigl( a^4 \left(3 j^2+j
   (-6-4 i \omega )-\omega ^2+6 i \omega +4\right)
+2 a^3 m (2 i j+\omega -3 i) \notag \\
&&~~~~~~~~~~~~~~~
-a^2 \left(m^2+2 \left(2 i (j-2) \omega
   +\omega ^2-1\right)\right)+2 a m (\omega -i)-\omega  (\omega -2 i)\Bigr) \notag \\
&&~~~~~~~
+j^2+2 j-3
\Bigr],
\end{eqnarray}
%
%==================================%
%============<Equation>=============%
%
\begin{eqnarray}
&&
c_{4}(\theta)=
-\frac{\sin^{2}{\theta}}{\left(1+a^2\right)^4 \cos{\theta}}
\Bigl[
\cos^{2}{\theta} \left(a^2 (-5 j+3 i \omega +2)-3 i a m+j+3 i \omega +5\right) \notag \\
&&~~~~~~~~~~~
+\cos^{4}{\theta} \left(a^4 (-2 j+3 i \omega +3)-3 i a^3 m+a^2 (j+2 i \omega +1)-i \omega
   \right) \notag \\
&&~~~~~~~~~~~
-i a^2
   \left(1+a^2\right) \omega  \cos^{6}{\theta}-3 j-5
\Bigr],
\end{eqnarray}
%
%==================================%
%============<Equation>=============%
%
\begin{eqnarray}
&&
c_{5}(\theta)=
\frac{\sin{\theta}}{\left(1+a^2\right)^4 \cos^{2}{\theta} \left(1+a^2 \cos^{2}{\theta}\right)}
\Bigl[
\cos^{2}{\theta}\Bigl(a^2 \left(10 j^2+2 j (8-3 i \omega )-7 i \omega -3\right) \notag \\
&&~~~~~~~~~
+i a (6 j+7) m-2
   j^2+j (-19-6 i \omega )-7 i \omega -9\Bigr) \notag \\
&&~~~~
+a^4 \left(a^2+1\right) \omega  \cos^{10}{\theta}\left(a^2 (2 i j+2 \omega
   -i)-2 a m+2 \omega +3 i\right) \notag \\
&&~~~~
+\cos^{4}{\theta} \Bigl(a^4 \left(12 j^2+j (-6-16 i \omega )-3 \omega ^2-i \omega -9\right)+a^3
   m (16 i j+6 \omega +i) \notag \\
&&~~~~~~~~~~~
-a^2 \left(6 j^2+4 j (7+3 i \omega )+3 m^2+6 \omega ^2-9 i \omega +2\right) \notag \\
&&~~~~~~~~~~~
+a m (-2 i j+6
   \omega -9 i) +j (5+4 i \omega )-3 \omega ^2+10 i \omega +4\Bigr) \notag \\
&&~~~~
+a^2 \cos^{8}{\theta} \Bigl(a^6 \left(j^2+j (-3-4 i \omega
   )-3 \omega ^2+4 i \omega +2\right)+2 a^5 m (2 i j+3 \omega -2 i) \notag \\
&&~~~~~~~~~~~
-a^4 \left(2 j^2+j (-2-4 i \omega )+3 m^2+2 \omega 
   (\omega +i)\right)+a^3 m (-2 i j+2 \omega +i) \notag \\
&&~~~~~~~~~~~
+a^2 \left(8 i j \omega +j+5 \omega ^2-5 i \omega -2\right)+a m (-4
   \omega +i)+\omega  (4 \omega +i)\Bigr) \notag \\
&&~~~~
+\cos^{6}{\theta}\Bigl(a^6 \left(6 j^2+j (-12-14 i \omega )-6 \omega ^2+10 i \omega
   +5\right)+2 a^5 m (7 i j+6 \omega -5 i) \notag \\
&&~~~~~~~~~~~
-a^4 \left(6 j^2+j (7+4 i \omega )+6 m^2+10 \omega ^2-15 i \omega
   -15\right)+2 a^3 m (-2 i j+5 \omega -4 i) \notag \\
&&~~~~~~~~~~~
+a^2 \left(2 j (3+5 i \omega )-2 \omega ^2+3 i \omega +6\right)+a m (-2
   \omega +i)+2 \omega  (\omega -i)\Bigr) \notag \\
&&~~~~
+3 j^2+13 j+5
\Bigr],
\end{eqnarray}
%
%==================================%
%============<Equation>=============%
%
\begin{eqnarray}
&&
d_{1}(\theta)=
-\frac{\cos{\theta} \sin^{3}{\theta}}{\left(1+a^2\right)^2  \left(1+a^2 \cos^{2}{\theta}\right)},
\end{eqnarray}
%
%==================================%
%============<Equation>=============%
%
\begin{eqnarray}
&&
d_{2}(\theta)=
-\frac{\sin^{2}{\theta}}{\left(1+a^2\right)^2  \left(1+a^2
   \cos^{2}{\theta}\right)^2}
\Bigl[
3 j+1\notag\\
&&~~~~~~~~
+\cos^{2}{\theta}\left(a^2
   (6 j-3 i \omega -4)+3 i a m-3 i \omega -1\right) \notag \\
&&~~~~~~~~
+a^2 \cos^{4}{\theta} \left(3 a^2 (j-i \omega -1)+3 i a m-3 i \omega +1\right) 
\Bigr],
\end{eqnarray}
%
%==================================%
%============<Equation>=============%
%
\begin{eqnarray}
&&
d_{3}(\theta)=
-\frac{\sin{\theta}}{\left(1+a^2\right)^2
   \cos{\theta} \left(1+a^2 \cos^{2}{\theta}\right)^3}
\Bigl[
\cos^{2}{\theta} \Bigl(a^2 \left(12 j^2-6 i j \omega +j+i \omega +3\right) \notag \\
&&~~
+i a (6 j-1) m-6 i j
   \omega -6 j+i \omega -1\Bigr)+a^4 \cos^{8}{\theta} \Bigl(a^4 \left(3 j^2-6 i j \omega -4 j-3 \omega ^2+5 i \omega
   +2\right) \notag \\
&&~~
+a^3 m (6 i j+6 \omega -5 i)+a^2 \left(-6 i j \omega +4 j-3 m^2-6 \omega ^2+2 i \omega -2\right)+3 a m (2
   \omega +i )\notag \\
&&~~
-3 \omega  (\omega +i)\Bigr)
+a^2 \cos^{6}{\theta}\Bigl(a^4 \left(12 j^2-18 i j \omega -17 j-6 \omega ^2+15 i
   \omega +7\right)+3 a^3 m (6 i j+4 \omega -5 i) \notag \\
&&~~
+a^2 \left(-18 i j \omega +2 j-6 m^2-12 \omega ^2+13 i \omega
   +5\right)+2 a m (6 \omega +i)-6 \omega ^2-2 i \omega +8\Bigr) \notag \\
&&~~
+\cos^{4}{\theta}\Bigl(a^4 \left(18 j^2+j (-17-18 i \omega )-3
   \omega ^2+11 i \omega -1\right)+a^3 m (18 i j+6 \omega -11 i)\notag \\
&&~~
-a^2 \left(2 j (4+9 i \omega )+3 m^2+6 \omega ^2-12 i
   \omega +10\right)+a m (6 \omega -i)+(-3 \omega +i) \omega \Bigr) \notag \\
&&~~
+3 j^2+5 j+1
\Bigr],
\end{eqnarray}
%
%==================================%
%============<Equation>=============%
%
\begin{eqnarray}
d_{5}(\theta)=
\frac{\sqrt{1+a^{2}}  \cos{\theta}^2 \sin^{2}{\theta}}{\left(1+a^2\right) \left(1+a^2 \cos^{2}{\theta}\right)},
\end{eqnarray}
%
%==================================%
%============<Equation>=============%
%
\begin{eqnarray}
d_{6}(\theta)=
\frac{ \cos{\theta} \sin{\theta} \left(\cos^{2}{\theta} \left(a^2 (j-i \omega -2)+i a m-i \omega
   \right)+j-1\right)}{\sqrt{1+a^{2}} \left(1+a^2 \cos^{2}{\theta}\right)}.
\end{eqnarray}
%
%==================================%
The coefficients $c_{6}(\theta)$ and $d_{4}(\theta)$ have very lengthy forms. They are written as
%============<Equation>=============%
%
\begin{eqnarray}
c_{6}(\theta)=\frac{1}{\left(a^2+1\right)^4 \cos^{3}{\theta} \left(1+a^2 \cos^{2}{\theta}\right)^2}\sum_{k=0}^{k=7}
\hat{c}_{6}^{(k)}\cos^{2k}{\theta},
\end{eqnarray}
%
%==================================%
and
%============<Equation>=============%
%
\begin{eqnarray}
d_{4}(\theta)=\frac{1}{\left(1+a^2\right)^2 \cos^{2}{\theta}
\left(1+a^2
   \cos^{2}{\theta}\right)^4}\sum_{k=0}^{k=6}
\hat{d}_{4}^{k}\cos^{2k}{\theta},
\end{eqnarray}
%
%==================================%
where
%============<Equation>=============%
%
\begin{eqnarray}
\hat{c}_{6}^{(0)}=j \left(j^2+8 j+12\right),
\end{eqnarray}
%
%==================================%
%============<Equation>=============%
%
\begin{eqnarray}
&&
\hat{c}_{6}^{(1)}=j \Bigl(a^2 \left(5 j^2-3 i j \omega +23 j-10 i \omega +16\right) \notag \\
&&~~~~~~~~~~~~~
+i a (3 j+10) m-j^2-3 i j (\omega -5 i)-10 i \omega
   -28\Bigr),
\end{eqnarray}
%
%==================================%
%============<Equation>=============%
%
\begin{eqnarray}
&&
\hat{c}_{6}^{(2)}=a^4 \left(10 j^3+j^2 (18-13 i \omega )-j \left(3 \omega ^2+19 i \omega +8\right)-2 \left(\omega ^2-2 i \omega
   +4\right)\right) \notag \\
&&~~
+a^3 m \left(13 i j^2+6 j \omega +19 i j+4 \omega -4 i\right)-a^2 \Bigl( 5 j^3+5 j^2 (9+2 i \omega) \notag \\
&&~~
+j \left(3 m^2+6 \omega ^2+44\right)+2 \left(m^2+2 \omega ^2-2 i \omega +4\right)\Bigr)
+a m \left(-2 i j^2+j (6
   \omega -17 i)+4 \omega \right) \notag \\
&&~~
+j^2 (7+3 i \omega )+j \left(-3 \omega ^2+19 i \omega +20\right)-2 \omega ^2,
\end{eqnarray}
%
%==================================%
%============<Equation>=============%
%
\begin{eqnarray}
&&
\hat{c}_{6}^{(3)}=a^6 \left(10 j^3+j^2 (-4-22 i \omega )-j \left(11 \omega ^2+14\right)+i \left(\omega ^3+8 \omega +8
   i\right)\right) \notag \\
&&~~~
+a^5 m \left(22 i j^2+22 j \omega -i \left(3 \omega ^2+8\right)\right)+a^4 \Bigl(-10 j^3+j^2 (-42-9
   i \omega ) \notag \\
&&~~~
+j \left(-11 m^2+(-19 \omega +39 i) \omega \right)+i \left(3 m^2+4\right) \omega +3 i \omega ^3+4 \omega
   ^2+8\Bigr) \notag \\
&&~~~
-i a^3 m \left(8 j^2+2 j (19+9 i \omega )+m^2+6 \omega ^2-8 i \omega -4\right)+a^2 \Bigl(j^2 (22+13 i
   \omega ) \notag \\
&&~~~
+j \left(m^2-5 \omega ^2+30 i \omega +34\right)+m^2 (4+3 i \omega )+3 i \omega ^3+8 \omega ^2-4 i \omega
   +16\Bigr) \notag \\
&&~~~
-a m (j (4 \omega -6 i)+(8+3 i \omega ) \omega )+3 j \omega ^2-9 i j \omega -4 j+i \omega ^3+4 \omega ^2,
\end{eqnarray}
%
%==================================%
%============<Equation>=============%
%
\begin{eqnarray}
&&
\hat{c}_{6}^{(4)}=a^8 \left(5 j^3+j^2 (-10-18 i \omega )-3 j \omega  (5 \omega -6 i)+3 i \omega ^3+7 \omega ^2+8\right) \notag \\
&&~~~
+a^7 m \left(18 i
   j^2+6 j (5 \omega -3 i)+(-14-9 i \omega ) \omega \right)+a^6 \Bigl(-10 j^3+j^2 (-6+4 i \omega ) \notag \\
&&~~~+j \left(-15 m^2-19
   \omega ^2+28 i \omega +18\right)+m^2 (7+9 i \omega )+8 i \omega ^3+15 \omega ^2-10 i \omega +32\Bigr) \notag \\
&&~~~
-i a^5 m
   \left(12 j^2+4 j (5+4 i \omega )+3 m^2+16 \omega ^2-21 i \omega -12\right)+a^4 \Bigl(j^2 (24+22 i \omega ) \notag \\
&&~~~
+j
   \left(3 m^2+7 \omega ^2-10 i \omega +8\right)+m^2 (6+8 i \omega )+6 i \omega ^3+7 \omega ^2-12 i \omega
   +16\Bigr) \notag \\
&&~~~
-a^3 m (14 j (\omega -i)+(4+5 i \omega ) \omega )+a^2 \Bigl(j \left(11 \omega ^2-20 i \omega
   -6\right)+m^2 (-1-i \omega ) \notag \\
&&~~~
-3 \omega ^2-2 i \omega -8\Bigr)+a m (3+2 i \omega ) \omega +(-2-i \omega ) \omega ^2,
\end{eqnarray}
%
%==================================%
%============<Equation>=============%
%
\begin{eqnarray}
&&
\hat{c}_{6}^{(5)}=a^2 \Biggl( a^8 \left(j^3+j^2 (-3-7 i \omega )+j \left(-9 \omega ^2+10 i \omega +2\right)
+\omega  \left(3 i \omega ^2+6
   \omega -4 i\right)\right) \notag \\
&&~~~
+a^7 m \left( 7 i j^2+2 j (9 \omega -5 i)-i (-3 \omega +2 i)^2\right)+a^6 \Bigl( -5 j^3+j^2
   (9+11 i \omega ) \notag \\
&&~~~
-3 j \left(3 m^2+\omega  (\omega +4 i)\right)+m^2 (6+9 i \omega )+6 i \omega ^3+\omega ^2-4 i
   \omega -8\Bigr) \notag \\
&&~~~
-i a^5 m \left(8 j^2-6 j+3 m^2+12 \omega ^2-i \omega -4\right)+a^4 \Bigl(2 j^2 (5+9 i \omega ) \notag \\
&&~~~
+j
   \left(3 m^2+21 \omega ^2-32 i \omega -6\right)+2 i \left(3 m^2-1\right) \omega -17 \omega ^2-24\Bigr) \notag \\
&&~~~
+a^3 m
   \left(-2 j (9 \omega -5 i)+3 i \omega ^2+14 \omega -4 i\right)+a^2 \Bigl(5 j (3 \omega -2 i) \omega +m^2 (-2-3 i
   \omega ) \notag \\
&&~~~
-6 i \omega ^3-13 \omega ^2-16\Bigr)+3 a m (1+2 i \omega ) \omega +\omega  \left(-3 i \omega ^2-\omega +2
   i\right)\Biggr),
\end{eqnarray}
%
%==================================%
%============<Equation>=============%
%
\begin{eqnarray}
&&
\hat{c}_{6}^{(6)}=a^4 \Biggl(a^8 \omega  \left(-i j^2-2 j \omega +i j+i \omega ^2+\omega \right)+a^7 m \left(i j^2+j (4 \omega -i)+(-2-3
   i \omega ) \omega \right) \notag \\
&&~~~
+a^6 \left(-j^3+j^2 (3+6 i \omega )-j \left(2 m^2-5 \omega ^2+12 i \omega +2\right)+m^2
   (1+3 i \omega )+(-7 \omega +6 i) \omega \right) \notag \\
&&~~~
-i a^5 m \left(2 j^2+j (-5-6 i \omega )+m^2+9 i \omega +4\right)+a^4
   \Bigl(j^2 (1+7 i \omega )+j \left(m^2+\omega  (16 \omega -9 i)\right) \notag \\
&&~~~
-2 m^2+\omega  \left(-6 i \omega ^2-13 \omega
   +4 i\right)\Bigr)+a^3 m \left(j (-10 \omega +2 i)+9 i \omega ^2+8 \omega -4 i\right) \notag \\
&&~~~
+a^2 \left(j \left(9 \omega
   ^2+4 i \omega +2\right)+m^2 (-1-3 i \omega )+\omega  \left(-8 i \omega ^2-\omega +2 i\right)\right) \notag \\
&&~~~
+3 i a m \omega 
   (2 \omega +i)+\omega  \left(-3 i \omega ^2+4 \omega +4 i\right)\Biggr),
\end{eqnarray}
%
%==================================%
%============<Equation>=============%
%
\begin{eqnarray}
&&
\hat{c}_{6}^{(7)}=a^6 \left(a^2+1\right) \omega  \Bigl(a^4 \left(i j^2+j (2 \omega -i)+(-1-i \omega ) \omega \right)+a^3 m (-2 j+2 i
   \omega +1) \notag \\
&&~~~
+a^2 \left(j (2 \omega +3 i)-i \left(m^2+2 \omega ^2+2 i \omega +2\right)\right)+a m (-3+2 i \omega ) \notag \\
&&~~~~
-i
   \omega ^2+3 \omega +2 i\Bigr),
\end{eqnarray}
%
%==================================%
%============<Equation>=============%
%
\begin{eqnarray}
\hat{d}_{4}^{(0)}=-j (j+2)^2,
\end{eqnarray}
%
%==================================%
%============<Equation>=============%
%
\begin{eqnarray}
&&
\hat{d}_{4}^{(1)}=j \Bigl(a^2 \left(-6 j^2+j (-11+3 i \omega )+2 i \omega -8\right)-i a (3 j+2) m \notag \\
&&~~~~~
+j (5+3 i \omega )+2 i \omega +8\Bigr),
\end{eqnarray}
%
%==================================%
%============<Equation>=============%
%
\begin{eqnarray}
&&
\hat{d}_{4}^{(2)}=a^4 \left(-15 j^3+j^2 (-3+15 i \omega )+j \left(3 \omega ^2-5 i \omega +2\right)-2 \left(\omega ^2-2 i \omega
   +4\right)\right) \notag \\
&&~~~~
+a^3 m \left(-15 i j^2-6 j \omega +5 i j+4 \omega -4 i\right)+a^2 \Bigl(j^2 (17+15 i \omega )+j
   \left(3 m^2+6 \omega ^2-10 i \omega +26\right) \notag \\
&&~~~~
-2 \left(m^2+2 \omega ^2-2 i \omega +4\right)\Bigr)+a m (4 \omega +j
   (-6 \omega +5 i))+j \left(3 \omega ^2-5 i \omega -4\right)-2 \omega ^2,
\end{eqnarray}
%
%==================================%
%============<Equation>=============%
%
\begin{eqnarray}
&&
\hat{d}_{4}^{(3)}=a^6 \left(-20 j^3+6 j^2 (3+5 i \omega )+2 j \left(6 \omega ^2-15 i \omega +5\right)-i \omega  \left(\omega ^2-10 i
   \omega -4\right)\right) \notag \\
&&~~~~
+a^5 m \left(-30 i j^2-6 j (4 \omega -5 i)+3 i \omega ^2+20 \omega -4 i\right)+a^4 \Bigl(6
   j^2 (3+5 i \omega ) \notag \\
&&~~~~
+4 j \left(3 m^2+6 \omega ^2-10 i \omega +6\right)+m^2 (-10-3 i \omega )-3 i \omega ^3-20 \omega
   ^2-8 i \omega +16\Bigr) \notag \\
&&~~~~
+a^3 m \left(j (-24 \omega +10 i)+i \left(m^2+6 \omega ^2-20 i \omega +12\right)\right)+a^2
   \Bigl(2 j \left(6 \omega ^2-5 i \omega -9\right) \notag \\
&&~~~~
-3 i \left(m^2+4\right) \omega-3 i \omega ^3-10 \omega
   ^2+16\Bigr)+3 i a m \omega ^2-i \omega ^3,
\end{eqnarray}
%
%==================================%
%============<Equation>=============%
%
\begin{eqnarray}
&&
\hat{d}_{4}^{(4)}=-a^2 \Biggl(a^6 \left(15 j^3+j^2 (-22-30 i \omega )-2 j \left(9 \omega ^2-20 i \omega +1\right)+3 i \omega ^3+16 \omega
   ^2-4 i \omega +8\right) \notag \\
&&~~~
+a^5 m \left(30 i j^2+4 j (9 \omega -10 i)-9 i \omega ^2-32 \omega +4 i\right)+a^4 \Bigl(j^2
   (-2-30 i \omega ) \notag \\
&&~~~
-2 j \left(9 m^2+18 \omega ^2-20 i \omega +2\right)+m^2 (16+9 i \omega )+9 i \omega ^3+30 \omega
   ^2+8 i \omega +16\Bigr) \notag \\
&&~~~
-a^3 m \left(-36 j \omega +3 i m^2+18 i \omega ^2+28 \omega +12 i\right)+a^2 \Bigl(j
   \left(26-18 \omega ^2\right)+m^2 (-2+9 i \omega ) \notag \\
&&~~~
+9 i \omega ^3+12 \omega ^2+4 i \omega +16\Bigr)+a m \left(-9 i
   \omega ^2+4 \omega +8 i\right)+3 i \omega ^3-2 \omega ^2-8 i \omega +8\Biggr),
\end{eqnarray}
%
%==================================%
%============<Equation>=============%
%
\begin{eqnarray}
&&
\hat{d}_{4}^{(5)}=a^4 \Biggl(a^6 \left(-6 j^3+3 j^2 (3+5 i \omega )+2 j \left(6 \omega ^2-10 i \omega -1\right)+\omega  \left(-3 i \omega
   ^2-10 \omega +4 i\right)\right) \notag \\
&&~~~
+a^5 m \left(-15 i j^2-4 j (6 \omega -5 i)+9 i \omega ^2+20 \omega -4 i\right)+a^4
   \Bigl(j^2 (-7+15 i \omega ) \notag \\
&&~~~
+2 j \left(6 m^2+\omega  (12 \omega -5 i)\right)+m^2 (-10-9 i \omega )-9 i \omega ^3-16
   \omega ^2+8\Bigr) \notag \\
&&~~~
+a^3 m \left(-2 j (12 \omega +5 i)+3 i m^2+18 i \omega ^2+12 \omega +4 i\right)+a^2 \Bigl(2 j
   \left(6 \omega ^2+5 i \omega -7\right) \notag \\
&&~~~
+m^2 (4-9 i \omega )-9 i \omega ^3-2 \omega ^2+4 i \omega +16\Bigr)+i a m
   \left(9 \omega ^2+8 i \omega -8\right) \notag \\
&&~~~
-3 i \omega ^3+4 \omega ^2+8 i \omega +8\Biggr),
\end{eqnarray}
%
%==================================%
%============<Equation>=============%
%
\begin{eqnarray}
&&
\hat{d}_{4}^{(6)}=-a^6 \Biggl(a^6 \left(j^3+j^2 (-1-3 i \omega )-3 j \omega  (\omega -i)+(2+i \omega ) \omega ^2\right) \notag \\
&&~
+a^5 m \left(3 i
   j^2+j (6 \omega -3 i)+(-4-3 i \omega ) \omega \right)+a^4 \Bigl(j^2 (3-3 i \omega ) \notag \\
&&~
-j \left(3 m^2+6 \omega ^2+2 i
   \omega +2\right)+m^2 (2+3 i \omega )+\omega  \left(3 i \omega ^2+2 \omega +4 i\right)\Bigr) \notag \\
&&~
-i a^3 m \left(j (-5+6
   i \omega )+m^2+6 \omega ^2+4\right)+a^2 \Bigl(j \left(-3 \omega ^2-5 i \omega +2\right) \notag \\
&&~
+i \left(m^2 (3 \omega +2
   i)+\omega  \left(3 \omega ^2+2 i \omega +4\right)\right)\Bigr)+a m (4-3 i \omega ) \omega +i \omega ^2 (\omega +2
   i)\Biggr).
\end{eqnarray}
%
%==================================%

\section{Explicit form of $\delta^{S}_{\omega}$} \label{C}

The QNM condition for the scalar type perturbation of the Myers-Perry black hole is
%============<Equation>=============%
%
\begin{eqnarray}
\delta^{S}_{\omega}=\ell-j. 
\end{eqnarray}
%
%==================================%
This condition can be rewritten as
%============<Equation>=============%
%
\begin{eqnarray}
\delta^{S}_{\omega}-\ell+j=0 \Leftrightarrow  \frac{F(\omega)}{G(\omega)}=0,
\end{eqnarray}
%
%==================================%
where
%============<Equation>=============%
%
\begin{eqnarray}
&&
G(\omega)=
\left(1+a^2\right)^3 \Bigl[
a^4 (j-i \omega )^2+2 a^3 m (\omega +i j) \notag \\
&&~~~~~~~~~~~~~
-a^2 \left(j (-2 \ell+2 i \omega +4)+2 i \ell \omega +m^2+2 \omega ^2-4 i\omega -4\right) \notag \\
&&~~~~~~~~~~~~~~~~~
+2 a m (i \ell+\omega -2 i)+(\ell-i \omega -2)^2
\Bigr],
\end{eqnarray}
%
%==================================%
and
%============<Equation>=============%
%
\begin{eqnarray}
F(\omega)=f_{5}\omega^{5}+f_{4}\omega^{4}+f_{3}\omega^{3}+f_{2}\omega^{2}+f_{1}\omega+f_{0}.
\end{eqnarray}
%
%==================================%
Each coefficient is
%============<Equation>=============%
%
\begin{eqnarray}
f_{5}=i \left(1+a^2\right)^5,
\end{eqnarray}
%
%==================================%
%============<Equation>=============%
%
\begin{eqnarray}
f_{4}=-\left(1+a^2\right)^4 \left(a^2 (5 j-2)+5 i a m+5 \ell-6\right),
\end{eqnarray}
%
%==================================%
%============<Equation>=============%
%
\begin{eqnarray}
&&
f_{3}=-i \left(1+a^2\right)^3 \Bigl(a^4 j (10 j-7)+4 i a^3 (5 j-2) m 
+a^2 \bigl(j (20 \ell-21) \notag \\
&&~~
-9 \ell-10 m^2+12\bigr)+2 i a (10 \ell-11) m+10 \ell^2-23
   \ell+12\Bigr),
\end{eqnarray}
%
%==================================%
%============<Equation>=============%
%
\begin{eqnarray}
&&
f_{2}=\left(1+a^2\right)^2 \Bigl(a^6 j^2 (10 j-9)+3 i a^5 j (10 j-7) m+a^4 \bigl(3 j^2 (10 \ell-9) \notag \\
&&~~
-6 j \left(4 \ell+5 m^2-5\right)+2 \left(\ell+6
   m^2-4\right)\bigr)-i a^3 m \bigl(j (57-60 \ell)+27 \ell \notag \\
&&~~
+10 m^2-30\bigr)+a^2 \left(30 j (\ell-1)^2-15 \ell^2-30 \ell m^2+34 \ell+30 m^2-16\right) \notag \\
&&~~
+3 i a
   \left(10 \ell^2-21 \ell+10\right) m+10 \ell^3-33 \ell^2+32 \ell-8\Bigr),
\end{eqnarray}
%
%==================================%
%============<Equation>=============%
%
\begin{eqnarray}
&&
f_{1}=
i \left(1+a^2\right) \Bigl( 5 a^8 (j-1) j^3+2 i a^7 j^2 (10 j-9) m+a^6 j \bigl(5 j^2 (4 \ell-3) \notag \\
&&~~
-3 j \left(7 \ell+10 m^2-8\right)+4 \ell+21
   m^2-12\bigr)-4 i a^5 m \bigl(-3 j^2 (5 \ell-4) \notag \\
&&~~
+j \left(12 \ell+5 m^2-12\right)-\ell-2 m^2+3\bigr)+a^4 \bigl(3 j^2 \left(10 \ell^2-17 \ell+8\right) \notag \\
&&~~
+j\left(-27 \ell^2+\ell \left(56-60 m^2\right)+51 m^2-24\right)+4 \ell^2+3 \ell \left(9 m^2-4\right)+m^2 \left(5 m^2-24\right)\bigr) \notag \\
&&~~
-2 i a^3 m
   \left(-6 j \left(5 \ell^2-9 \ell+4\right)+15 \ell^2+2 \ell \left(5 m^2-14\right)-9 m^2+12\right) \notag \\
&&~~
+a^2 \left(j \left(20 \ell^3-57 \ell^2+52 \ell-12\right)-11
   \ell^3+\ell^2 \left(32-30 m^2\right)+3 \ell \left(19 m^2-8\right)-24 m^2\right) \notag \\
&&~~
+4 i a \left(5 \ell^3-15 \ell^2+13 \ell-3\right) m+\ell \left(5 \ell^3-21 \ell^2+28
   \ell-12\right)\Bigr),
\end{eqnarray}
%
%==================================%
and
%============<Equation>=============%
%
\begin{eqnarray}
&&
f_{0}=-\left(a^2 j+i a m+\ell\right)^2 \Bigl(a^6 (j-1) j^2+3 i a^5 (j-1) j m \notag \\
&&~~
+a^4 \left(3 j^2 (\ell-1)+j \left(-4 \ell-3 m^2+6\right)+2
   \left(\ell+m^2-2\right)\right) \notag \\
&&~~
-i a^3 m \left(j (7-6 \ell)+5 \ell+m^2-6\right)+a^2 \bigl(j \left(3 \ell^2-8 \ell+6\right)-3 \ell^2 \notag \\
&&~~
-3 \ell m^2+10 \ell+4
   m^2-8\bigr)+3 i a \left(\ell^2-3 \ell+2\right) m+(\ell-2)^2 (\ell-1)\Bigr).
\end{eqnarray}
%
%==================================%

%%%%%%%%%%%%%%%%%%%%%%%%%%%%%%%%%%%%%%%%%

\end{document}